\documentclass[numberedappendix]{emulateapj} 

\usepackage{epsfig,color,pifont}

\usepackage{natbib}
\bibpunct{(}{)}{;}{a}{}{,}
\def\spose#1{\hbox to 0pt{#1\hss}}
\def\ltsimm{\mathrel{\spose{\lower 3pt\hbox{$\sim$}}
        \raise 2.0pt\hbox{$<$}}}
\def\gtsimm{\mathrel{\spose{\lower 3pt\hbox{$\sim$}}
        \raise 2.0pt\hbox{$>$}}}

\def\km{{\rm\thinspace km}}
\def\cm{{\rm\thinspace cm}}

\def\s{{\rm\thinspace s}}
\def\yr{{\rm\thinspace yr}}
\def\g{{\rm\thinspace g}}
\def\kmps{\hbox{${\rm\km\s^{-1}\,}$}}

\def\erg{{\rm\thinspace erg}}
\def\Hz{{\rm\thinspace Hz}}

\def\ster{{\rm\thinspace ster}}
\def\ergps{\hbox{${\rm\erg\s^{-1}\,}$}}
\def\Rsol{\hbox{${\rm\thinspace R_{\odot}}$}}
\def\Msol{\hbox{${\rm\thinspace M_{\odot}}$}}

\def\Msolpyr{\hbox{${\rm\Msol\yr^{-1}\,}$}}
\def\pcm{\hbox{${\rm\cm^{-1}\,}$}}
\def\pcm2{\hbox{${\rm\cm^{-2}\,}$}}
\def\pcm3{\hbox{${\rm\cm^{-3}\,}$}}
\def\ergpscm3Hz{\hbox{${\rm\ergps\cm^{-3}\Hz^{-1}\,}$}}
\def\ergpscm3Hzster{\hbox{${\rm\ergps\cm^{-3}\Hz^{-1}\ster^{-1}\,}$}}
\def\gpcm3{\hbox{${\rm\g\cm^{-3}\,}$}}
\def\ergpcm2{\hbox{${\rm\erg\cm^{-2}\,}$}}
\def\ergpcm3{\hbox{${\rm\erg\cm^{-3}\,}$}}
\def\phpscm2{\hbox{${\rm photons\s^{-1}\cm^{-2}\,}$}}

\def\etacar{$\eta\thinspace\rm{Car}\;$}

\def\XMM{\textit{XMM-Newton}}

\def\aap{{\rm A\&A}}
\def\apj{{\rm ApJ}}
\def\apjs{{\rm ApJS}}
\def\aj{{\rm AJ}}
\def\mnras{{\rm MNRAS}}

\def\araa{{\rm ARA\&A}}

\def\cpc{{\rm Comp.~Phys.~Comm.}}

\def\aspc{{\rm ASPC}}

\begin{document} 

\title{Spiralling out of control: 3D hydrodynamical modelling of the
  colliding winds in $\eta\thinspace$Carinae}

\shorttitle{3D hydrodynamical modelling of $\eta\thinspace$Carinae}

\author{E.~R.~Parkin\altaffilmark{1,2,6}\thanks, J.~M.~Pittard\altaffilmark{2}, M.~F.~Corcoran\altaffilmark{3,4}, and K.~Hamaguchi\altaffilmark{3,5}}
\affil{$^{1}$Institut d'Astrophysique et de G\'{e}ophysique, Universit\'{e} de
    Li\`{e}ge, 17, All\'{e}e du 6 Ao\^{u}t, B5c, B-4000 Sart Tilman,
    Belgium} 
\affil{$^{2}$School of Physics and Astronomy, The University of
    Leeds, Woodhouse Lane, Leeds LS2 9JT, UK}
\affil{$^{3}$CRESST and X-ray Astrophysics Laboratory, NASA/GSFC, 
  Greenbelt, MD 20771, USA}
\affil{$^{4}$Universities Space Research Association, 10211 Wisconsin Circle,
  Suite 500 Columbia, MD 21044, USA}
\affil{$^{5}$Department of Physics, University of Maryland, Baltimore County, 
  1000 Hilltop Circle, Baltimore, MD 21250, USA}
\affil{$^{6}$Current address: Research School of Astronomy and
  Astrophysics, Mount Stromlo Observatory, Australian National
  University, Cotter Road, Weston Creek, ACT 2611, Australia}

\email{parkin@mso.anu.edu.au}

\shortauthors{E.~R.~Parkin, J.~M.~Pittard, M.~F.~Corcoran, \& K.~Hamaguchi}





\label{firstpage}

\begin{abstract}
Three dimensional (3D) adaptive-mesh refinement (AMR) hydrodynamical
simulations of the wind-wind collision between the enigmatic
super-massive star \etacar and its mysterious companion star are
presented which include radiative driving of the stellar winds,
gravity, optically-thin radiative cooling, and orbital
motion. Simulations with static stars with a periastron passage
separation reveal that the preshock companion star's wind speed is
sufficiently reduced that radiative cooling in the postshock gas
becomes important, permitting the runaway growth of non-linear thin
shell (NTSI) instabilities which massively distort the WCR. However,
large-scale simulations which include the orbital motion of the stars,
show that orbital motion reduces the impact of radiative inhibition,
and thus increases the acquired preshock velocities. As such, the
postshock gas temperature and cooling time see a commensurate
increase, and sufficient gas pressure is preserved to stabilize the
WCR against catastrophic instability growth. We then compute synthetic
X-ray spectra and lightcurves and find that, compared to previous
models, the X-ray spectra agree much better with {\it XMM-Newton}
observations just prior to periastron. The narrow width of the 2009
X-ray minimum can also be reproduced. However, the models fail to
reproduce the extended X-ray mimimum from previous cycles. We conclude
that the key to explaining the extended X-ray minimum is the rate of
cooling of the companion star's postshock wind. If cooling is rapid
then powerful NTSIs will heavily disrupt the WCR. Radiative inhibition
of the companion star's preshock wind, albeit with a stronger
radiation-wind coupling than explored in this work, could be an
effective trigger.
\end{abstract}


\maketitle

\section{Introduction}
\label{sec:intro}
Of the known massive stars in our galaxy, \etacar is possibly the
largest and finest example of a pre-hypernova candidate, presenting a
rare but exceptional opportunity to test our current understanding of
stellar evolution in the upper Hertzsprung-Russell diagram. Yet, this
gem is not without its flaws as, since the ``Great Eruption'' which
formed the Homunculus nebula and the more recent eruption which formed
the Little Homunculus \citep{Ishibashi:2003}, \etacar has been
enshrouded by a dusty cocoon which complicates observations in the UV
and optical. Fortunately, \etacar is extremely bright at X-ray
wavelengths which suffer less from extinction, thus providing an
invaluable probe of the inner nebula.

The exceptional monitoring of \etacar at X-ray wavelengths
\citep{Corcoran:2001, Corcoran:2004, Corcoran:2005, Corcoran:2010,
  Hamaguchi:2007, Henley:2008, Leyder:2008, Leyder:2010, Pian:2009}
has characterised the periodic variability as indicative of a highly
eccentric ($e\simeq 0.9$), long-period ($\sim5.54\;$yr) binary system
(Table~\ref{tab:system_parameters}). The binary hypothesis appears to
be well supported by theoretical models \citep{Pittard:1998,
  Ishibashi:1999, Pittard:2002, Akashi:2006, Henley:2008,
  Okazaki:2008, Parkin:2009, Kashi:2009b}, as well as by observations
at infrared \citep{Whitelock:1994, Whitelock:2004, Nielsen:2009},
radio \citep{Duncan:2003, Abraham:2005b}, optical
\citep{Damineli:1996, Daminelli:2000, Damineli:2008a, Damineli:2008b,
  vanGenderen:2003, Nielsen:2007}, far ultra-violet
\citep{Iping:2005}, and $\gamma-$ray wavelengths \citep[][- although
  see \citeauthor{Ohm:2010}~\citeyear{Ohm:2010} and
  \citeauthor{Abdo:2010}~\citeyear{Abdo:2010}]{Tavani:2009,
  Walter:2010}. In this scenario the X-ray emission originates from
the shock heated plasma generated by the fast wind of a companion star
which ploughs into the slow dense wind of \etacar\footnote{For the
  remainder of this paper we refer to the larger star \citep[commonly
    referred to as a luminous blue variable (LBV) star - although see
  ][]{Mehner:2010} and the smaller companion star \citep[estimated to
    be of early O type or a WR star - ][]{Pittard:2002, Verner:2005,
    Teodoro:2008, Mehner:2010} as the primary and companion star,
  respectively.}  \citep[e.g. ][]{Pittard:1998}.

The observed X-ray lightcurve displays the general characteristics
expected from a high orbital eccentricity colliding winds binary, but
has some puzzling features. For instance, the most recent X-ray
minimum was considerably shorter than the extended minima observed in
previous cycles \citep{Corcoran:2001, Corcoran:2005,
  Corcoran:2010}. The exact cause of the X-ray minimum is a matter of
debate \citep[for an overview of some of the possibilities
  see][]{Parkin:2009}, yet it is now clear that models must also be
able to account for the observed cycle-to-cycle variation.

The nature of the wind-wind collision region (WCR) has recently been
investigated with 3D models. Using a Smoothed-Particle Hydrodynamics
(SPH) model of the wind-wind collision, \cite{Okazaki:2008} found that
they could match the extended X-ray minimum. In contrast, adopting the
3D dynamical model of \cite{Parkin:2008}, \cite{Parkin:2009} showed
that when the spatial extent of the X-ray emission region and energy
dependence of the emission and absorption were taken into
consideration, the width of the observed X-ray minimum could not be
reproduced by an eclipse of the X-ray emitting plasma
alone. Furthermore, models with the preshock stellar winds at terminal
velocity over-predicted the observed X-ray emission in the 7-10 keV
band by an order of magnitude, indicating that a reduction in the
preshock speed of the companion star's wind is required. 

Such a reduction is plausible, as the highly eccentric orbit causes
the WCR to enter into the wind acceleration region of the companion
star around periastron passage. Considering the immense luminosity of
the primary star, the acceleration of the companion's wind may also be
significantly inhibited. A reduction in the preshock velocity will
cause radiative cooling to become increasingly important in the
postshock companion's wind, and will affect the stability of the WCR
\citep{Davidson:2002, Parkin:2009}. In fact, a ram pressure balance
between the stellar winds may be lost and the shock may collapse onto
the companion star \citep[][]{Parkin:2009}. In this case the 7-10~keV
flux which originates predominantly from the apex of the WCR would be
quenched, a feature which is necessary in models aiming to explain the
extended X-ray minimum. However, it is unclear why a collapse of the
WCR (if one occured) would be of a much shorter duration in the most
recent periastron passage whilst being extended in previous cycles.

In this paper we describe our results from a suite of three
dimensional (3D) simulations in which we have implemented radiative
driving, gravity, orbital motion, and optically thin radiative
cooling, allowing the r\^{o}le of wind acceleration, interacting
radiation fields, and instabilities in the wind-wind collision region
(WCR) on the gas dynamics and resulting X-ray emission to be
explored. We first present simulations with static stars at a
separation corresponding to periastron passage. The turbulent, highly
unstable nature of the wind-wind collision region (WCR) is revealed,
and results show that when the acceleration regions of the stars are
considered the conditions in the postshock gas permit the growth of
powerful non-linear thin-shell instabilities \citep[NTSI -
][]{Vishniac:1994}. Subsequent vigorous oscillations lead to the
collision of dense fragments of the WCR against the companion
star. Importantly, the 7-10 keV X-ray luminosity shows a marked
reduction when compared to an equivalent simulation which neglects the
wind acceleration. We then perform large-scale, high resolution
simulations which include the orbital motion of the stars, and find
that the rapid orbital motion of the stars around periastron acts to:
reduce the degree of radiative inhibition of the preshock companion
star's wind. Consequently, the catastrophic disruption of the WCR seen
in the static stars simulation is not reproduced. 

We conclude that the behaviour of the observed X-ray emission from
\etacar around periastron is tied to the rate of radiative cooling in
the postshock companion's wind. The suppression of the preshock
companion's wind by radiative inhibition (albeit with a stronger
radiation-wind coupling than explored in this work) could be an
important trigger for rapid cooling. Furthermore, the differences
between the observed short and long duration X-ray minima may be due
to the companion's postshock gas conditions lying close to the
dividing line between instability growth which merely perturbs the
WCR, and far more vigorous NTSIs which destroy it (possibly causing
the WCR to collapse onto the companion star in the process). The
latter case may explain the longer observed minima.

The remainder of this paper is structured as follows: a description of
the hydrodynamic model and the X-ray emission calculations are given
in \S~\ref{sec:model}. In \S~\ref{sec:theory} we review some
theoretical background relevant to our current investigation. The
results of the hydrodynamic simulations are presented in
\S~\ref{sec:periastron} (static stars at periastron separation), and
\S~\ref{sec:orbit} (large-scale orbit). A discussion and suggestions
for future directions are given in \S~\ref{sec:discussion}, and we
close with our conclusions in \S~\ref{sec:conclusions}.

\begin{table}
\begin{center}
\caption[]{Adopted system parameters for
  $\eta\;$Car.} \label{tab:system_parameters}
\begin{tabular}{lll}
\hline
Parameter & Value & Reference \\
\hline
Orbital period (d) & 2024 & 1 \\
Eccentricity ($e$) & 0.9 & 1 \\
$a$ (au) & 16.64 & 2 \\
$d_{\rm sep}(\phi = 0.0)$ ($10^{13}\;$cm) & 2.48 & $-$ \\
$d_{\rm sep}(\phi = 0.5)$ ($10^{13}\;$cm) & 47.3 & $-$ \\ 
Distance (kpc) & 2.3 & 3 \\
ISM + nebula column ($10^{22}\rm{cm}^{-2}$) & 5 & 4 \\
\hline
\end{tabular}
\tablecomments{$a$ is the semi-major axis of the orbit and $d_{\rm
    sep}(\phi)$ is the binary separation as a function of orbital
  phase. References are as follows: 1 = \cite{Damineli:2008b}, 2 =
  \cite{Hillier:2001}, 3 = \cite{Davidson:1997}, 4 =
  \cite{Hamaguchi:2007}.}
\end{center}
\end{table}

\section{The model}
\label{sec:model}

\subsection{Hydrodynamic modelling}
\label{subsec:hydromodel}

The wind-wind collision is modelled by numerically solving the
time-dependent equations of Eulerian hydrodynamics in a 3D Cartesian
coordinate system. The relevant equations for mass, momentum, and
energy conservation are:
\begin{eqnarray}
\frac{\partial\rho}{\partial t} + \nabla \cdot \rho {\bf v} &  =  & 0, \\
\frac{\partial\rho{\bf v}}{\partial t} + \nabla\cdot\rho{\bf vv} + \nabla P & = & \rho{\bf f},\\
\frac{\partial\rho E}{\partial t} + \nabla\cdot[(\rho E + P){\bf v}] & =& \left(\frac{\rho}{m_{\rm H}}\right)^{2}\Lambda(T) + \rho {\bf f}\cdot {\bf v}.
\end{eqnarray}

\noindent Here $E = \epsilon + \frac{1}{2}|{\bf v}|^{2}$, is the total
gas energy, $\epsilon$ is the specific internal energy, ${\bf v}$ is
the gas velocity, $\rho$ is the mass density, $P$ is the pressure, $T$
is the temperature, and $m_{\rm H}$ is the mass of hydrogen. ${\bf f}$
is the force per unit mass and includes gravity and radiative driving
terms. We use the ideal gas equation of state, $P = (\gamma -
1)\rho\epsilon$, where the adiabatic index $\gamma = 5/3$.

The radiative cooling term, $\Lambda(T)$, is calculated from the
\textsc{MEKAL} thermal plasma code \citep{Mewe:1995,Kaastra:1992}
distributed in \textsc{XSPEC} (v11.2.0). The temperature of the
unshocked winds is assumed to be maintained at $\approx 10^{4}\;$K via
photoionization heating by the stars. Throughout this work solar
abundances are assumed \citep{Anders:1989}.

\begin{figure}
  \begin{center}
    \begin{tabular}{c}
\resizebox{80mm}{!}{\includegraphics{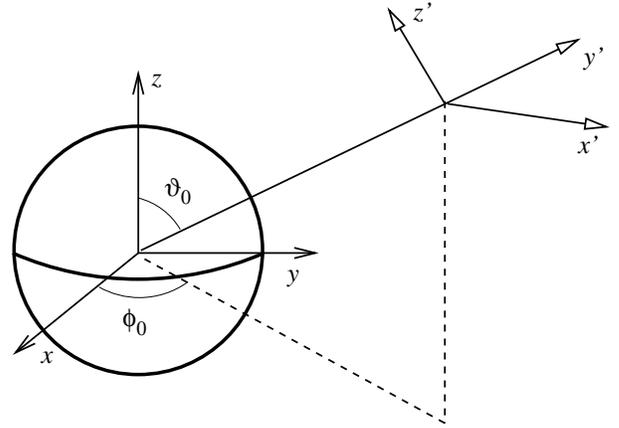}} \\
   \end{tabular}
    \caption{Schematic diagram illustrating the transformation from
      the ``wind centred'' to the ``star centred' coordinate system.}
    \label{fig:geometry}
  \end{center}
\end{figure}

The body forces acting on each hydrodynamic cell are the vector
summation of gravitational forces from each star, and continuum and
line driving forces from the stellar radiation fields. The computation
of the line acceleration is based on a local \cite{Sobolev:1960}
treatment of the line transport, following the standard
\cite*{Castor:1975} \citep[hereafter][]{Castor:1975} formalism
developed for single OB star winds. To account for the correction to
the line force due to the finite size of the stellar disk
\citep[i.e. the finite disk correction factor -][]{Castor:1974,
  Pauldrach:1986} we follow \cite{Cranmer:1995} and define a ``wind
centred'' coordinate system ($x',y',z'$) to calculate the local
velocity gradient (see Fig.~\ref{fig:geometry}) and then move to the
``star centred'' coordinate system ($x,y,z$) to determine the
projected velocity gradient along the direction vector $\hat{\bf n}$
using the following transformation,
\begin{eqnarray}
  x = & (x'\cos\theta_0 + z'\sin\theta_0)\cos\phi_0 - y'\sin\phi_0 \nonumber \\
  y = & y'\cos\phi_0 + (x'\cos\theta_0 + z'\sin\theta_0)\sin\phi_0 \\
  z = & z'\cos\theta_0 - x'\sin\theta_0 \nonumber
\end{eqnarray}

\noindent The vector radiative force per unit mass, ${\bf g}_{\rm
  rad}$, is then computed from a numerical Gaussian integration (using
8 directional vectors) of the intensity $I({\rm \hat{\bf n}})$ times
$\hat{\bf n}\cdot \nabla (\hat{\bf n}\cdot v)$ over the solid angle
covering the stellar disk,
\begin{equation}
{\bf g}_{\rm rad} = \frac{\sigma_{c}^{1-\alpha}k}{c} \oint I(\hat{\bf n})\left[\frac{\hat{\bf n}\cdot \nabla (\hat{\bf n}\cdot v)}{\rho v_{\rm th}}\right]^{\alpha}\hat{\bf n}d{\bf \Omega}.
\end{equation}
\noindent $\alpha$ and $k$ are the standard \cite{Castor:1975}
parameters, $\sigma_{\rm e}$ is the specific electron opacity due to
Thomson scattering, and $v_{\rm th}$ is a fiducial thermal velocity
calculated for hydrogen. Shadowing by the companion star is accounted
for in our calculations, and in such cases only the visible part of
the stellar disk contributes to the radiative driving force. The line
driving is set to zero in cells with temperatures above $10^{6}\;$K,
since this plasma is mostly ionized. Further details concerning the
line force calculations can be found in \cite{Cranmer:1995} and
\cite{Gayley:1997}.

The stellar winds are initiated in the instantaneously accelerated and
radiatively driven stellar winds simulations in two slightly different
ways. In the former, appropriate hydrodynamic variables (i.e. $\rho,
P, {\bf v}$) are mapped into cells residing within a radial distance
$R_{\rm map}$ which is a function of stellar separation (see
Appendix~\ref{sec:remap}). Thus,
\begin{equation}
  \rho_{\rm i} = \frac{\dot{M_{\rm i}}}{4\pi r_{\rm i}^{2} |v_{\infty \rm{i}}|},
\end{equation}
\noindent where $r_{\rm i}$ is the distance of the cell from the
respective star and the subscript i refers to the wind being
considered. In contrast, to initiate radiatively driven stellar winds
we map hydrodynamic variables determined from single star wind
profiles (calculated with the parameters noted in
Table~\ref{tab:stellar_parameters}) within a stellar radius of
$\simeq1.1~R_{\ast}$ after every time step. As noted by
\cite{Pittard:2009}, it is necessary to resolve the region above the
stellar surface (i.e $0.1~R_{\ast}$) with at least 3 cells. To
distinguish between the stellar winds in the simulations we include an
advected scalar variable in the hydrodynamics calculations.

It should be noted that the primary is an LBV, and therefore it is
unclear to what level using the \cite{Castor:1975} formalism is
appropriate to describe the wind driving from this star. Furthermore,
\cite{Smith:2003a} inferred a higher rate of mass-loss at the poles
based on the observation of varying absorption as a function of
latitude in the reflected emission from the Homunculus nebula, which
may be related to the rotation rate of the primary star, though it may
also be related to the presence of the companion star
\citep[][]{Groh:2010b}. Since we do not consider a latitude
dependence, the mass-loss rates which we infer are likely indicative
of the equatorial wind (assuming the equatorial and orbital planes are
aligned).

Adopting solar abundances may also affect the inferred mass-loss
rates. For instance, \cite{Hillier:2001} found N to be significantly
enhanced, while C and O could be as low as $1/25$ and $1/50$ solar,
respectively. Although these differences will alter the opacity of the
primary wind, we do not anticipate that the observed X-ray spectrum at
$E>1\;$keV will be significantly affected.

\subsection{The hydrodynamic code}
\label{subsec:hydrocode}

The simulations presented in this work were performed using version
3.1.1 of the \textsc{FLASH} \citep{Fryxell:2000, Dubey:2009}
hydrodynamical code. This code operates with a block-structured AMR
grid \citep[e.g.][]{Berger:1989} using the \textsc{PARAMESH} package
\citep{MacNeice:2000} under the message-passing interface (MPI)
architecture. The piecewise-parabolic method of \cite{Colella:1984} is
used to solve the hydrodynamic equations. Customized units have been
implemented into the \textsc{FLASH} code for radiative cooling for
optically-thin plasma using the explicit method described by
\cite{Strickland:1995}, radiative driving, gravity, and orbital
motion. Simulations are performed with the stars fixed at periastron
(\S~\ref{sec:periastron}) and as they move through an orbit
(\S~\ref{sec:orbit}), and a description of the grids used can be found
in the relevant section (i.e. size, shape, resolution, and refinement
criteria).

We note that as a goal of our investigation is to explore the r\^{o}le
of instabilities in the WCR, a grid-based hydrodynamic code is
advantageous for this purpose in comparison to an SPH code
\citep[][]{Agertz:2007}.

\begin{table}
\begin{center}
  \caption[]{Parameters used to calculate the line driving of the
    stellar winds.} \label{tab:stellar_parameters}
\begin{tabular}{llll}
\hline
Parameter & Primary  & Secondary & Reference\\
\hline
$M$ (M$_{\odot}$) & 120 & 30   & 1 \\ 
$R_{\ast}$ (R$_{\odot}$) & 100 & 20 & 2 \\
$T_{\rm{cs}}$ (K)& 25,800 & 30,000 & 3 \\
$L_{\ast} $ ($10^{6} {\rm L_{\odot}}$) & $4$ & $0.3$ & 3 \\
$k$ & 0.30 &  0.50 & $-$ \\
$\alpha$ & 0.52 & 0.68 & $-$ \\
$\dot{M}\;(\Msolpyr)$ & $4.8\times10^{-4}$ & $1.4\times10^{-5}$ & 3 \\
$v_{\infty}\;(\kmps)$ &  500 & 3000 & 4 \\
$\eta$ & \multicolumn{2}{c}{0.18} & $-$ \\
\hline
\end{tabular}
\tablecomments{$M$ is the stellar mass, $R_{\ast}$ is taken to be the
  radius of the gravitationally bound core of the star (which for the
  primary star is taken to be at a radius of $100\Rsol$, i.e. the
  photosphere will exist somewhere in the stellar wind), $T_{\rm{cs}}$
  is the temperature at $R_{\ast}$, $L_{\ast}$ is the stellar
  luminosity, $\dot{M}$ is the stellar mass-loss rate, $v_{\infty}$ is
  the wind terminal velocity, and $\eta = \dot{M}_{2}v_{\infty 2}/
  \dot{M}_{1}v_{\infty 1}$ is the wind momentum ratio. $k$ and
  $\alpha$ are the \cite{Castor:1975} line driving parameters, where
  subscripts 1 and 2 are used to define the coupling between the winds
  and the radiation fields of the primary and secondary star
  respectively. Note that single star mCAK calculations
  \citep{Pauldrach:1986} were performed to determine values of $k$ and
  $\alpha$ required to produce the desired $\dot{M}$ and $v_{\infty}$
  for each wind. References are as follows: 1 = \cite{Hillier:2001}, 2
  = \cite{Corcoran:2007}, 3 = \cite{Parkin:2009}, 4 =
  \cite{Pittard:2002}.}
\end{center}
\end{table}

\subsection{X-ray emission}
\label{subsec:xray_emission}

To calculate the X-ray emission from the simulation we use
emissivities for optically thin gas in collisional ionization
equilibrium obtained from look-up tables calculated from the
\textsc{MEKAL} plasma code containing 200 logarithmically spaced
energy bins in the range 0.1-10 keV, and 101 logarithmically spaced
temperature bins in the range $10^{4}-10^{9}\;$K. When calculating the
emergent flux we use energy dependent opacities calculated with
version C~$08.00$ of Cloudy \citep[][see also
  \citeauthor{Ferland:1998}~\citeyear{Ferland:1998}]{Ferland:2000}. The
advected scalar is used to separate the X-ray emission contributions
from each wind. 

Based on the findings of \cite{Parkin_Pittard:2010}, we note that {\it
  numerical} heat conduction from the hot, low density postshock
companion star wind to the cooler, higher density postshock gas of the
primary star's wind will occur in the simulations. This effect is
purely numerical in origin, and one must be careful when interpreting
the derived X-ray emission. Therefore, as a cautionary measure we only
consider the X-ray emission contributed by the postshock companion
star's wind.

\section{Theoretical background of radiatively driven winds}
\label{sec:theory}

\subsection{Radiatively driven winds of single hot stars}

For the stellar atmosphere to be driven by the radiation field there
is the simple requirement that $GM/r^{2} < g_{\rm rad}$. The key to
accurately describing the motion of the wind comes in the form of the
$g_{\rm rad}$ term, and significant progress has been made using the
force multiplier, $\mathfrak{M}(t) = kt^{-\alpha}$, originally
proposed by \cite{Castor:1975},
\begin{eqnarray}
  g_{\rm rad} = \frac{\sigma_{e}F k t^{-\alpha}}{c},
\label{eqn:cak_g_rad}
\end{eqnarray}
\noindent where $F$ is the radiative flux, and $t= \sigma_{\rm
  e}v_{\rm th}\rho[\hat{\bf n}\cdot \nabla (\hat{\bf n}\cdot v)]^{-1}$
is the Sobolev optical depth parameter. The force multiplier
parameters $\alpha$ and $k$ are determined from a power law fit to
$\mathfrak{M}(t)$. $k$ can be interpreted as the fraction of flux that
would be blocked if all lines were optically thick. $\alpha$ is the
ratio of the line acceleration from optically thick lines to the total
one.

\subsection{Finite disk correction factor}
\label{subsec:FDCF}

A simplifying assumption adopted in early works on radiatively driven
winds \citep[e.g.][]{Lucy:1970, Castor:1975} was that of radially
streaming photons, i.e. a point source of radiation. In this case the
projected velocity gradient becomes $\partial v_{\rm r} / \partial r$
where $v_{\rm r}$ is the radial velocity, and angle integrals simplify
to purely radial terms. Unfortunately, the assumption of radially
streaming photons is a poor one close to the star where the wind is
rapidly accelerated. To circumvent this problem it has become common
place to incorporate a multiplicative factor to correct for the finite
size of the stellar disk, $K$, which is attained by adopting the exact
optical depth rather than the radial one,
\begin{eqnarray}
  K(r,v,dv/dr) & \equiv & \frac{{\bf g}_{\rm rad}({ \rm finite~disk})}{{\bf g}_{\rm rad}({\rm point~source})} \\
  & = & \frac{(1 + \sigma)^{1 + \alpha} - (1 +
    \sigma\mu_{\ast}^2)^{1 + \alpha}}{\sigma(1 + \alpha)(1 +
    \sigma)^{\alpha}(1 - \mu_{\ast}^{2})} \label{eqn:dynamical_fdcf}
\end{eqnarray}

\noindent where $\theta$ is the angle subtended between the radial
direction and a point on the stellar disk, $\mu = \cos\theta$,
$\mu_{\ast} = \sqrt{1 - (R_{\ast}/r)^2}$, and $\sigma = (d \ln v/d \ln
r) - 1$ \citep{Castor:1970}. $K$ is commonly referred to as the
\textit{finite disk correction factor} (FDCF) \citep{Castor:1974,
  Castor:1975, Pauldrach:1986}. $\mathfrak{M}(t)_{\rm FDCF}$ is
related to the standard (under the assumption of radially streaming
photons) force multiplier, $\mathfrak{M}(t)$, via,
\begin{equation}
   \mathfrak{M}(t)_{\rm FDCF} = K \mathfrak{M}(t)
\end{equation}

\begin{figure}
  \begin{center}
    \begin{tabular}{c}
\resizebox{80mm}{!}{\includegraphics{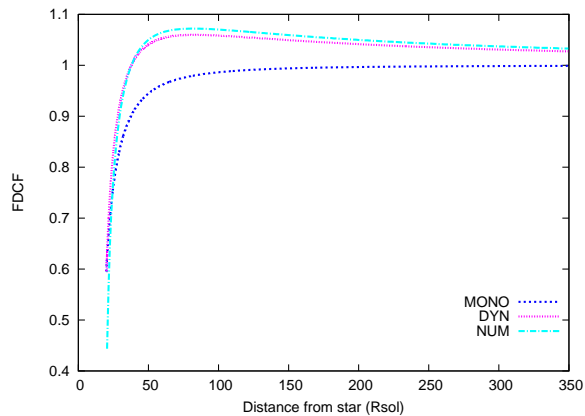}} \\
    \end{tabular}
    \caption{Comparison of finite disc correction factors (FCDFs)
      calculated assuming monotonic flow \citep[MONO,
        e.g. ][]{Stevens:1994}, with dynamical consistence (DYN,
      i.e. including velocity gradient terms - see
      Eq.~\ref{eqn:dynamical_fdcf}), and via a numerical evaluation
      (NUM - see \S~\ref{subsec:hydromodel}). The FDCFs are calculated
      using the companion (single) star wind velocity profile (see
      Fig.~\ref{fig:vel_comparison}).}
    \label{fig:fdcf_comparison}
  \end{center}
\end{figure}

In certain instances \citep[e.g.][]{Pauldrach:1986, Stevens:1994} the
FDCF is approximated as a purely radial, monotonic function
(i.e. $K(r,v,dv/dr)\sim K(r)$). Fig.~\ref{fig:fdcf_comparison} shows a
comparison of monatonic and dynamically consistent FDCFs; a monotonic
FDCF always stays below 1, whereas a dynamically consistent FDCF rises
above 1 in the wind acceleration region. For comparison we have also
plotted the numerical evaluation of the FDCF which we have implemented
into \textsc{FLASH} - there is a good agreement with the analytical
result.

\subsection{Interacting radiation fields in binary systems}

In the standard \cite{Castor:1975} model for radiatively driven winds
the outwards acceleration is produced by the radiation field of the
star. If one introduces an opposing radiation field two effects become
apparent: radiative inhibition \citep{Stevens:1994} and radiative
braking (\citeauthor{Gayley:1997}~\citeyear{Gayley:1997} - see also
\citeauthor{Owocki:1995}~\citeyear{Owocki:1995}). The former refers to
the reduction of the net rate of acceleration of a stellar wind due to
the opposing radiation field, whereas the latter refers to the
(sudden) deceleration of a stellar wind before it reaches the WCR. In
the \cite{Stevens:1994} formulation of radiative inhibition, the basic
premise is that the line force (Eq.~\ref{eqn:cak_g_rad}) becomes
\begin{equation}
  g_{\rm rad} = \frac{\sigma_{\rm e}\mathfrak{M}(t)}{c}\{F_1 K_1 - F_2 K_2\} \label{eqn:RI}
\end{equation}
\noindent where $F_1$, $F_2$ are the radiative fluxes, and $K_1$,
$K_2$ are the finite disk correction factors for stars 1 and 2,
respectively. The radiative inhibition calculations presented by
\cite{Stevens:1994} adopted a monotonic FDCF and, as noted by
\cite{Gayley:1997}, it was for this reason that they did not see
radiative braking in their calculations.

Radiative inhibition is expected to be important if the stars are
close enough that their opposing radiation fields reduce the net force
within the acceleration region of the winds, whereas radiative braking
is important in systems where the imbalance in wind momenta is such
that the stronger wind approaches the companion star close enough for
its radiation pressure to exceed the wind ram pressure. Depending on
the stellar and system parameters adopted both effects may be
important for $\eta$~Car \citep{Parkin:2009}.

\subsection{Considerations for \etacar}

The binary orbit of \etacar is highly eccentric. Therefore, for a
large part of the orbit the stars are well separated, the winds have
accelerated to terminal velocity when they reach the WCR, and the
effects of radiative inhibition and radiative braking are
negligible. However, an accurate description of wind acceleration
(particularly of the companion's wind) is essential around periastron
when the WCR enters into the wind acceleration regions.

\begin{figure}
  \begin{center}
    \begin{tabular}{c}
\resizebox{80mm}{!}{\includegraphics{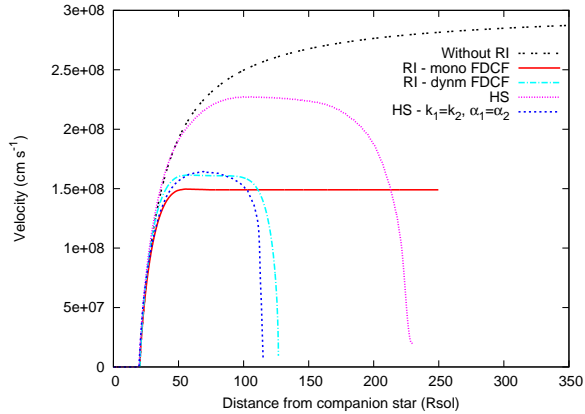}} \\
    \end{tabular}
    \caption{Companion star wind velocity along the line of centres
      between the stars. The primary star is situated at a distance of
      359\Rsol (corresponding to periastron separation when
      $e=0.9$). The following solutions are shown: without radiative
      inhibition (RI)(i.e. single star), RI using a monotic FDCF, RI
      using a dynamically consistent FDCF, and a numerical
      hydrodynamic simulation (HS) with separate \cite{Castor:1975}
      parameters for each wind, and a HS but with $k_{1}$ and
      $\alpha_{1}$ used to drive both winds. The stellar and line
      driving parameters used to perform these calculations are noted
      in Table~\ref{tab:stellar_parameters}.}
    \label{fig:vel_comparison}
  \end{center}
\end{figure}

Compared to a monotonic FDCF, a dynamically consistent FDCF will
provide additional acceleration close to the star driving the wind
(see \S~\ref{subsec:FDCF}). Therefore, one may wonder what difference
using dynamically consistent FDCFs in Eq.~\ref{eqn:RI} has on the
level of radiative inhibition?  In Fig.~\ref{fig:vel_comparison} we
show the results from static two-star radiative driving calculations
for \etacar at periastron. Using a monotonic FDCF and with
$k_{1}=k_{2}$ and $\alpha_{1}=\alpha_{2}$ (i.e. the coupling between
the primary's radiation and the companion's wind uses $k=0.50\;$and
$\alpha=0.68$) we see that the terminal velocity attained by the
companion star wind is dramatically reduced from $3000\;$km~s$^{-1}$
(without RI) to $\simeq1500\;$km~s$^{-1}$ (RI - mono FDCF). If we
include a dynamically consistent FDCF the level of radiative
inhibition is lessened and the maximum velocity attained now increases
to $\simeq1600\;$km~s$^{-1}$ (RI - dynm FDCF). Interestingly, the
velocity profile now displays radiative braking. Comparing the
radiative driving calculations against those from a numerical
hydrodynamic simulation calculated using the code described in
\S~\ref{subsec:hydromodel} (HS), we see good agreement, albeit with a
minor difference in the position of the braking radius (the radius at
which rapid deceleration of the wind occurs).

In both the radiative inhibition \citep{Stevens:1994} and braking
\citep{Owocki:1995, Gayley:1997} theories the line force is assumed to
take the form of Eq.~\ref{eqn:RI}, and in so doing the force
multiplier parameters used to accelerate/decelerate the stellar wind
are assumed to be equivalent (i.e. $k_{1}=k_{2}$ and
$\alpha_{1}=\alpha_{2}$ within the wind of star~1). At the alternative
extreme, the line force could be described as\footnote{It is currently
  unclear what the correct choice for the coupling between the
  radiation field of a star with the opposing star's wind is. We
  merely note that previous works on the interaction of radiation
  fields in binary star systems \citep[e.g.][]{Pittard:1998,
    Gayley:1997, St-Louis:2005, Parkin:2009} suggest that the actual
  coupling lies somewhere between the two cases presented in
  Eqs.~\ref{eqn:RI} and \ref{eqn:RI2}.}
\begin{equation}
  g_{\rm rad} = \frac{\sigma_{\rm e}}{c}\{\mathfrak{M}_{1}(t) F_1 K_1 - \mathfrak{M}_{2}(t) F_2
  K_2\} \label{eqn:RI2}
\end{equation}

\noindent where $\mathfrak{M}_{\rm w}(t)=k_{\rm w}t^{- \alpha_{\rm
    w}}$ and the subscript $w$ denotes the wind which the line driving
parameters are describing. Fig.~\ref{fig:vel_comparison} shows the
result of a hydrodynamic calculation (with radiatively driven winds)
where $g_{\rm rad}$ is equivalent to that of Eq.~\ref{eqn:RI2}
(i.e. the coupling between the primary's radiation and the companion's
wind uses $k=0.30\;$and $\alpha=0.52$). The resulting level of
radiative inhibition is lower and the companion's wind reaches a
maximum velocity of 2200~km~s$^{-1}$ (HS - $k_{1}=k_{2}$, $\alpha_1 =
\alpha_2$). Furthermore, the braking radius has increased by a factor
of $\sim2$. The explanation for this can be found by considering the
line driving parameters of each individual wind. The significantly
higher luminosity of the primary star means that to achieve its huge
mass-loss rate we actually require lower values of $k$ and $\alpha$
than the companion star (with its lower luminosity) requires to drive
a weaker wind. In essence, the coupling between the primary's
radiation field and its wind is weaker than the respective case for
the companion star. Therefore, if the primary's radiation field is
more strongly coupled to the companion's wind (i.e $k_{1}=k_{2}$ and
$\alpha_{1}=\alpha_{2}$), the decelerating force, and thus the net
level of radiative inhibition, will be greater. Clearly, this result
has implications for the dynamics of the WCR, and the influence of
radiative inhibition on the X-ray emission from \etacar
\citep{Parkin:2009}. In the following work we conservatively adopt the
assumption of weaker coupling (Eq.~\ref{eqn:RI2}), but note that it
could be stronger in reality. If this were the case one would expect
lower preshock velocities for the companion's wind, and reduced X-ray
emission, particularly at higher energies (e.g. $E>2\;$keV).

\begin{figure}
  \begin{center}
    \begin{tabular}{cc}
\resizebox{25mm}{!}{\includegraphics{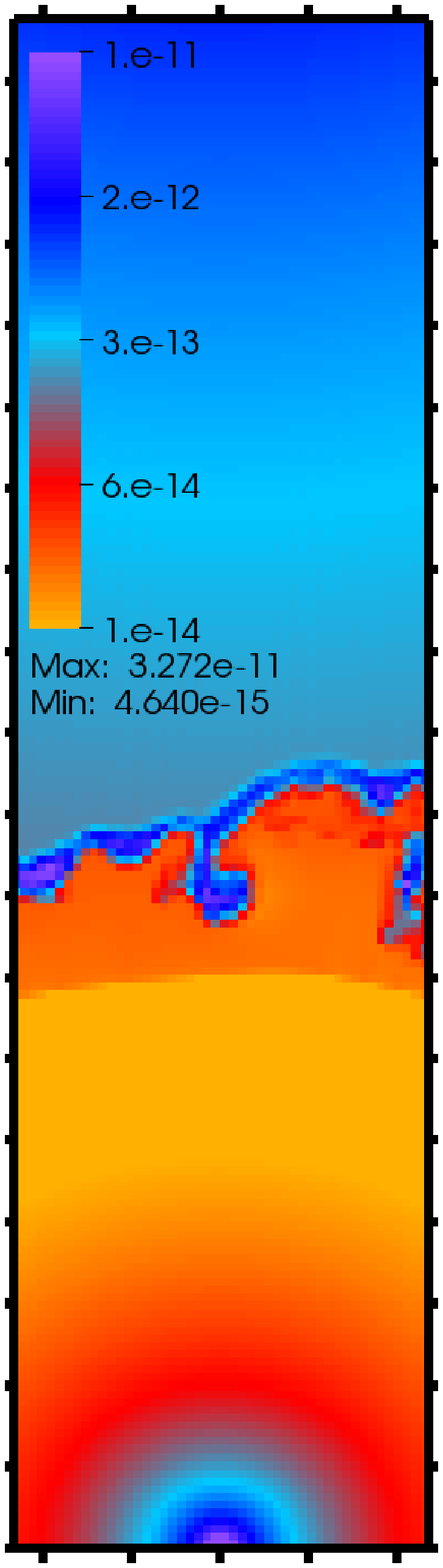}} &
\resizebox{25mm}{!}{\includegraphics{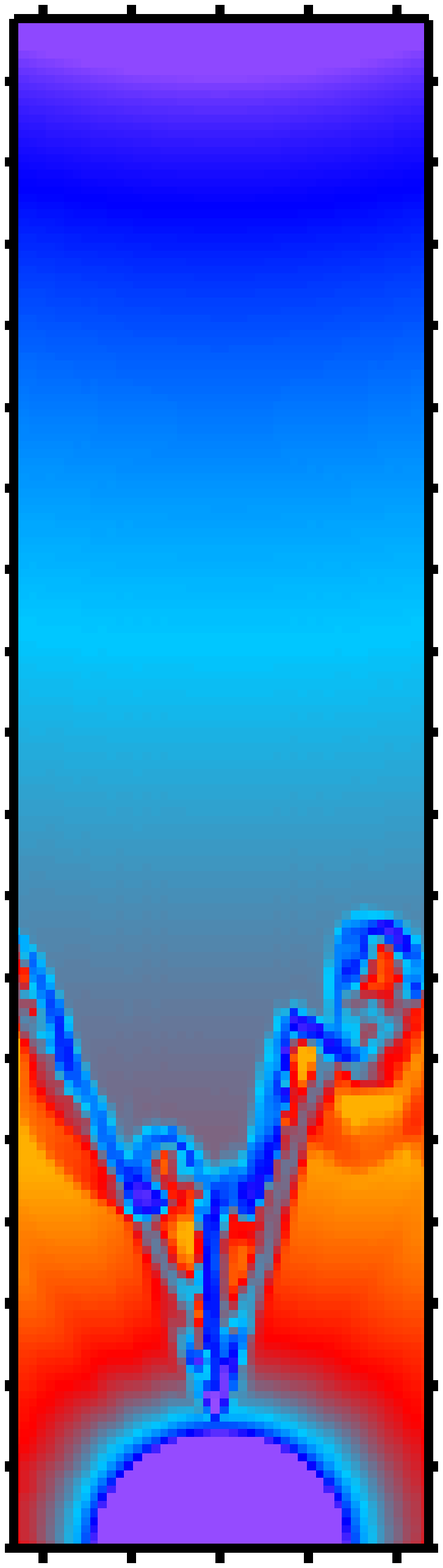}} \\

\resizebox{25mm}{!}{\includegraphics{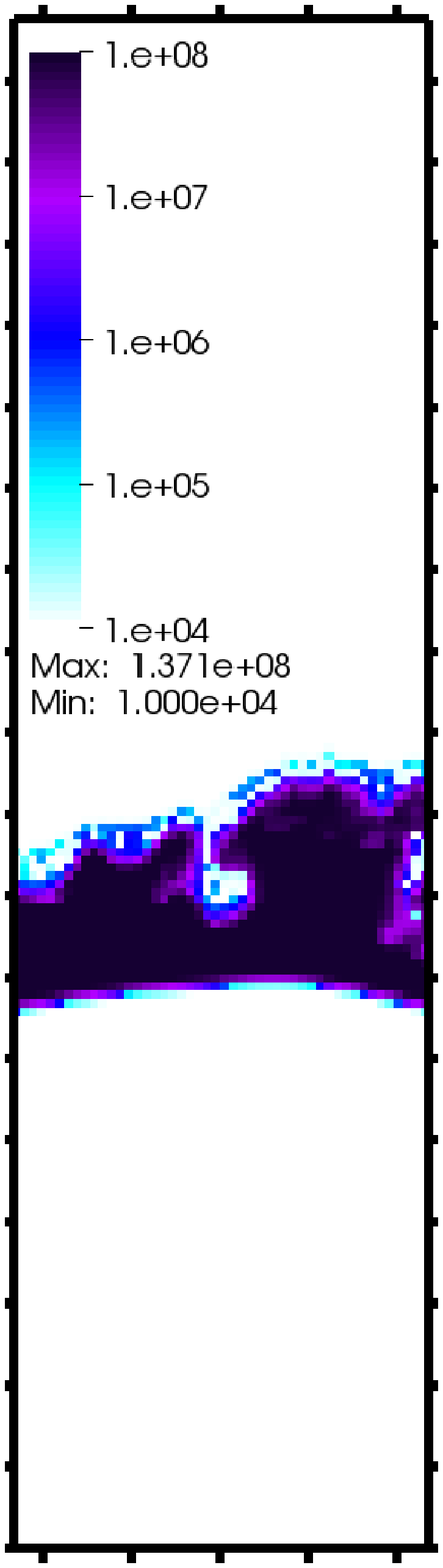}} &
\resizebox{25mm}{!}{\includegraphics{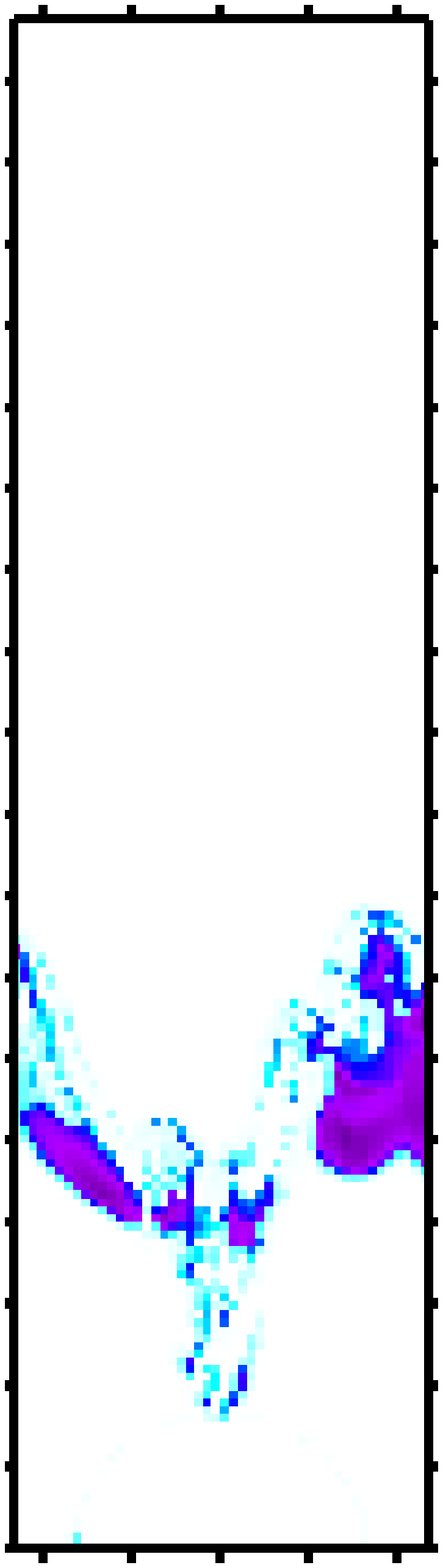}} \\
    \end{tabular}
    \caption{Density (upper row) and temperature (lower row) snapshots
      of the $x-y\;$plane from models Peri-IA (left column) and
      Peri-RD (right column) at a simulation time of
      $2.4\times10^{6}\;$s. The companion star is situated at the
      lower boundary and the primary star is situated beyond the
      opposite end of the box. All plots show a region of
      $x=\pm2.34\times10^{12}\;$cm and $y=(0-1.88\times10^{13}\;)$cm -
      tick marks correspond to a distance of $1\times10^{12}\;$cm. The
      hot postshock companion wind in model Peri-IA acts as a
      ``cushion'' against thin-shell instability. In contrast, NTSIs
      are permitted to grow by the lower temperature postshock
      companion's wind in model Peri-RD. A volume rendering of the
      model Peri-RD density distribution is shown in
      Fig.~\ref{fig:peri_collision}.}
    \label{fig:peri_collision_slices}
  \end{center}
\end{figure}

\begin{figure}
  \begin{center}
    \begin{tabular}{c}
\resizebox{80mm}{!}{\includegraphics{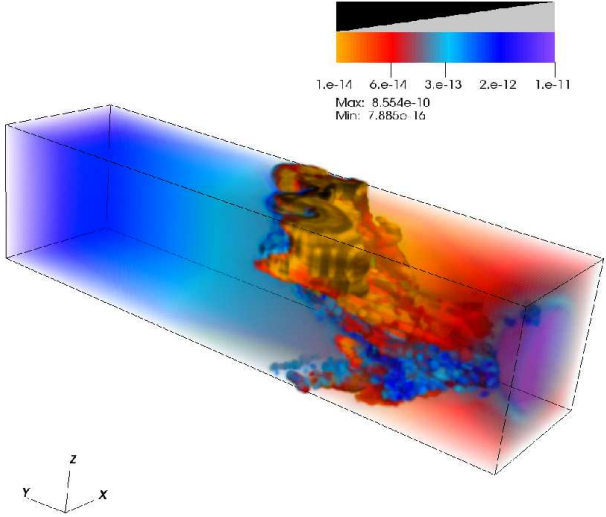}} \\
    \end{tabular}
    \caption{Density volume rendering of model Peri-RD at a
      simulation time of $2.4\times10^{6}\;$s. The companion star is
      situated to the right of the image (the end face of the box cuts
      through its centre) and the primary star is situated beyond the
      opposite end of the box such that only a small area of its
      surface fits within the box. Driven by instabilities in the WCR,
      fragments of cold, dense postshock primary wind gas (blue) can
      be seen colliding against the companion star (purple).}
    \label{fig:peri_collision}
  \end{center}
\end{figure}

\section{Periastron simulations}
\label{sec:periastron}

As the stars move towards periastron passage their separation
contracts, their orbital velocities increase, and the preshock and
postshock gas density (and velocity in the accelerating winds case)
undergo a considerable rate of change as the WCR moves closer to the
stars. To form a basis for qualitative and quantitative predictions
for the large-scale simulations that will follow (\S~\ref{sec:orbit}),
we begin with simulations focused on periastron passage where orbital
motion is not included. A fixed resolution grid with cell size of
$\simeq 9.8\times10^{10}\;$cm is used (i.e. no AMR) and the simulation
box extends from $x = z = \pm2.34\times10^{12}\;$cm and $y = 0 -
1.88\times10^{13}\;$cm (corresponding to $x \times y \times z = 32
\times 192 \times 32\;$cells). The companion star is situated at
($x,y,z$) = (0,0,0) and the primary star is situated at ($x,y,z$) =
(0, $2.5\times10^{13}\;$cm, 0). Zero-gradient boundary conditions are
used on all faces of the simulation box. The calculations were evolved
for a time of $4\times10^{6}\;$s, corresponding to $\simeq 0.02$ in
orbital phase, and therefore sufficiently long to examine the growth
of instabilities over the characteristic time that the stars spend at
closest approach. For comparison we have performed simulations with
instantaneously accelerated winds (model Peri-IA) and radiatively
driven winds (model Peri-RD). We now proceed with a discussion of the
results of these models, including calculations of the intrinsic
7-10~keV X-ray luminosity and its variability, and also the results of
additional tests performed with varying grid resolutions and at
separations corresponding to orbital phases around periastron.

\begin{figure}
  \begin{center}
    \begin{tabular}{c}
\resizebox{80mm}{!}{\includegraphics{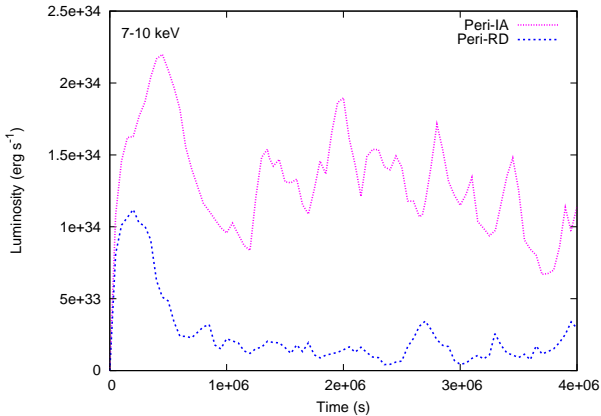}} \\
    \end{tabular}
    \caption{Variability of the intrinsic 7-10 keV X-ray luminosity
      for models Peri-IA and Peri-RD. Note that the simulation grid
      for models Peri-IA and Peri-RD is focused on the region between
      the stars, therefore the total X-ray emission from the entire
      WCR cannot be calculated. However, our comparison here
      highlights important differences between the flux levels of the
      two models.}
    \label{fig:peri_lx}
  \end{center}
\end{figure}

\begin{figure}
  \begin{center}
    \begin{tabular}{c}
\resizebox{80mm}{!}{\includegraphics{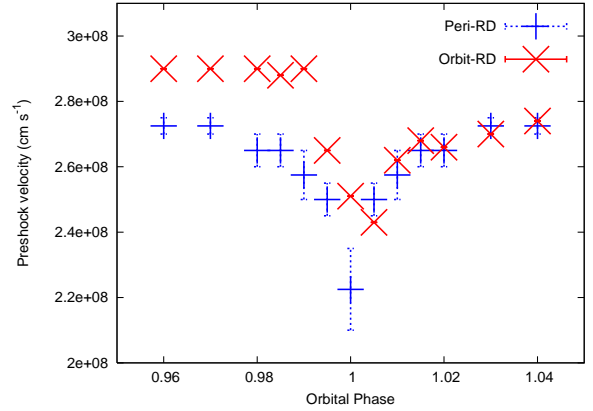}} \\
    \end{tabular}
    \caption{Preshock companion wind speed measured along the line of
      centres as a function of orbital phase for simulations Peri-RD
      and Orbit-RD.}
    \label{fig:preshock_velocity}
  \end{center}
\end{figure}

The periastron simulations are initialized with two opposing flows
separated by a contact discontinuity at the ram pressure balance point
between the winds. As model Peri-IA is evolved through time the winds
collide and shocks are formed. The primary's relatively slow and dense
wind cools rapidly to form a thin dense layer, whereas the faster,
more tenuous wind of the companion star behaves quasi-adiabatically
\citep[][]{Pittard:1998, Pittard:2002, Parkin:2009}. Consequently,
there is a large density and temperature contrast between gas on
either side of the contact discontinuity and the growth of
Rayleigh-Taylor (RT) instabilities ensues. Subsequent perturbations to
the contact discontinuity seed the linear thin-shell instability
\citep{Vishniac:1983}. However, the hot, quasi-adiabatic postshock
companion's wind has a sufficiently fast sound speed, and thus rate of
response to perturbations in the contact discontinuity by the dense
thin layer, to render it linearly stable and the regime of non-linear
instability growth is never reached \citep{Vishniac:1994}. Therefore,
the thermal pressure of the hot postshock companion's wind acts like a
cushion and prevents an instability-driven collapse/collision of the
WCR against the companion star (Fig.~\ref{fig:peri_collision_slices}).

However, the inclusion of the radiative driving of the stellar winds
in model Peri-RD, and thus of the wind acceleration regions,
introduces some important differences to this picture. Firstly, by a
simulation time of $t=10^{5}\;$s the primary's radiation field has
reduced the preshock companion's wind velocity from $\simeq
2500\;$km~s$^{-1}$ to $\simeq 2200\;$km~s$^{-1}$, and radiative
inhibition of the initial (single star) velocity profile (see
Fig.~\ref{fig:vel_comparison}) is evident. As such, the postshock gas
temperature is lower. Secondly, the WCR moves slightly closer to the
companion star to attain a new ram pressure balance and in so doing
the preshock (and thus postshock) gas densities are higher. The
combination of these factors causes radiative cooling to become
important for the postshock companion wind, and as the gas cools it
loses the thermal pressure support which ensures its stability. When
this occurs NTSIs begin to grow and by a simulation time of
$7\times10^{5}\;$s oscillations in the WCR push a dense clump of
postshock primary wind towards the companion star. This clump then
fragments from the WCR and passes deep into the companion's wind
acceleration region, narrowly missing the star. Driven by the NTSI,
the position of the WCR continues to oscillate throughout the
simulation causing the mean preshock velocity of the companion's wind
between the stars to vary in the range $\sim
2100-2350\;$km~s$^{-1}$. At a time of $2.4\times10^{6}\;$s multiple
fragmented clumps of postshock primary wind collide against the
companion star (see Figs.~\ref{fig:peri_collision_slices} and
\ref{fig:peri_collision}). Such collisions occur at multiple times
throughout the simulation, interspersed by brief recoveries as the
shock oscillates away from the companion star.

An obvious question is why do collisions occur in model Peri-RD but
not in model Peri-IA? As stated above, when radiative cooling becomes
important for the postshock companion's wind in model Peri-RD linear
thin-shell instabilities can grow non-linearly with devastating
results for the WCR. Hence, we can quantify the difference between
models Peri-RD and Peri-IA using the cooling parameter
\citep{Stevens:1992}
\begin{equation}
  \chi = \frac{t_{\rm cool}}{t_{\rm flow}} =
  \frac{v_{8}^4d_{12}}{\dot{M}_{-7}},
\end{equation}
where $t_{\rm cool}$ is the cooling time, $t_{\rm flow}=
10^{12}d_{12}/s_{\rm ps}$ is the flow time, $s_{\rm ps}$ is the
postshock sound speed along the line of centres, $v_{8}$ is the
preshock gas velocity (in $10^{8}\;$cm~s$^{-1}$), $d_{12}$ is a
characteristic distance (in $10^{12}\;$cm), and $\dot{M}_{-7}$ is the
mass-loss rate (in $10^{-7}\Msolpyr$). Here we take $d_{12}$ to be the
distance of the companion shock from the star as we are interested in
the importance of cooling close to the WCR apex. As a rule of thumb
$\chi > 1$ indicates quasi-adiabatic gas whereas $\chi\ltsimm 1$
indicates that gas cools rapidly. Examining $\chi$ for the companion's
wind we find values of 3.6 and 1 for models Peri-IA and Peri-RD,
respectively. Therefore, on the timescales that we are interested in
(i.e. the closest approach of the stars around periastron passage) the
differences in postshock gas conditions between models Peri-RD and
Peri-IA are sufficient to trigger rapid radiative cooling and the
subsequent growth of catastrophic instabilities in the
former. Interestingly, the differences between postshock gas
conditions in models Peri-RD and Peri-IA are relatively minor, but
because of the strong dependence on the preshock velocity there is a
fine line between a WCR which is only plagued by instabilities and one
in which the instabilities ultimately destroy it. The differences in
the X-ray minima (extended in 1998 and 2003.5 versus the quick
recovery seen in 2009), may therefore reflect differences in the
behaviour of the WCR caused by a small change in one or both winds. We
note again that our adopted radiation-wind coupling essentially
provides a lower limit to the degree of radiative inhibition of the
companion's wind. If a stronger coupling were adopted the companion
wind would collide at lower speeds and radiate more efficiently, and a
WCR disruption (potentially even more ferocious) as in model Peri-RD
would occur more readily.

\cite{Akashi:2010} have recently presented numerical simulations akin
to model Peri-IA. However, while the radiation fields were ignored,
the gravitational influence from {\it only} the companion star was
included. This results in the balance point of the WCR shifting closer
to the companion star, thereby increasing the cooling rate of the
shocked gas from the companion's wind, and making it easier for
instabilities in the WCR to strike the companion star. Clearly, one
must include all relevant forces acting on the shocked gas.

\subsection{X-ray emission}

Previous models of energy dependent X-ray emission from \etacar have
revealed that if the preshock companion's wind is assumed to be at
terminal velocity at orbital phases coinciding with the extended X-ray
minimum there is an overestimate of roughly an order of magnitude in
the 7-10~keV flux when compared to that observed by
\XMM~\citep{Parkin:2009}. To examine whether the differences between
models Peri-IA and Peri-RD can provide clues to alleviating this
discrepency we have performed X-ray calculations on the simulation
output (Fig.~\ref{fig:peri_lx}). The X-ray luminosity rapidly rises as
the winds start to collide in the simulation and then reaches a peak
before declining. In model Peri-RD the first minimum corresponds to
the first occurence of a fragment of dense postshock primary star wind
coming into close proximity and/or colliding with the companion
star. In contrast, no such collisions occur in model Peri-IA, and the
variability in the 7-10~keV flux instead corresponds to oscillations
in the shock front. The lower preshock velocities, and thus lower
postshock gas temperatures (see Fig.~\ref{fig:peri_collision_slices}),
in model Peri-RD combined with the shock obliquity created by the
NTSI results in an average intrinsic 7-10~keV luminosity which is
roughly a factor of 8 lower than that calculated for model
Peri-IA. Therefore, it seems that a substantial disruption of the WCR
can provide a plausible explanation for the observed behaviour of the
hard X-ray flux \citep{Hamaguchi:2007}.

\subsection{Resolution dependence}

The growth rate of instabilities are dependent on the wavelength, with
smaller wavelengths producing faster growth rates. To examine the
dependence on grid resolution we have performed tests\footnote{Note
  that to accurately model the acceleration of the stellar wind
  requires that the resolution not decrease in regions of sharp
  velocity gradient (i.e. within a radius of $\sim3~R_{\ast}$). Hence
  in the resolution tests the companion star wind acceleration region
  is covered with cells of side $\ltsimm 9.8\times10^{10}\;$cm and we
  only vary the cell size for {\it postshock} gas using the AMR
  capability of the code.} with cell sizes of
$(4.9-39.2)\times10^{10}\;$cm. Encouragingly, the general
characteristics appear consistent. In each case instabilities grow and
break-up the WCR causing stochastic collisions between dense clumps
and the companion star. There does not appear to be any correlation
between the grid resolution and the frequency of collisions.

Additional tests were performed using simulations boxes with a cell
size of $\simeq 9.8\times10^{10}\;$cm and a width up to $x = z = \pm
3.5\times10^{12}\;$cm. Consistent with model Peri-RD, when the
acceleration (and radiative inhibition) of the companion's wind are
considered the WCR becomes disrupted. Furthermore, the significantly
lower 7-10 keV X-ray flux illustrated by the comparison between models
Peri-RD and Peri-IA is reproduced.

\subsection{Variation of preshock velocity with orbital phase}

The separation of the stars changes rapidly around periastron passage
and, therefore, so does the level of radiative inhibition, and the
preshock companion star wind velocity. To examine this orbital phase
dependence we have repeated model Peri-RD with stellar separations
corresponding to orbital phases in the range $\phi=0.96-1.04$, the
results of which are plotted in
Fig.~\ref{fig:preshock_velocity}. Error bars have been included to
indicate the range of preshock velocities attained due to oscillations
in the position of the companion shock. As the stellar separation
tends towards its periastron value the instabilities become more
vigorous, perturbing the shock front in the wind acceleration region
(where the velocity gradient is high) and thus broadening the range of
sampled velocities.

Comparing Fig.~\ref{fig:preshock_velocity} against figure 21 of
\cite{Parkin:2009} it is evident that the conservative assumption of
weak coupling adapted in the present paper leads to a reduced level of
radiative inhibition and noticeably higher preshock velocities (even
though fragments of the WCR stochastically collide with the companion
star at $\phi=1.0$ an average preshock velocity of $\simeq
2200\;$km~s$^{-1}$ is attained)\footnote{In this work the winds are
  initiated by forcing an outflow in a spherical shell around each
  star. This prevents accretion from over-powering the wind, which may
  shut down the companion's wind for a significant period of time
  \citep{Soker:2005, Akashi:2006, Kashi:2009}}.

\section{Orbit simulations}
\label{sec:orbit}

While the periastron simulations presented in \S~\ref{sec:periastron}
provide some interesting insight, to fully understand the flow
dynamics and resulting X-ray emission from $\eta$~Car we must perform
simulations which include the motion of the stars. For this purpose we
have performed two large-scale simulations: one in which the winds are
assumed to be instantaneously accelerated at the surface of the star
(model Orbit-IA), and another where the winds are radiatively driven
(model Orbit-RD). Our adopted orbital and stellar parameters are
noted in Tables~\ref{tab:system_parameters} and
\ref{tab:stellar_parameters}. The mapping of the winds in model
Orbit-IA is described in Appendix~\ref{sec:remap}.

To model the orbit of the \etacar binary system we use a simulation
domain which extends from $x = y = \pm 2\times10^{15}\;$cm and $z = (0
- 2\times10^{15})$cm. Outflow boundary conditions are used for all
faces of the simulation box except the lower $z$ boundary which is a
symmetry boundary. The grid is initialized with $x \times y \times z =
16 \times 16 \times 8$ cubic blocks each containing $8^{3}$ cells. We
allow for 9 levels of refinement, which results in an effective
resolution on the finest grid level of $x \times y \times z = 32768
\times 32768 \times 16384\;$(i.e. a cell size of
$1.22\times10^{11}\;$cm or $1.75\;{\rm R_{\odot}}$). The refinement of
the grid is determined using a second-derivation error check
\citep{Fryxell:2000} on $\rho$ and the requirement of an effective
number of cells between the stars to accurately describe the WCR
dynamics (see Appendix~\ref{sec:refinement}). The former identifies
cells for refinement (and derefinement), whereas the latter controls
the maximum resolution (for postshock gas) at a given orbital phase.

In the following we discuss the dynamics from simulations Orbit-IA and
Orbit-RD, focusing first on the large-scale dynamics and then
concentrating on the occurrences around periastron passage. We then
present the results of X-ray calculations performed on the simulation
output and detailed comparisons against observed lightcurves and
spectra obtained with {\it RXTE} and \XMM.

\subsection{Dynamics}
\label{subsec:dynamics}

\subsubsection{Large-scale dynamics}

\begin{figure}
  \begin{center}
    \begin{tabular}{cc}
\resizebox{40mm}{!}{\includegraphics{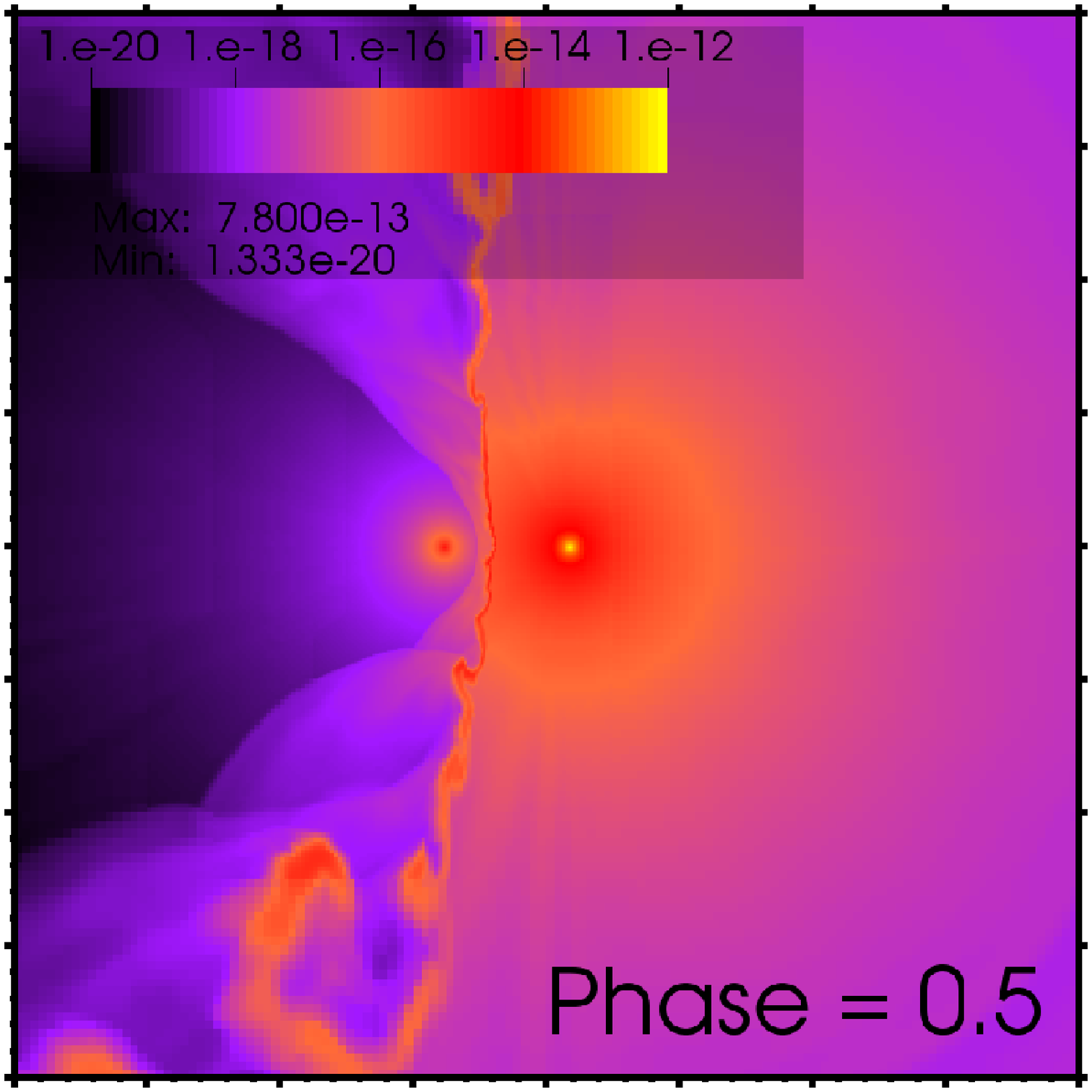}} & 
\resizebox{40mm}{!}{\includegraphics{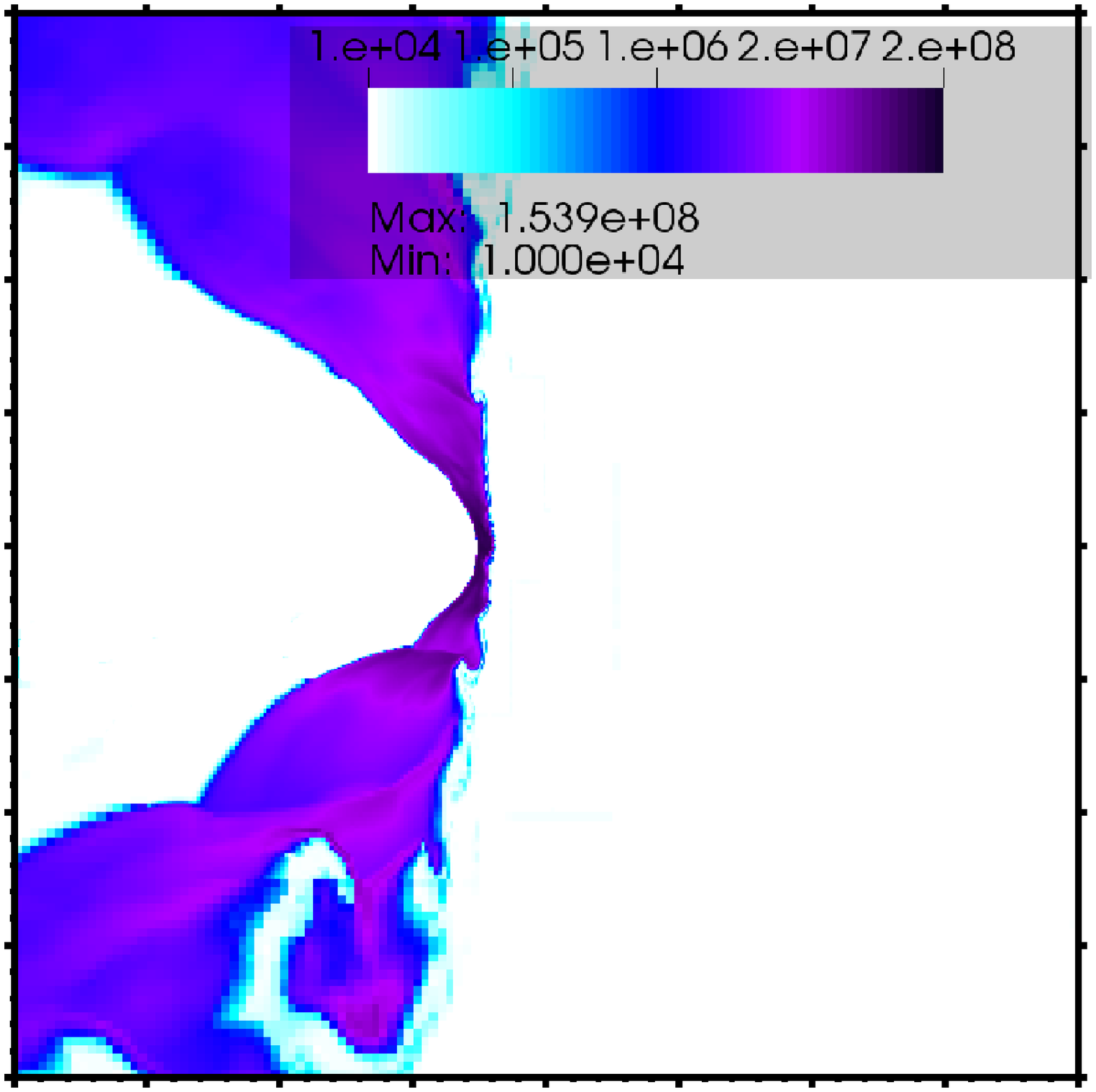}} \\ 

\resizebox{40mm}{!}{\includegraphics{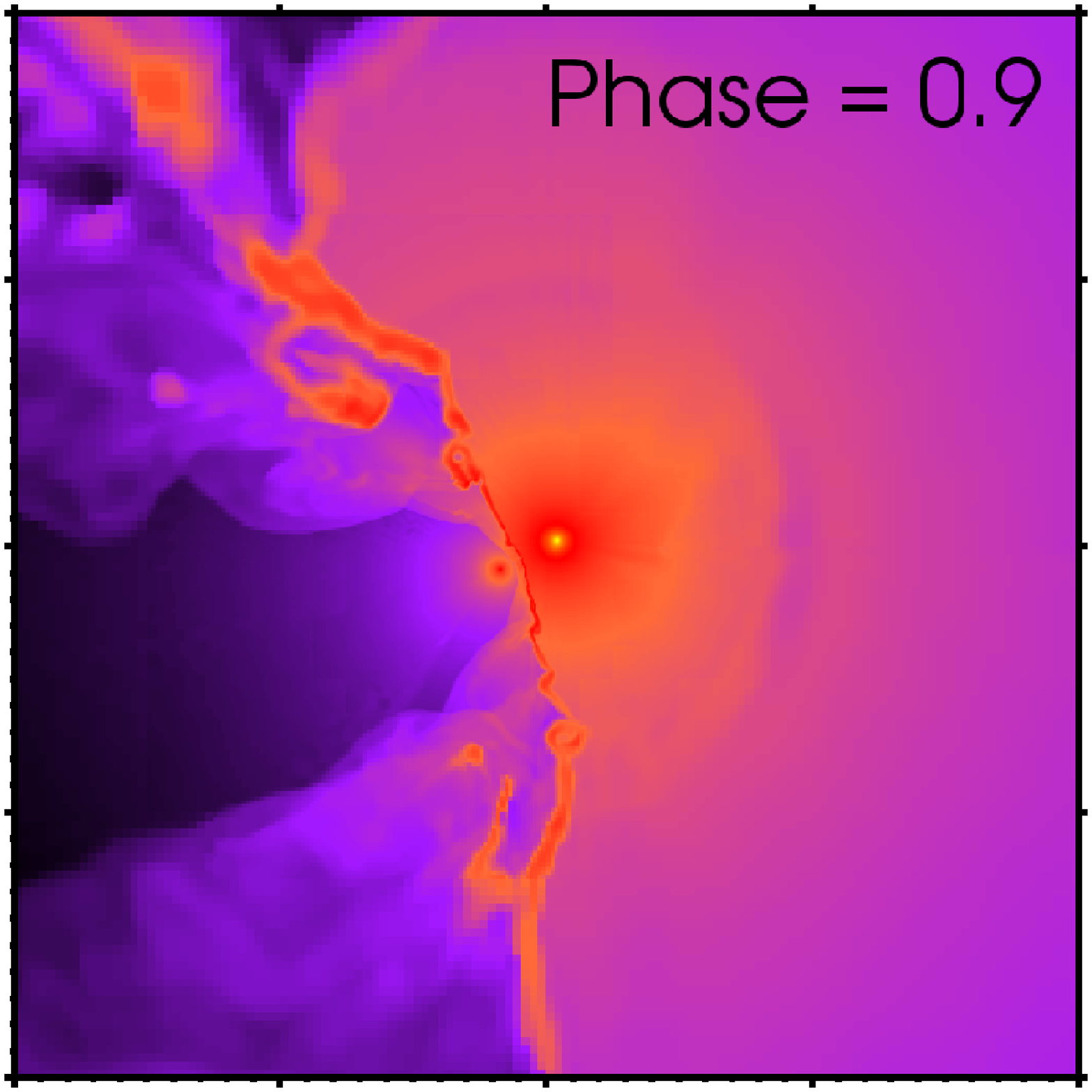}} &
\resizebox{40mm}{!}{\includegraphics{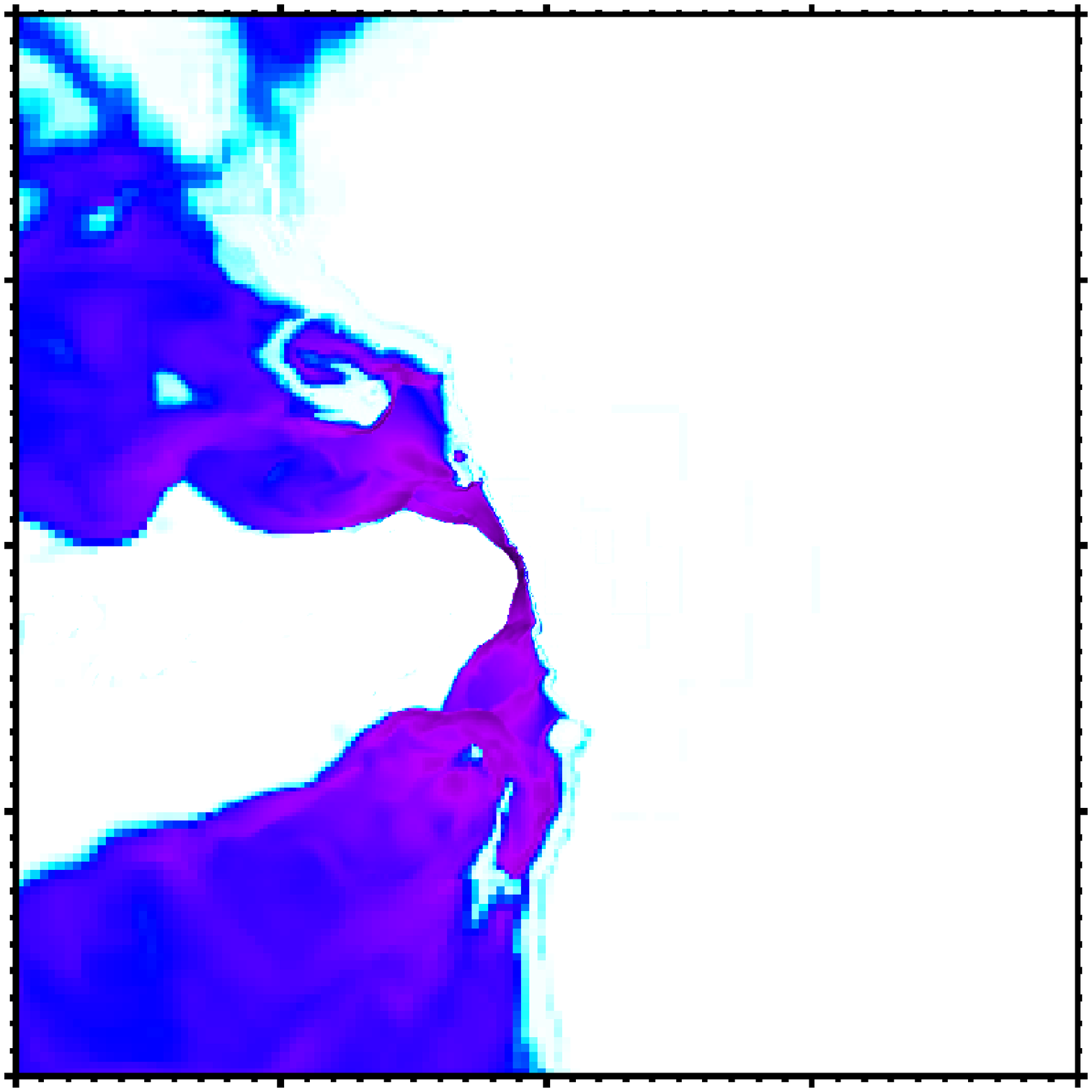}} \\ 

\resizebox{40mm}{!}{\includegraphics{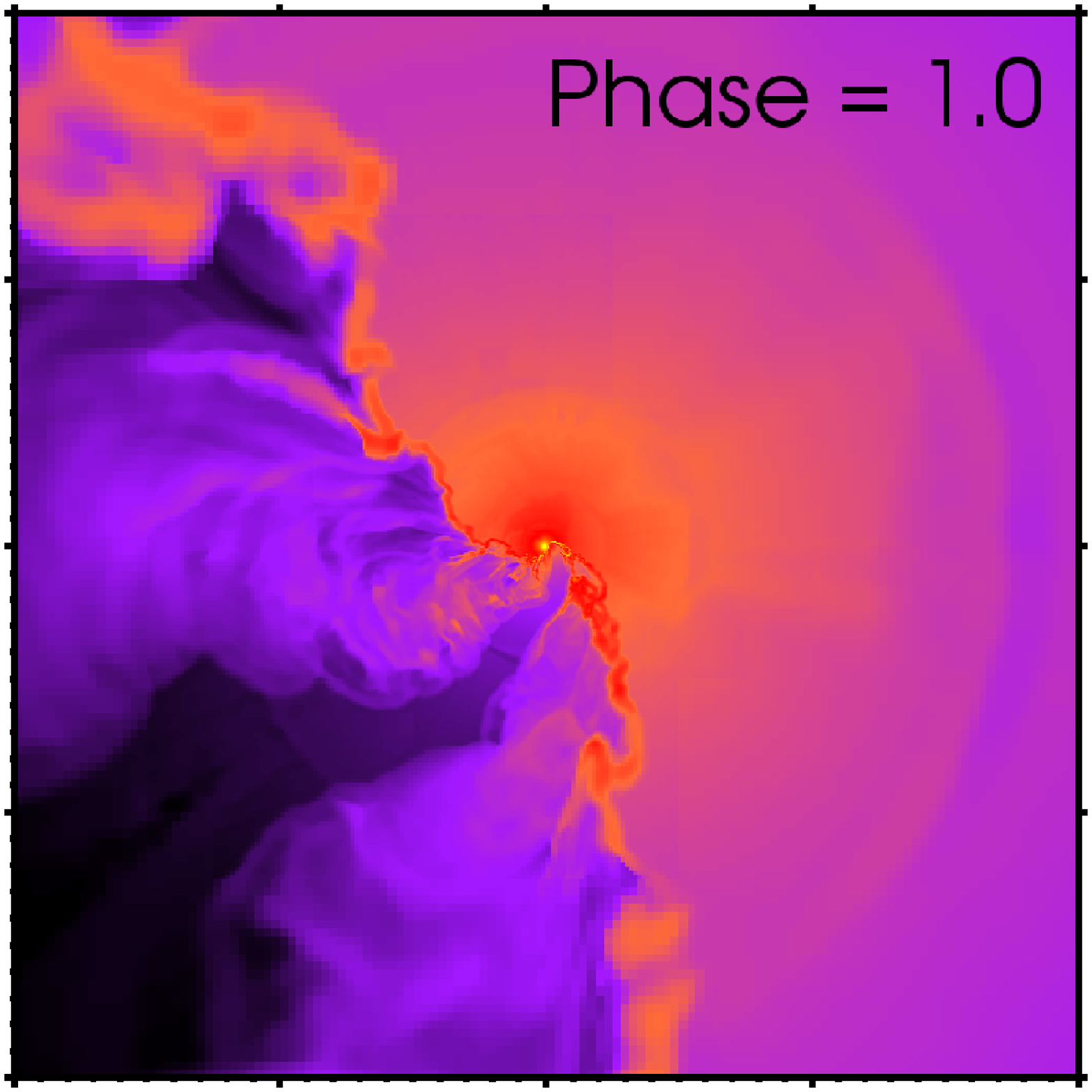}} &
\resizebox{40mm}{!}{\includegraphics{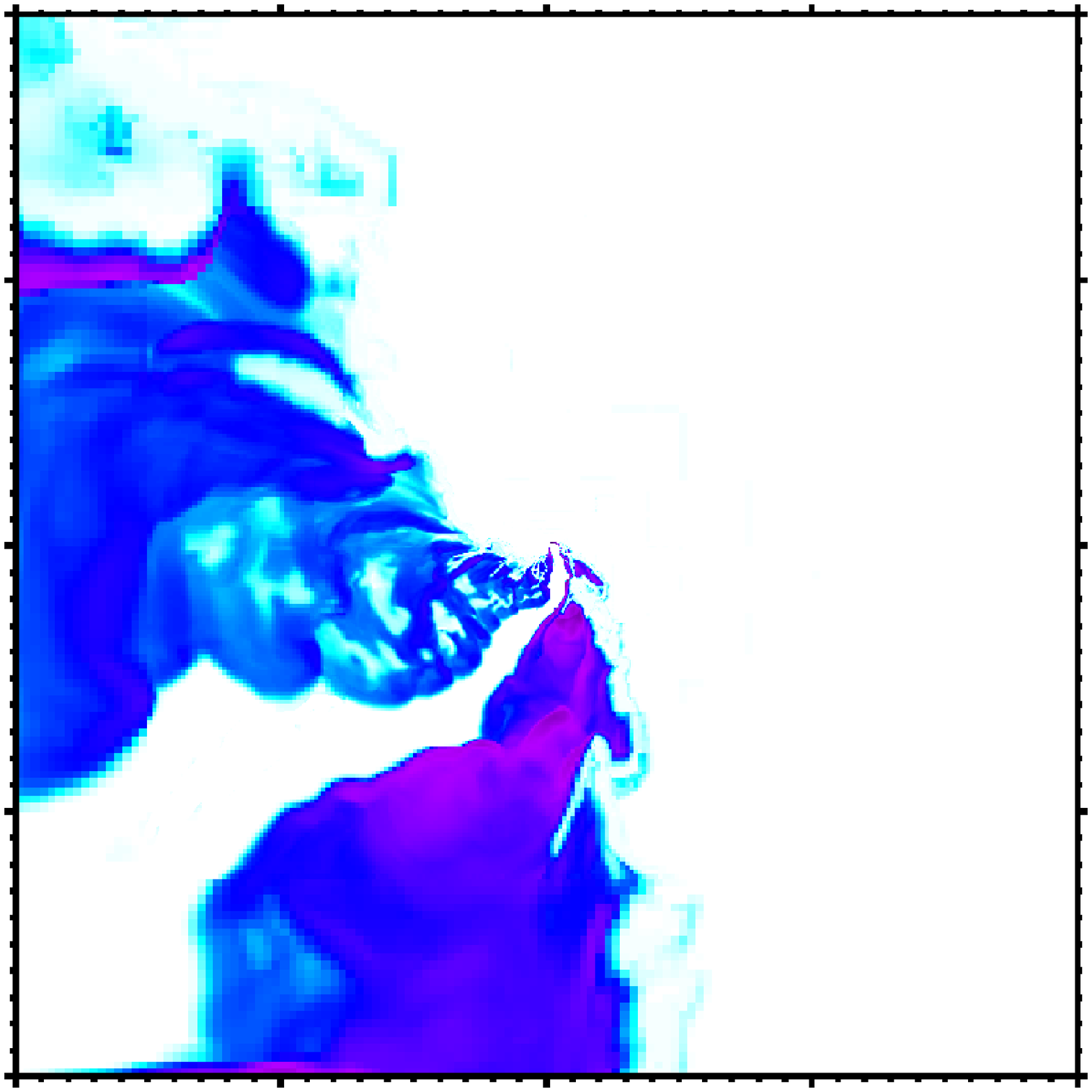}} \\ 

\resizebox{40mm}{!}{\includegraphics{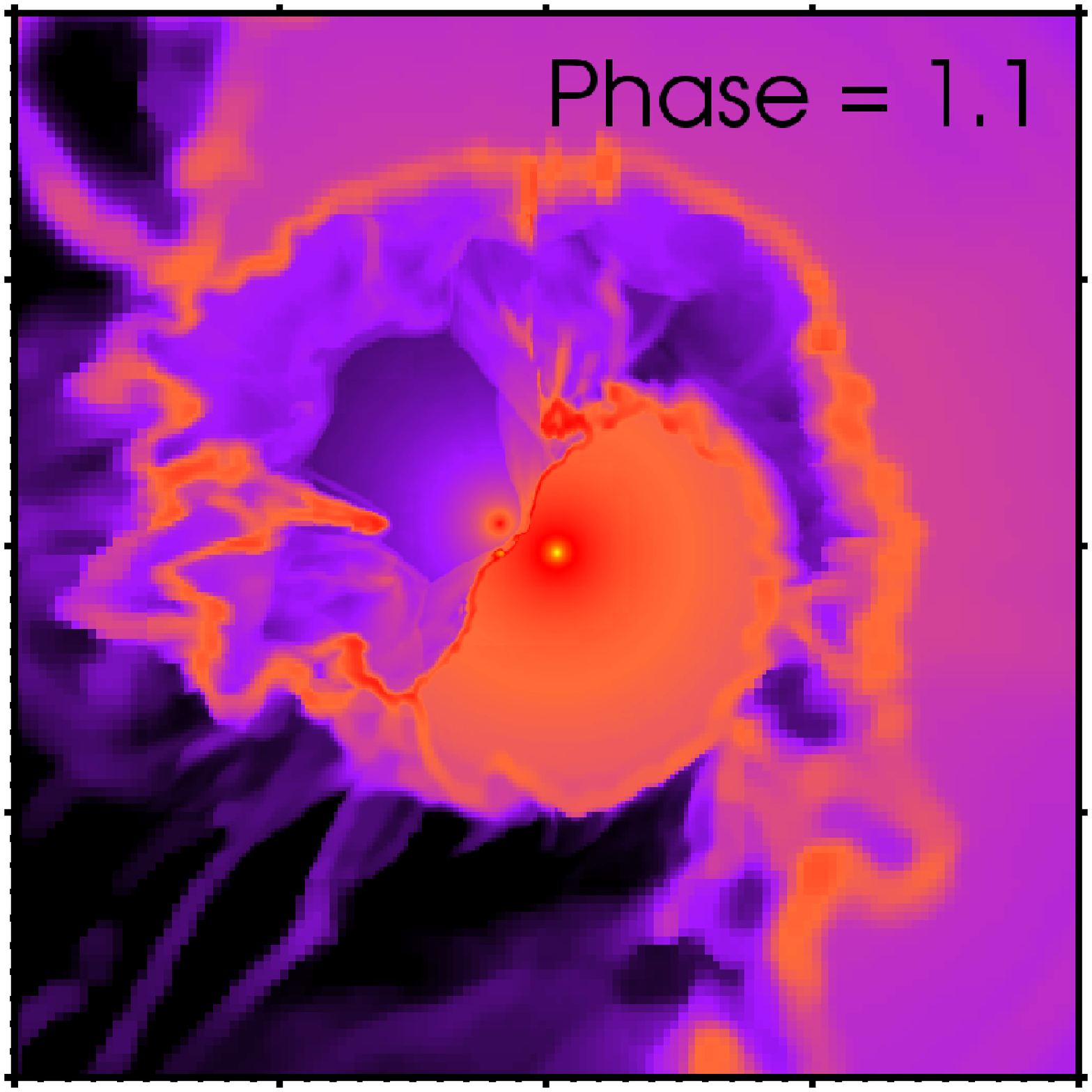}} & 
\resizebox{40mm}{!}{\includegraphics{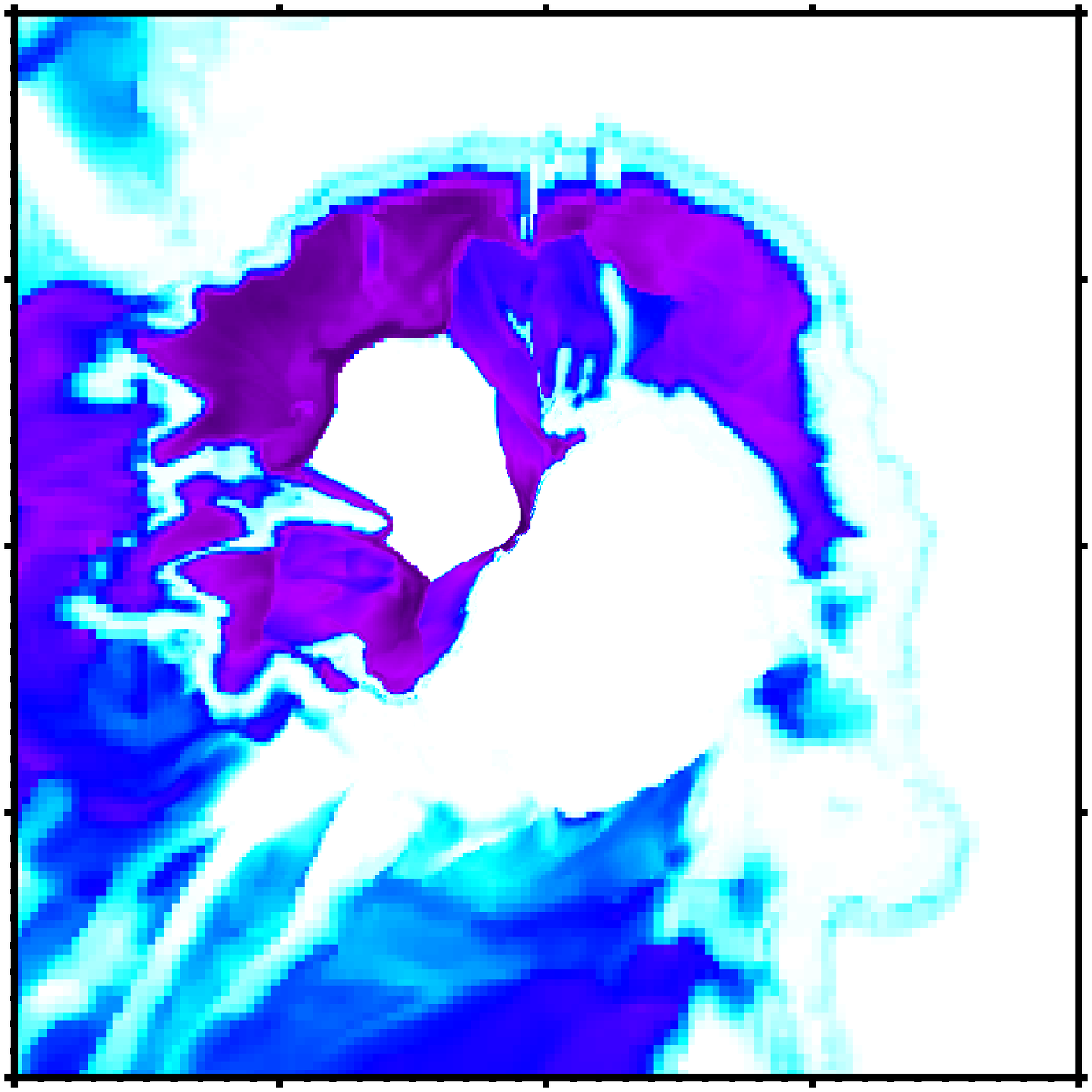}} \\ 
    \end{tabular}
    \caption{Snapshots of the gas density (left column) and
      temperature (right column) in the orbital ($x-y$) plane from
      model Orbit-IA at $\phi = 0.5\;$(top row), 0.9 (upper
      middle row), 1.0 (lower middle row), and 1.1 (bottom row). The
      orbital motion of the stars is calculated in the centre of mass
      frame. At apastron ($\phi =0.5$) the primary star is to the
      right, and the companion star is to the left, of the image
      centre. The motion of the stars proceeds in an anti-clockwise
      direction. All plots show a region of $\pm2\times10^{15}\;$cm -
      large axis tick marks correspond to a distance of
      $1\times10^{15}\;$cm.}
    \label{fig:vterm_big_images}
  \end{center}
\end{figure}

\begin{figure}
  \begin{center}
    \begin{tabular}{cc}
\resizebox{40mm}{!}{\includegraphics{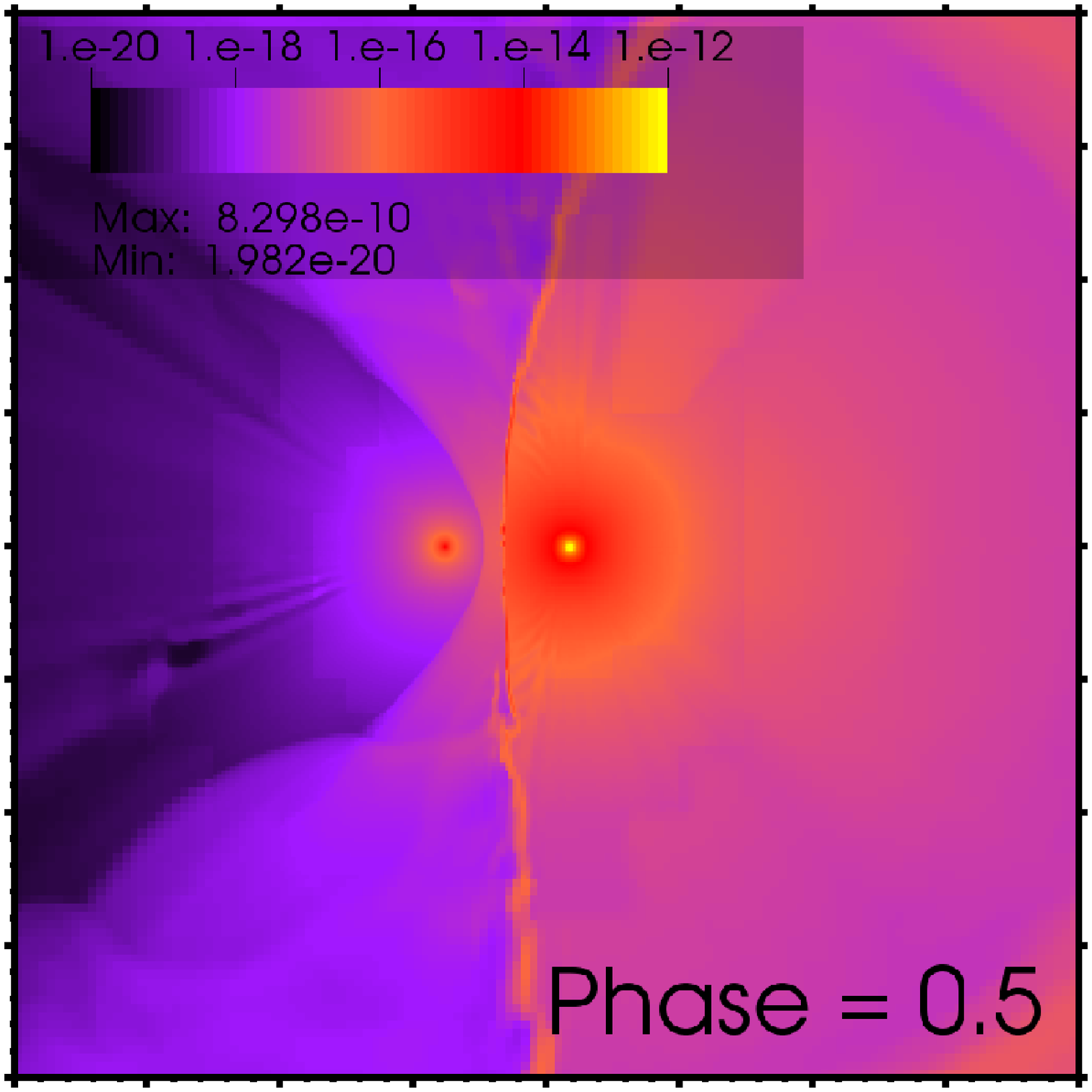}} & 
\resizebox{40mm}{!}{\includegraphics{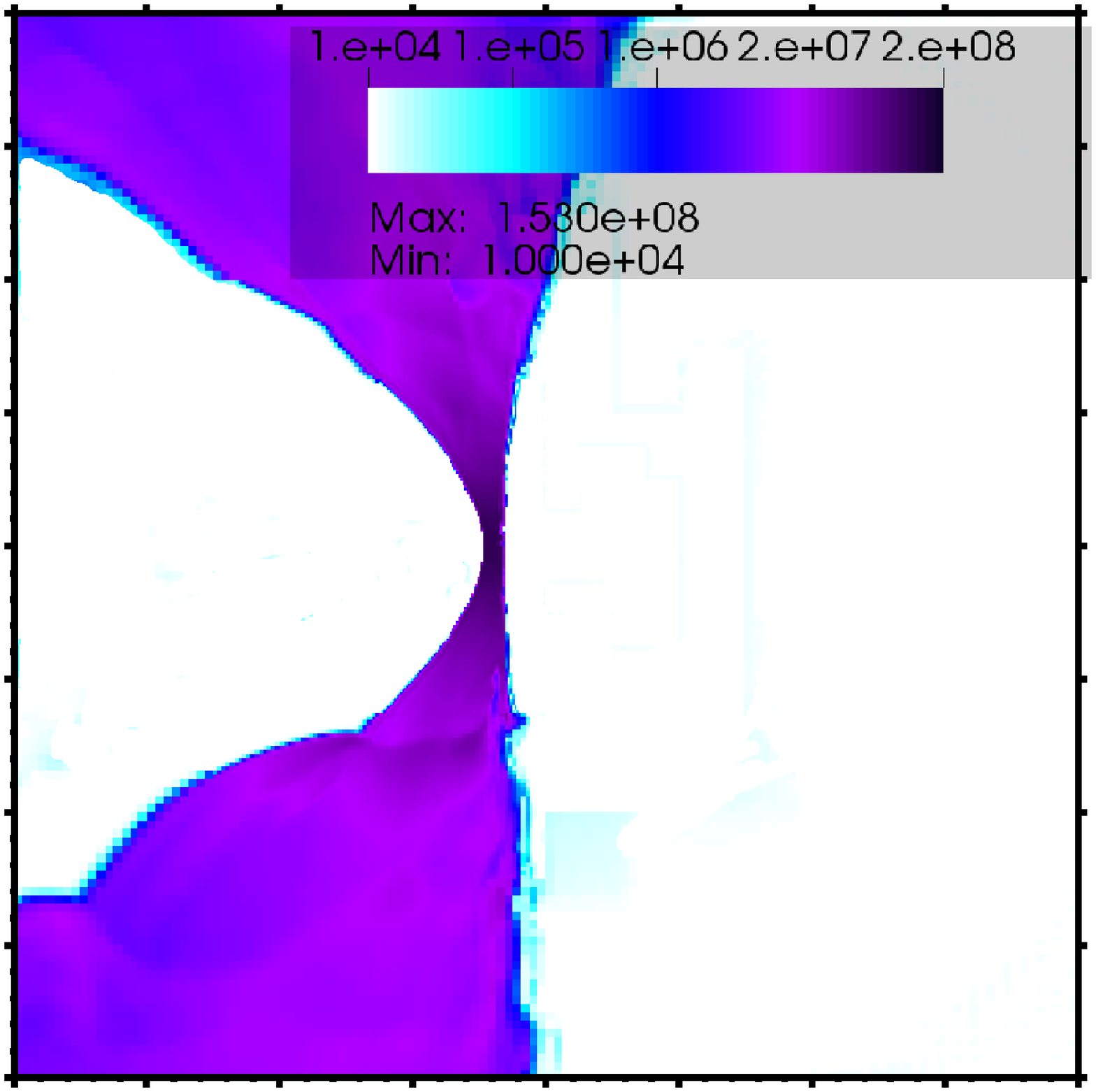}} \\ 

\resizebox{40mm}{!}{\includegraphics{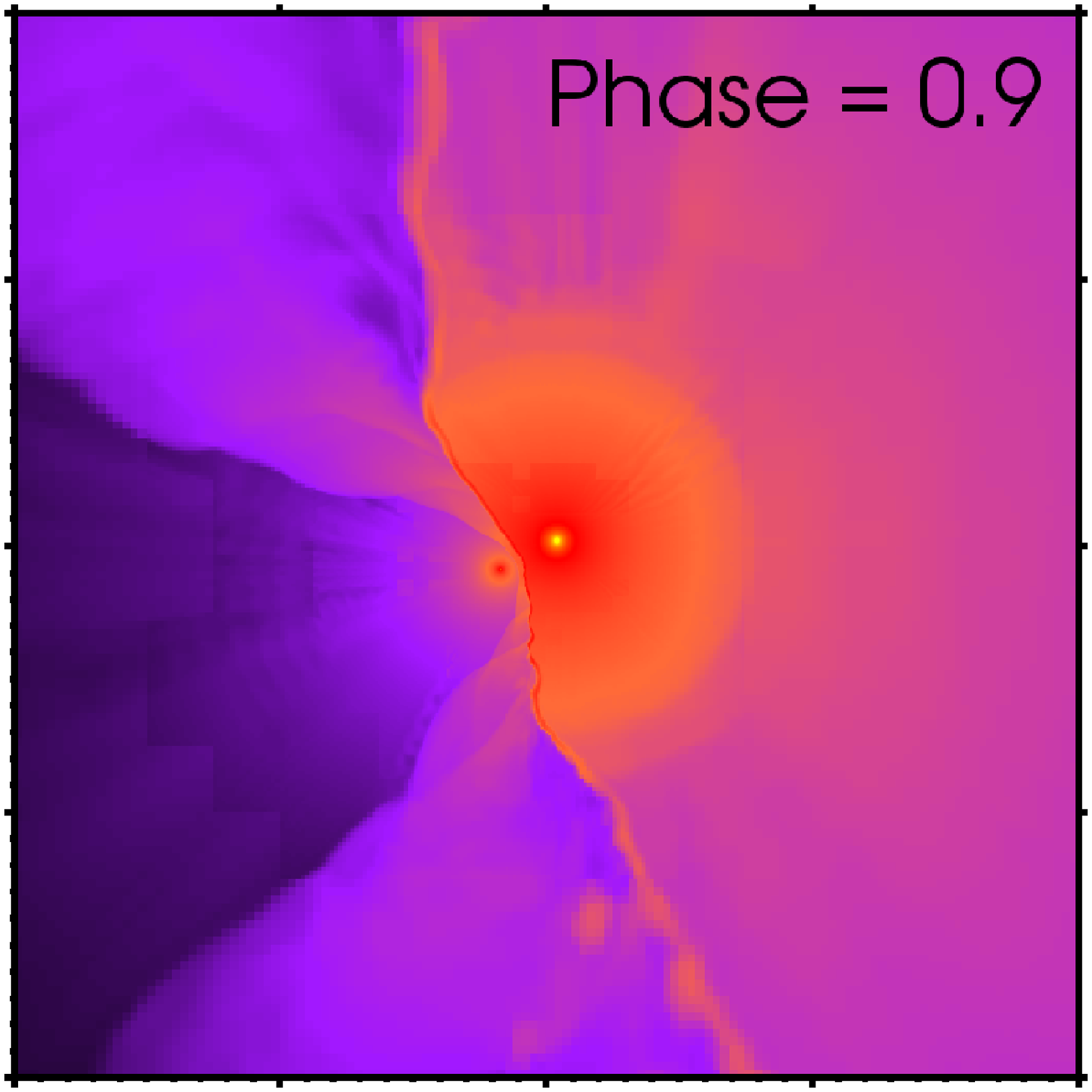}} &
\resizebox{40mm}{!}{\includegraphics{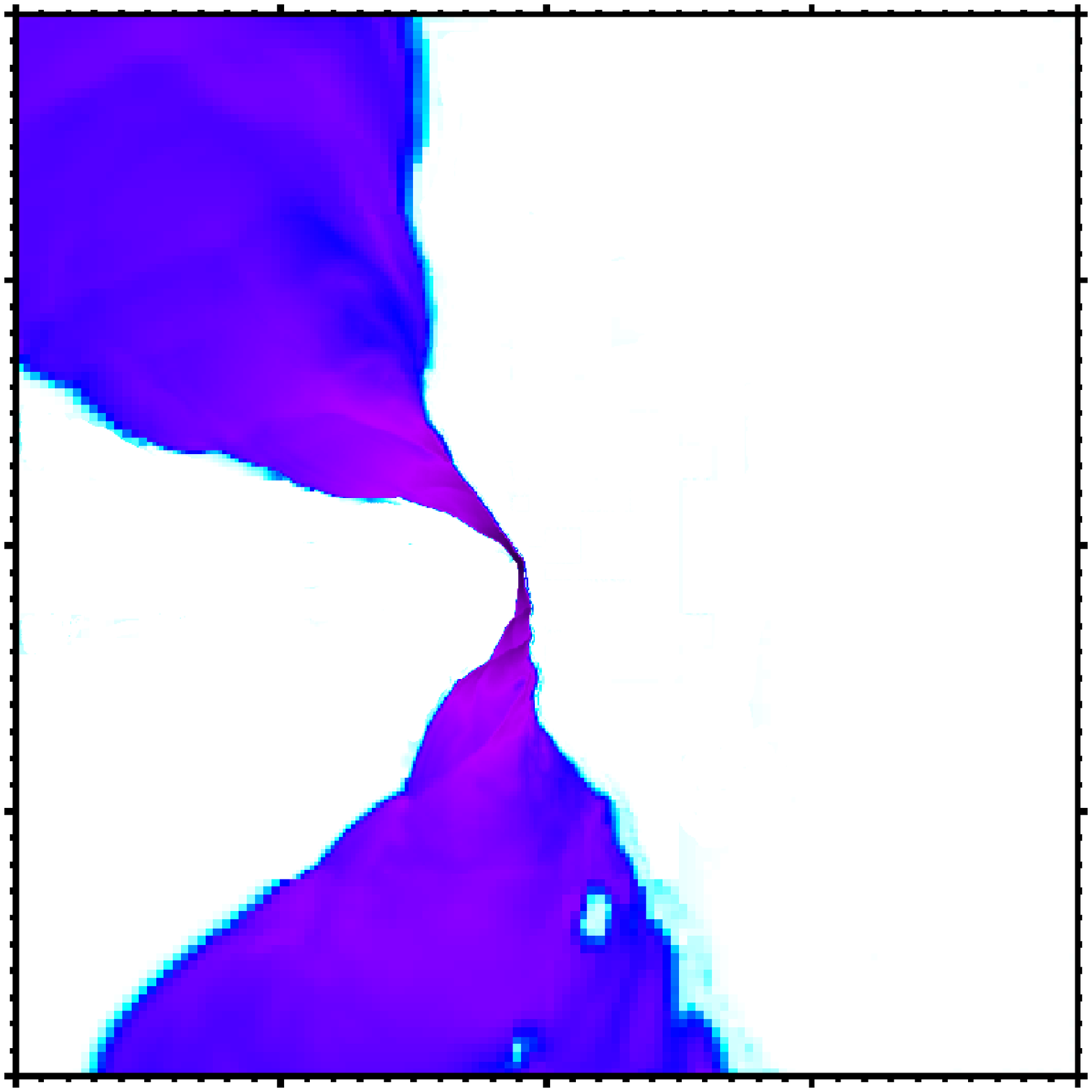}} \\ 

\resizebox{40mm}{!}{\includegraphics{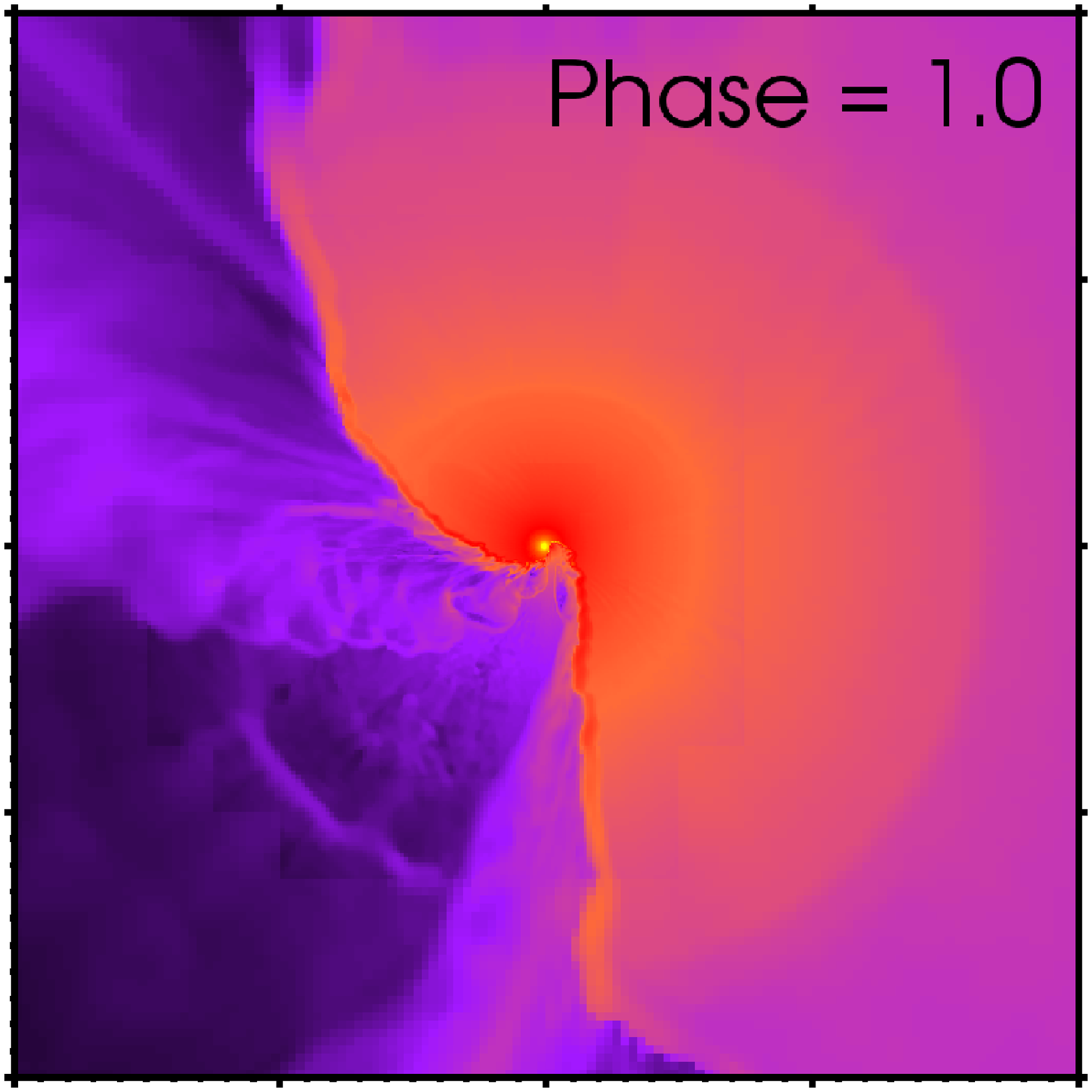}} &
\resizebox{40mm}{!}{\includegraphics{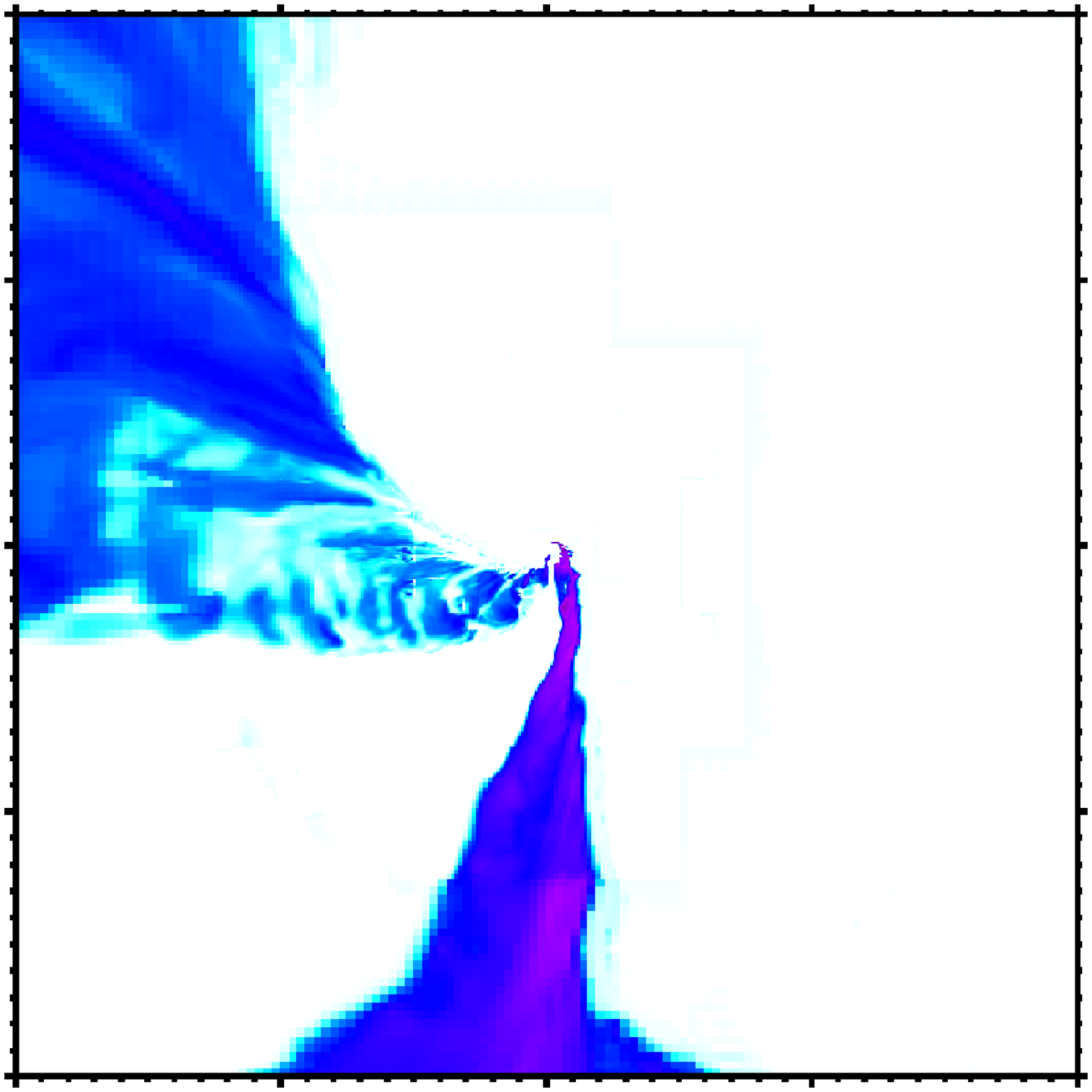}} \\ 

\resizebox{40mm}{!}{\includegraphics{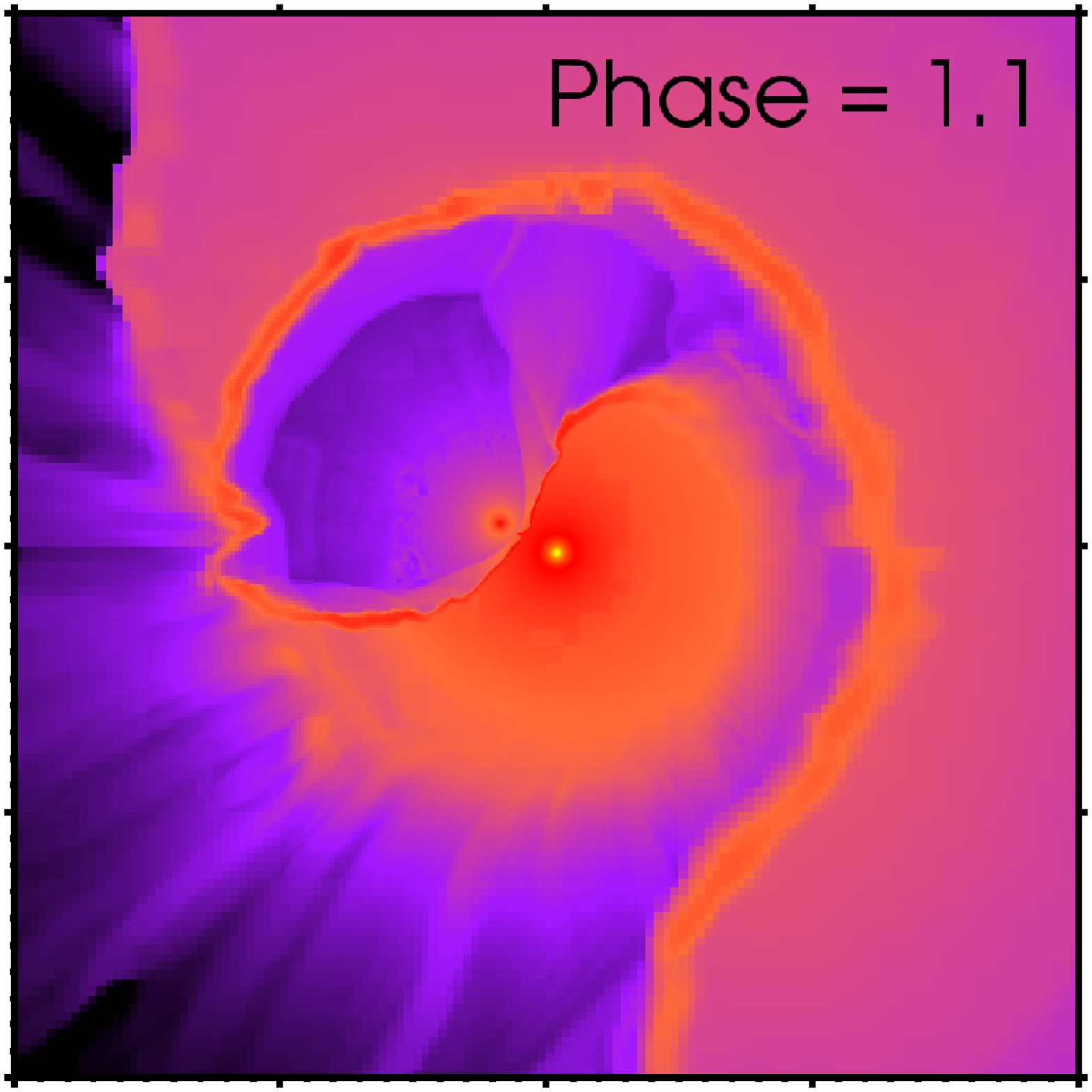}} & 
\resizebox{40mm}{!}{\includegraphics{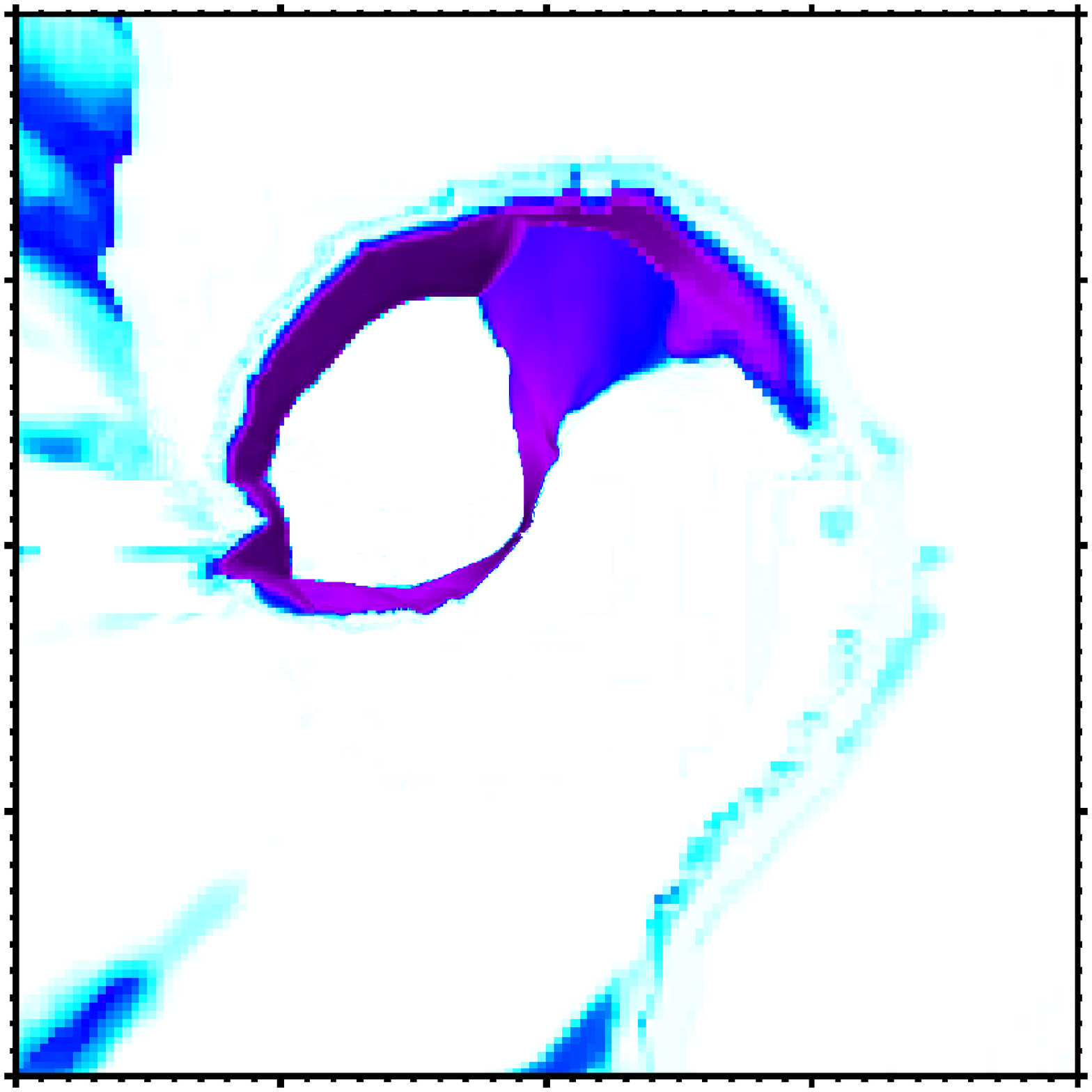}} \\ 
    \end{tabular}
    \caption{Same as Fig.~\ref{fig:vterm_big_images} except model
      Orbit-RD is shown.}
    \label{fig:driven_big_images}
  \end{center}
\end{figure}

For most of the orbit the stars advance relatively slowly and the
effects of orbital motion on the gas dynamics are modest. At these
times the separation of the stars is sufficiently large for the
stellar winds to be at terminal velocity when they reach their
respective shock. However, around periastron passage the orbital
velocities and stellar separation change rapidly as the stars career
past each other. The rapid motion of the stars around periastron
passage acts to contort the WCR into a spiral-like shape which
subsequently expands outwards with the flow \citep{Okazaki:2008,
  Parkin:2009}. Figs.~\ref{fig:vterm_big_images} and
\ref{fig:driven_big_images} show a series of gas density and
temperature snapshots from simulations Orbit-IA and Orbit-RD,
respectively, from which the salient features of the wind-wind
collision and the effects of orbital motion can be seen. The contact
discontinuity separates the winds, being abutted by a thin dense layer
of postshock primary wind on one side, and a thicker ``puffed-up''
region of high temperature quasi-adiabatic postshock companion wind on
the other. The density contrast and velocity shear across the contact
discontinuity subject it to Kelvin-Helmholtz (KH) and RT instabilities.

At $\phi=0.5$, the stars are at their largest separation, moving at
their slowest orbital velocities, and it is apparent that a reasonable
approximation to the X-ray emitting part of the WCR can be attained
from a 2D model which neglects orbital motion
\citep[][]{Pittard:2002}. The apex of the WCR is closer to the primary
star in model Orbit-RD than in model Orbit-IA due to the primary's
wind ram pressure being slightly lower at an equivalent radius in the
former - this is a numerical artifact resulting from the different
ways by which the winds are initiated in the models. By $\phi=0.9$ the
stars have moved around in their orbits and their separation has
reduced, yet the general shape of the large scale WCR has not changed
considerably. However, the rapid motion of the stars during a brief
period around periastron causes the WCR close to the stars to become
very distorted (see Figs.~\ref{fig:vterm_peri_images} and
\ref{fig:driven_peri_images}), although the large scale WCR is not
impacted upon by periastron passage at $\phi=1.0$. At $\phi=1.1$ the
contortion of the WCR has had time to advect out in the flow, and the
spiral structure of the WCR visible in previous 3D models is
reproduced \citep{Okazaki:2008, Parkin:2009}. Interestingly, the
increasing influence of orbital motion as the stars approach
periastron results in an increasingly asymmetric temperature
distribution in the postshock gas, with gas in the leading arm of the
WCR clearly at a higher temperature than gas in the trailing arm
(Figs.~\ref{fig:vterm_big_images} and \ref{fig:driven_big_images}). In
model Orbit-IA the postshock gas in the trailing arm is far from
smooth; there are thin, dense filaments separated by low density,
rarefied gas.

\begin{figure*}
  \begin{center}
    \begin{tabular}{cc}
\resizebox{70mm}{!}{\includegraphics{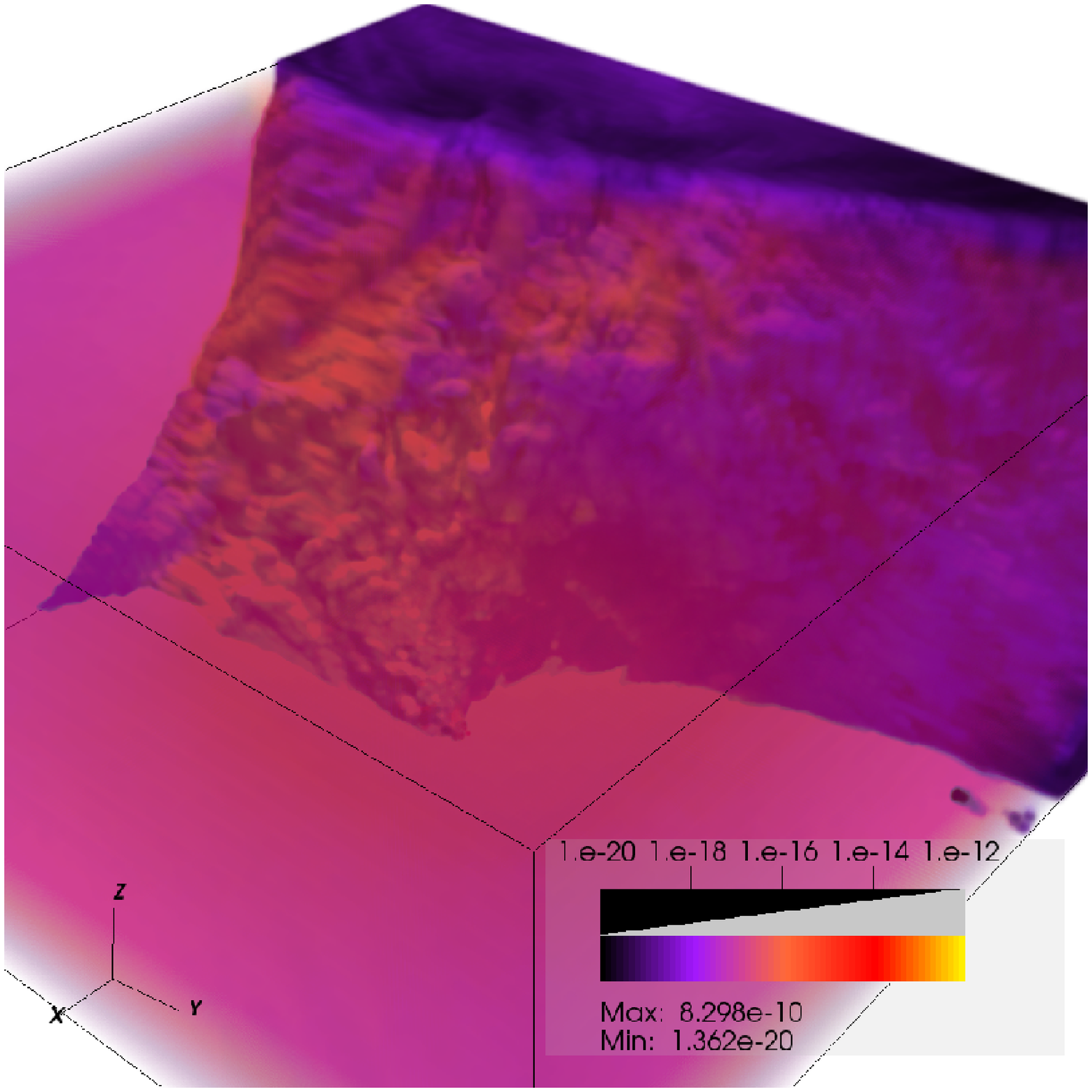}} &
\resizebox{70mm}{!}{\includegraphics{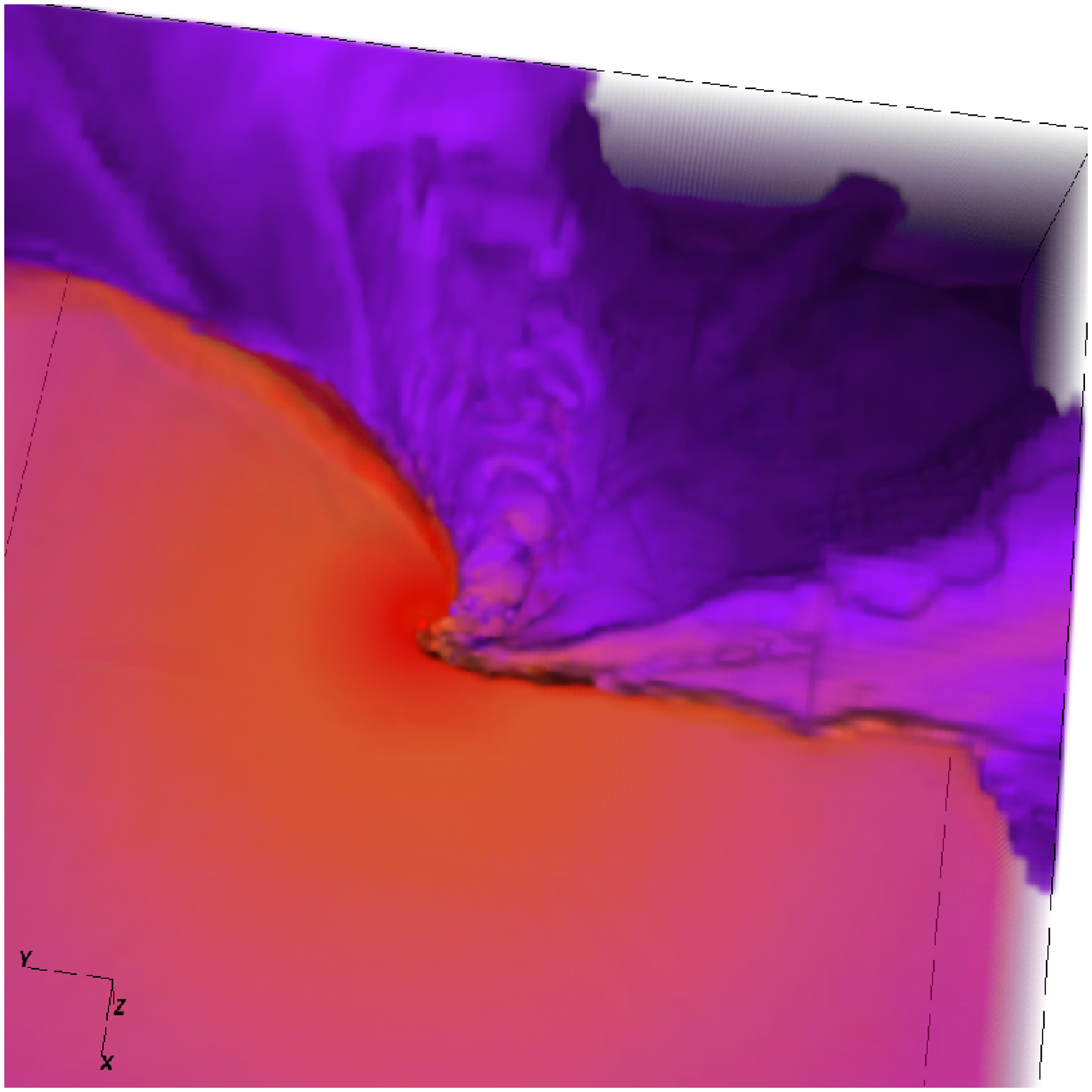}} \\
    \end{tabular}
    \caption{3D gas density volume rendering of simulation Orbit-RD
      at periastron passage ($\phi=1.0$) viewed from above (left
      panel) and below (right panel) the simulation box. The entire
      simulation domain is shown.}
    \label{fig:density_peri_volume}
  \end{center}
\end{figure*}

Fig.~\ref{fig:density_peri_volume} displays a snapshot of the 3D
structure of the WCR at periastron in simulation Orbit-RD. The
clumpy/rippled texture to the WCR is clearly evident, as is the
immense difference in scale between the large scale structure of the
WCR and the region between the stars where the most ferocious activity
occurs.

The snapshots of simulations Orbit-IA and Orbit-RD in
Figs.~\ref{fig:vterm_big_images} and \ref{fig:driven_big_images}
reveal a number of noticeable differences. Firstly, the opening angles
of the primary star and companion star shocks are slightly larger in
model Orbit-RD when compared to model Orbit-IA. This occurs due to
the radiation fields of the stars; off-axis the preshock speed is
greater than the on-axis preshock speed. The radiation fields of the
stars also cause differences in the postshock gas between the two
simulations. For instance, there are less instabilities in the WCR in
model Orbit-RD when compared to model Orbit-IA. Examining the layer
of postshock primary wind in the two models we find that the gas
velocity (density) is higher (lower) in model Orbit-RD. Noting that
the growth rate for KH instability is $\propto \sqrt{\rho_1
  \rho_2}/(\rho_1 + \rho_2)$, the lower density and higher velocity in
model Orbit-RD results in slower growth rates. Therefore, the flow
exits the grid with fewer noticeable perturbations. But why does this
difference occur? Consider the motion of postshock gas away from the
stagnation point. In model Orbit-IA postshock gas is accelerated by
the pressure gradient. However, in model Orbit-RD the postshock
primary wind is of a low enough temperature to be acted upon by the
radiation fields of the stars. Therefore, the higher velocity of
postshock primary's wind in model Orbit-RD is due to radiative
acceleration.

\begin{figure}
  \begin{center}
    \begin{tabular}{cc}
\resizebox{40mm}{!}{\includegraphics{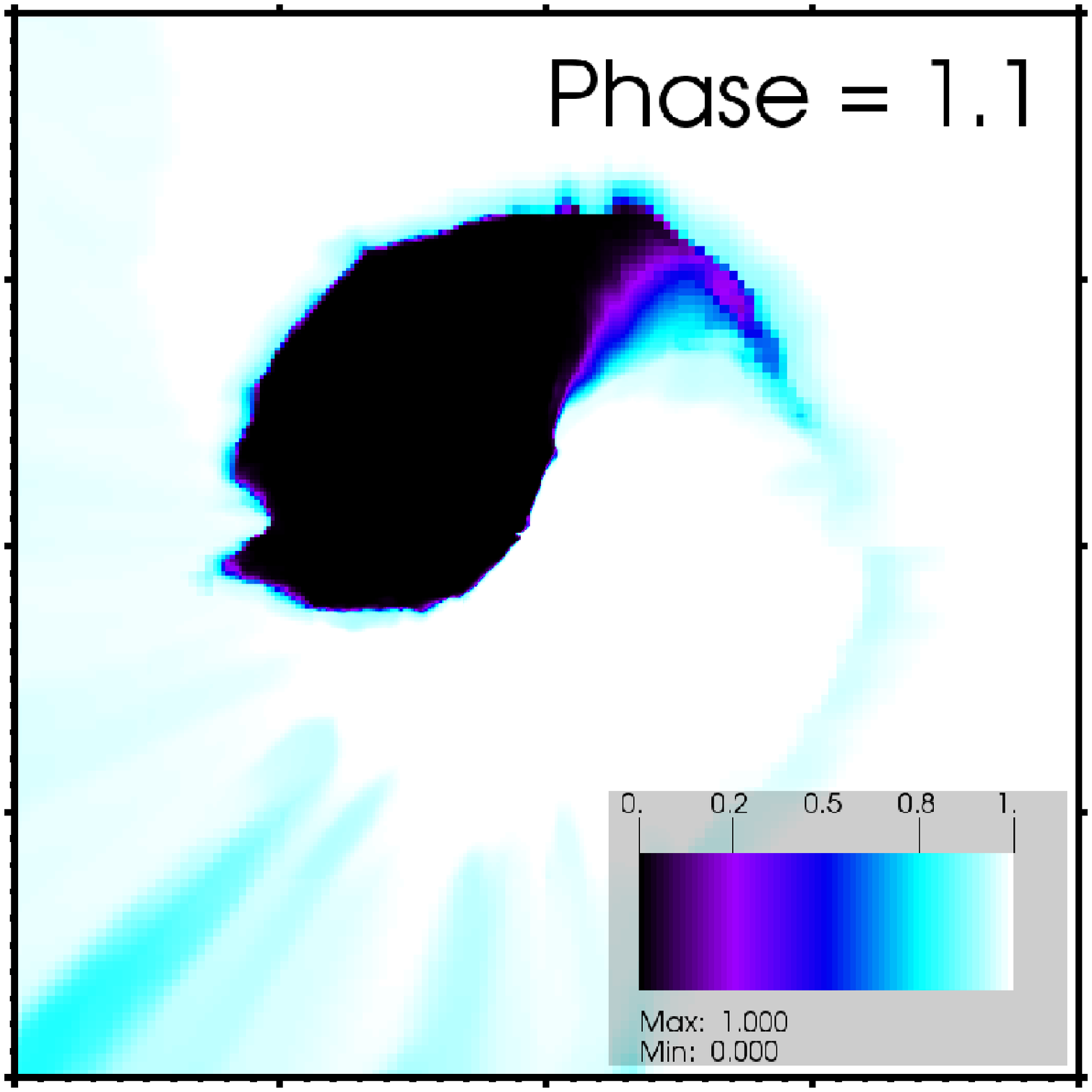}} &
\resizebox{40mm}{!}{\includegraphics{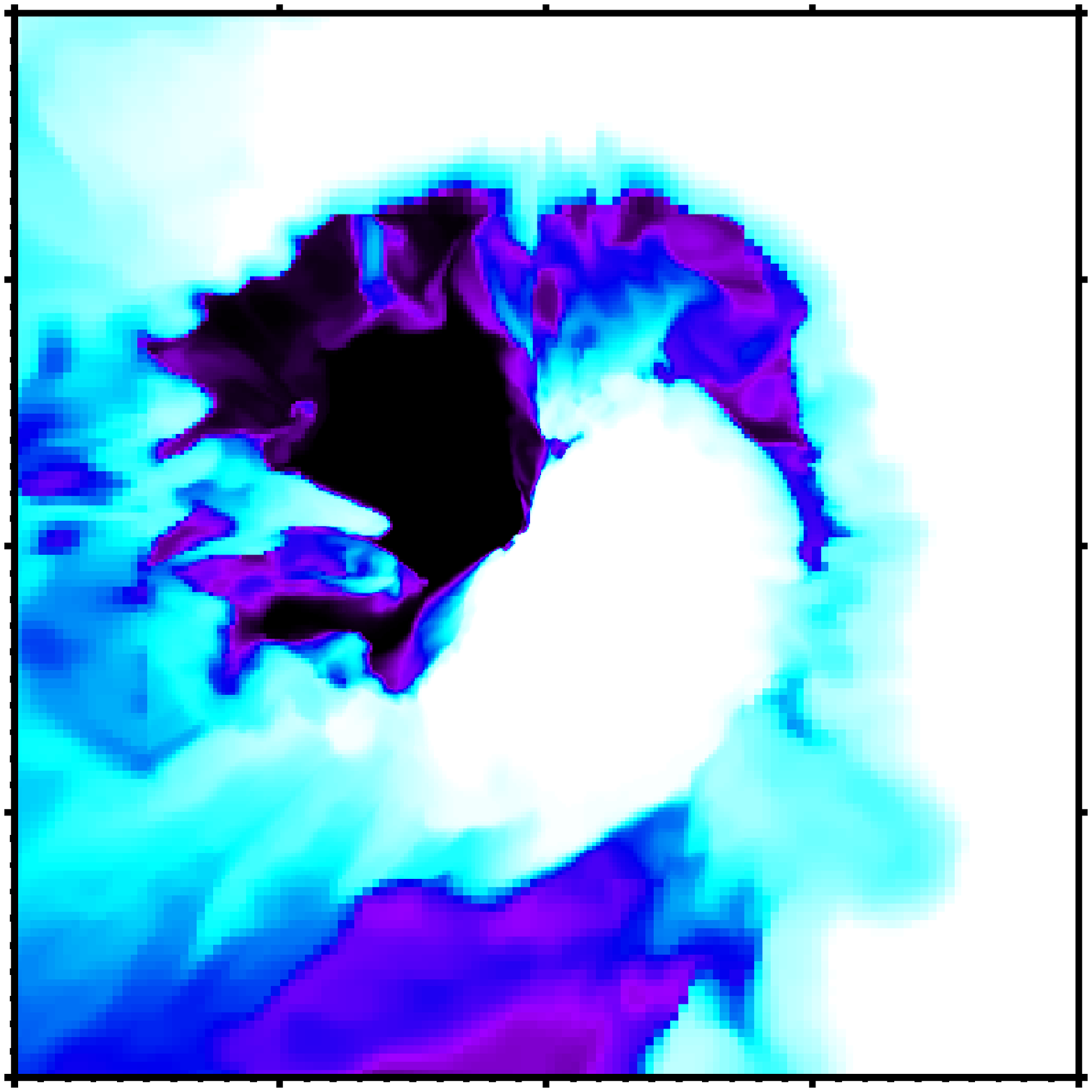}} \\
    \end{tabular}
    \caption{Snapshots of orbital ($x-y$) plane from simulation
      Orbit-RD (left panel) and Orbit-IA (right panel) at $\phi =
      1.1$ showing the fluid dye variable.  Values of 1 and 0
      correspond to a cell consisting entirely of primary or companion
      wind material, respectively, and intermediate values indicate a
      mixture. The plots show a region of $\pm2\times10^{15}\;$cm -
      large axis tick marks correspond to a distance of
      $1\times10^{15}\;$cm. For the corresponding density and
      temperature images see Figs.~\ref{fig:vterm_big_images} and
      \ref{fig:driven_big_images}.}
    \label{fig:colour_images}
  \end{center}
\end{figure}

Examining the leading arm of the WCR the layer of postshock primary
wind appears thicker in model Orbit-RD when compared to model
Orbit-IA. Interestingly, at phases close to periastron when this gas
resides close to the stars, its thermal pressure is lower than the
radiation pressure, which provides resistance against contraction and
thus widens the layer. However, comparing the snapshots at $\phi=1.1$
we see that at an equivalent distance from the stars the density of
the unshocked winds is slightly higher in model Orbit-RD, which means
the mass in the swept-up shell is greater. The inertia of the swept-up
mass is therefore greater in model Orbit-RD, which accounts for the
smaller distance that the spiral has travelled to by $\phi=1.1$ in
this model compared to model Orbit-IA. 

The width of the dense layer clearly affects the growth of
instabilities in the expanding spiral-shaped shell - in model Orbit-IA
the shell appears to be subject to the NTSI, whereas in model
Orbit-RD the additional thickness to the layer renders it
stable. This is unsurprising as the stability of an expanding shell
depends on the shock thickness \citep{Vishniac:1983,
  Wunsch:2010}. This raises questions about the fate of the expanding
shell in each simulation. As its outwards acceleration is decreased by
an increasing amount of swept up mass its Mach number will decrease
and as the shocks dissipate it will gradually mix with the bubble of
companion wind which it encases. However, this only appears to be the
case for model Orbit-RD in certain directions. In the trailing arm of
the WCR the dense layer of postshock primary wind is photo-ablated by
the radiation fields. Yet, this only occurs in the trailing arm of the
WCR because the dense layer can be driven into a more tenuous cavity
made by the companion's wind as it swings through periastron. In
contrast, the leading arm is bordered by unshocked primary wind, and
therefore the dense layer is merely widened by the radiation
pressure. Examining the fluid dye variable (which tracks the quantity
of each wind in a given cell) the photo-ablation by the radiation
fields of the stars appears to be a very effective mixing agent and
(unlike in model Orbit-IA) smooth out any filamentary
structure. Comparing models Orbit-RD and Orbit-IA at $\phi=1.1$ we
see that in the former the photo-ablation of the dense layer snips the
tail of companion wind gas in the WCR
(Fig.~\ref{fig:colour_images}). 

Following the expansion of the shell as it exits the system, we see
that the shocked gas in the current spiral crashes through the remnant
of the WCR from prior to periastron passage, and Richtmyer-Meshkov
(RM) instabilities are formed at the interface between the two. The
additional mixing provided by photo-ablation and the collision of
successive expanding spiral-like shells will have implications for
models of observations directly related to the postshock primary
star's wind, e.g. forbidden line emission \citep[][]{Gull:2009},
episodic dust formation \citep{Smith:2010}, and high-velocity
absorption features \citep{Groh:2010a}.

\begin{figure*}
  \begin{center}
    \begin{tabular}{ccc}
\resizebox{40mm}{!}{\includegraphics{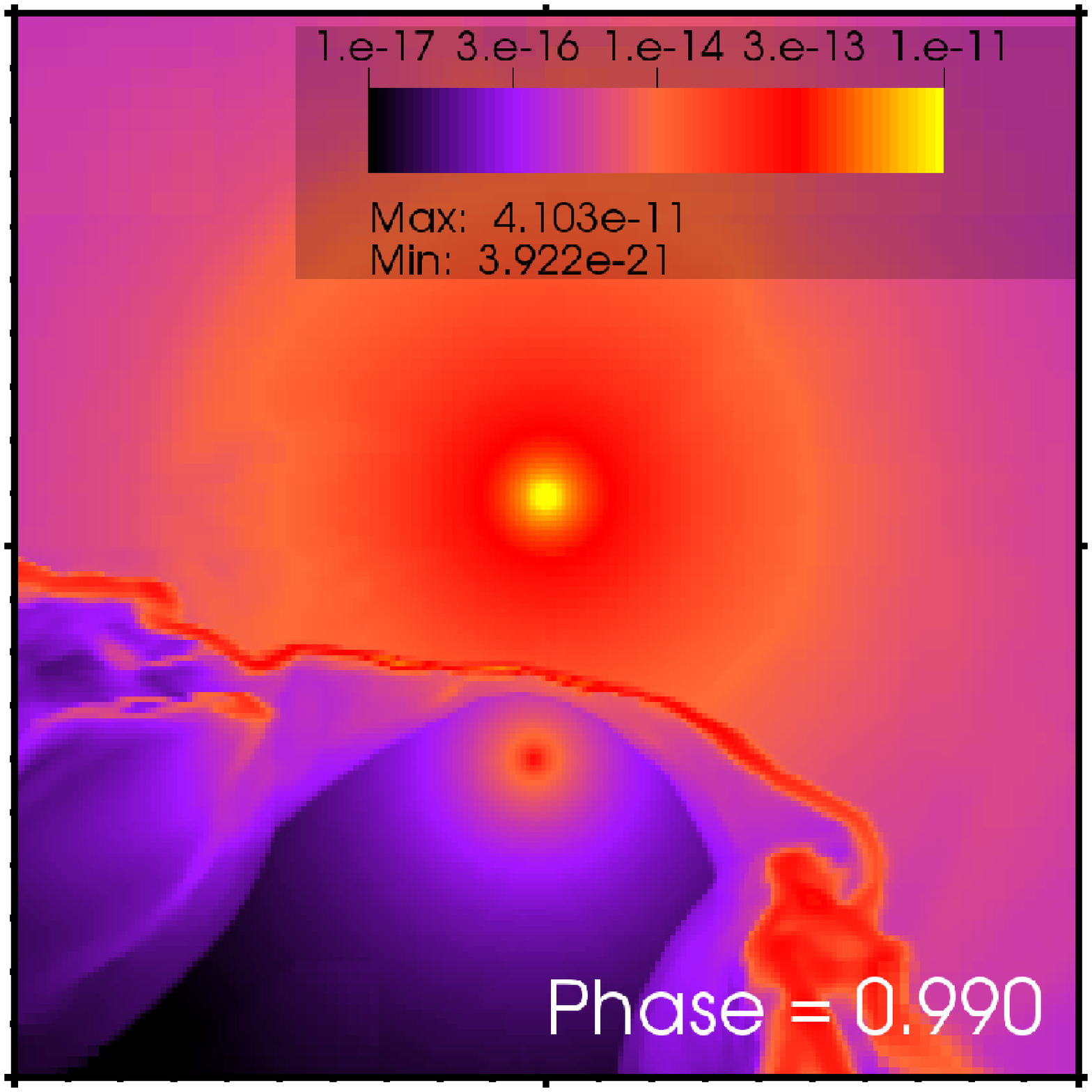}} & 
\resizebox{40mm}{!}{\includegraphics{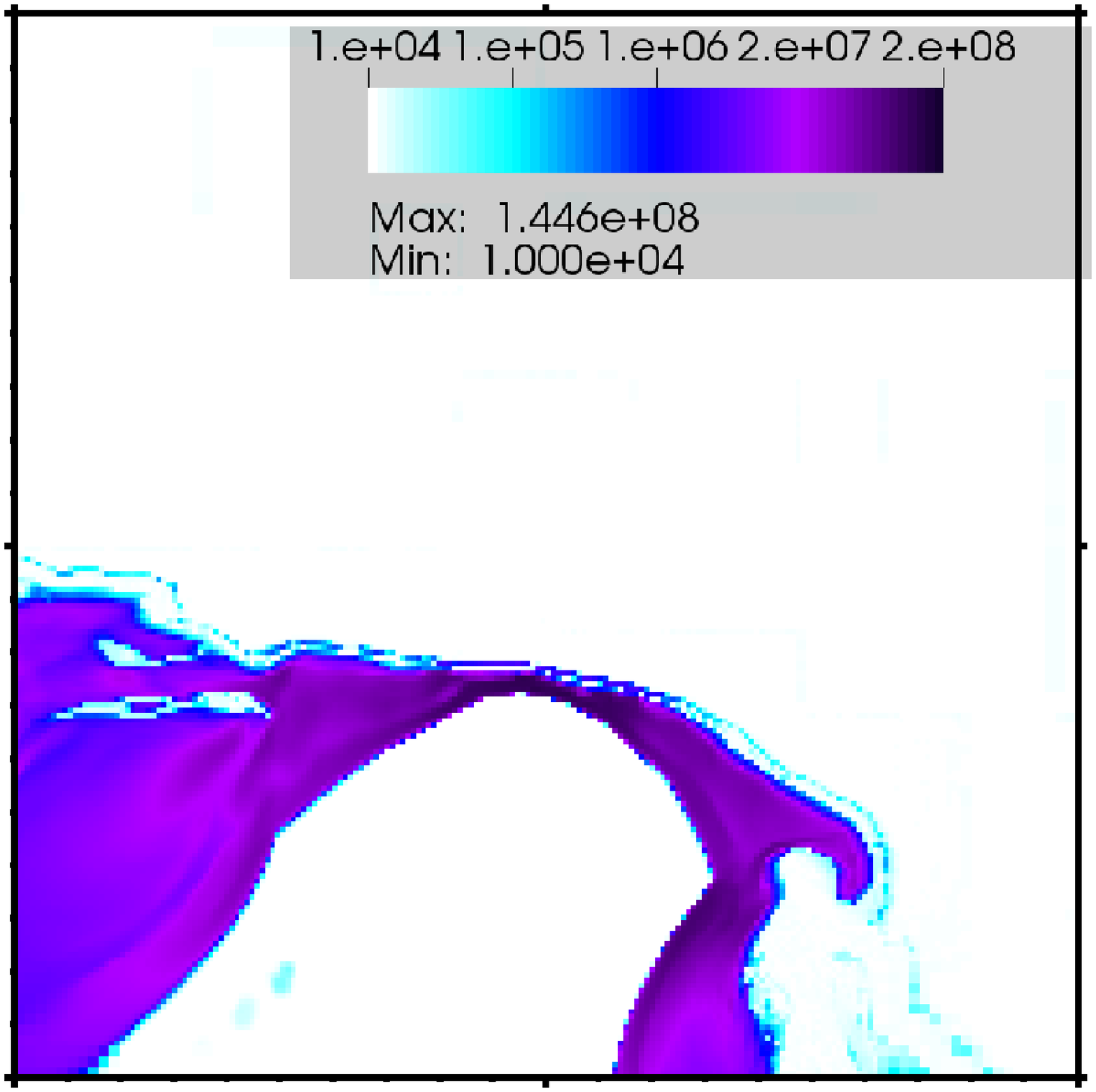}} & 
\resizebox{40mm}{!}{\includegraphics{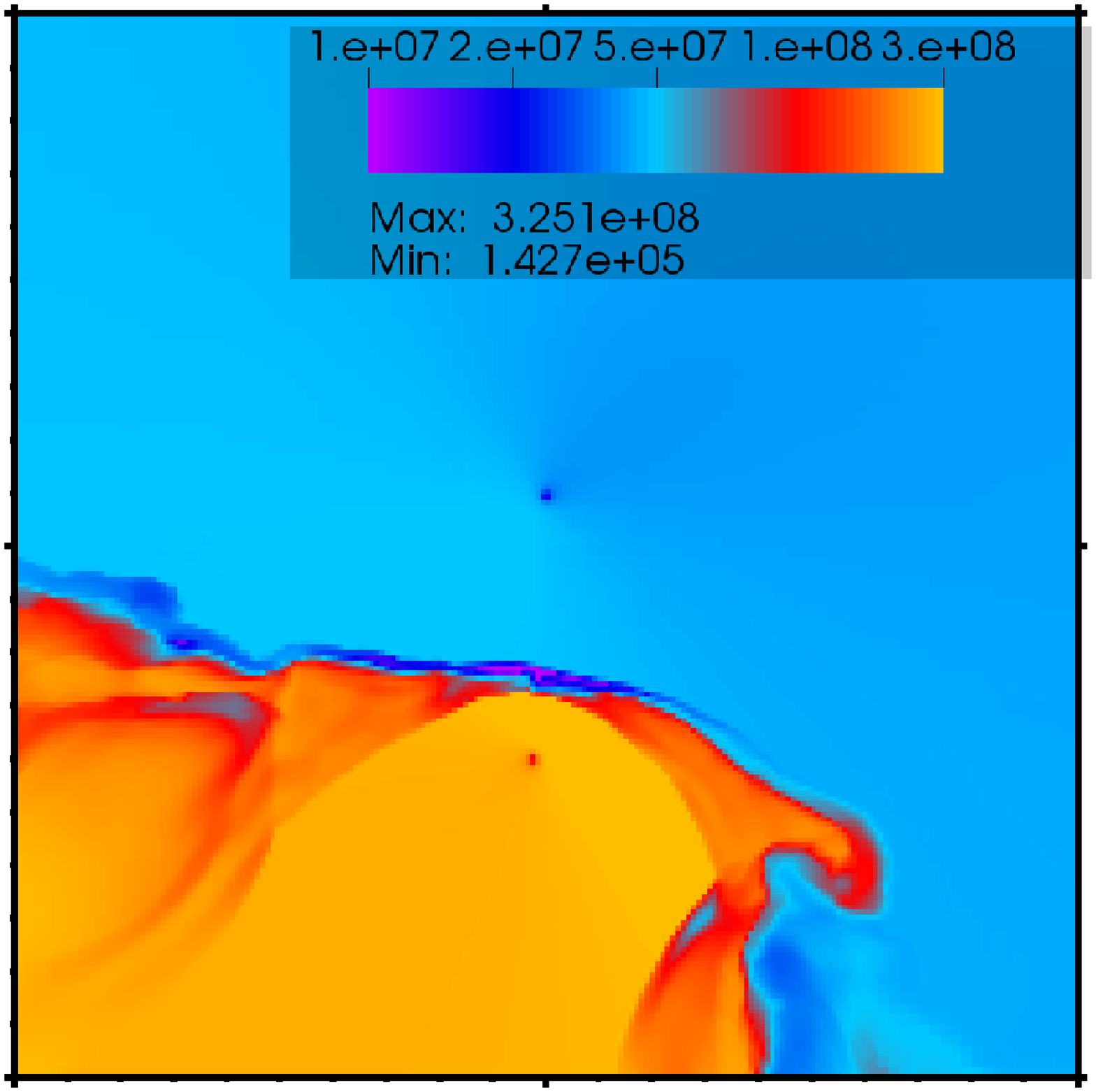}} \vspace{-1mm} \\ 

\resizebox{40mm}{!}{\includegraphics{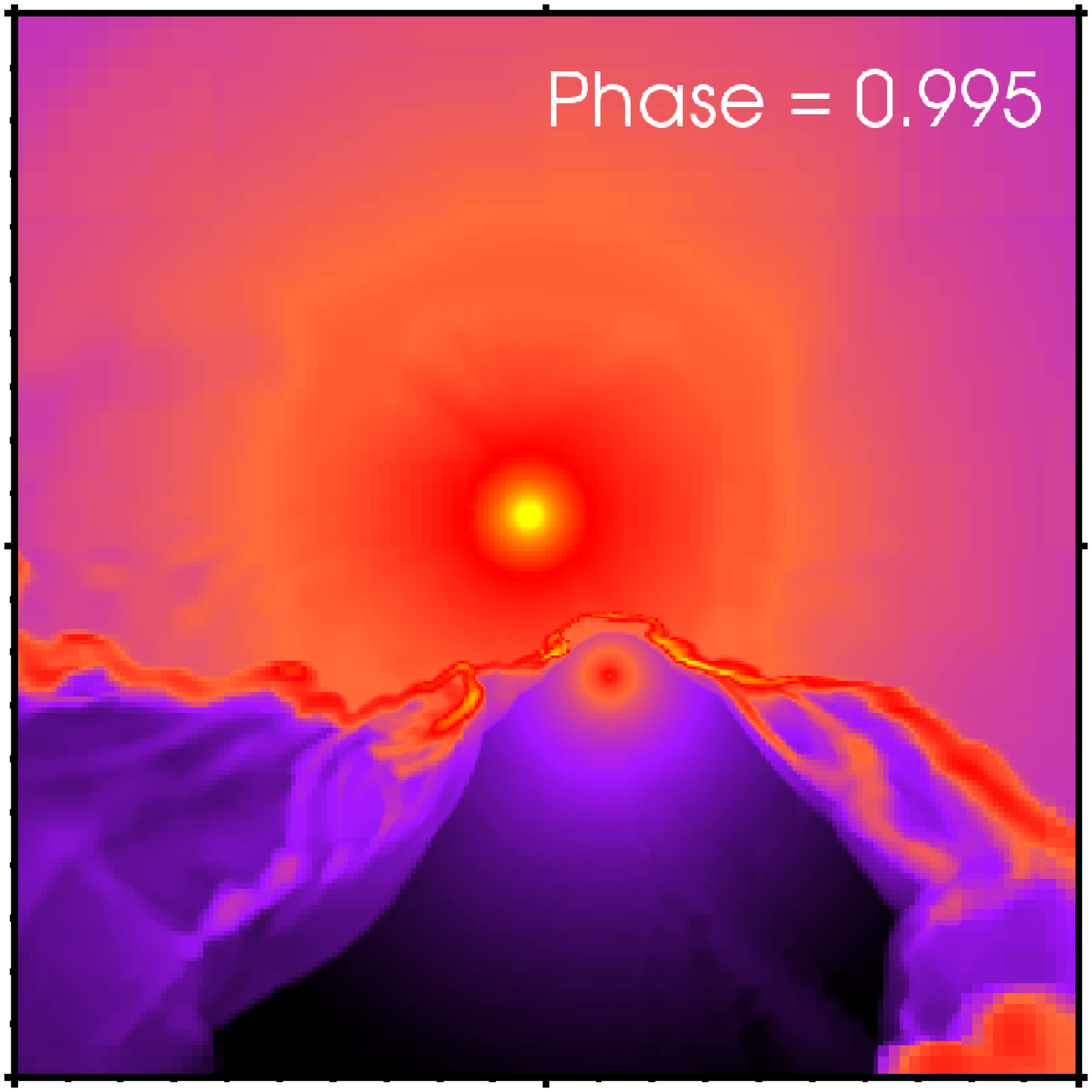}} & 
\resizebox{40mm}{!}{\includegraphics{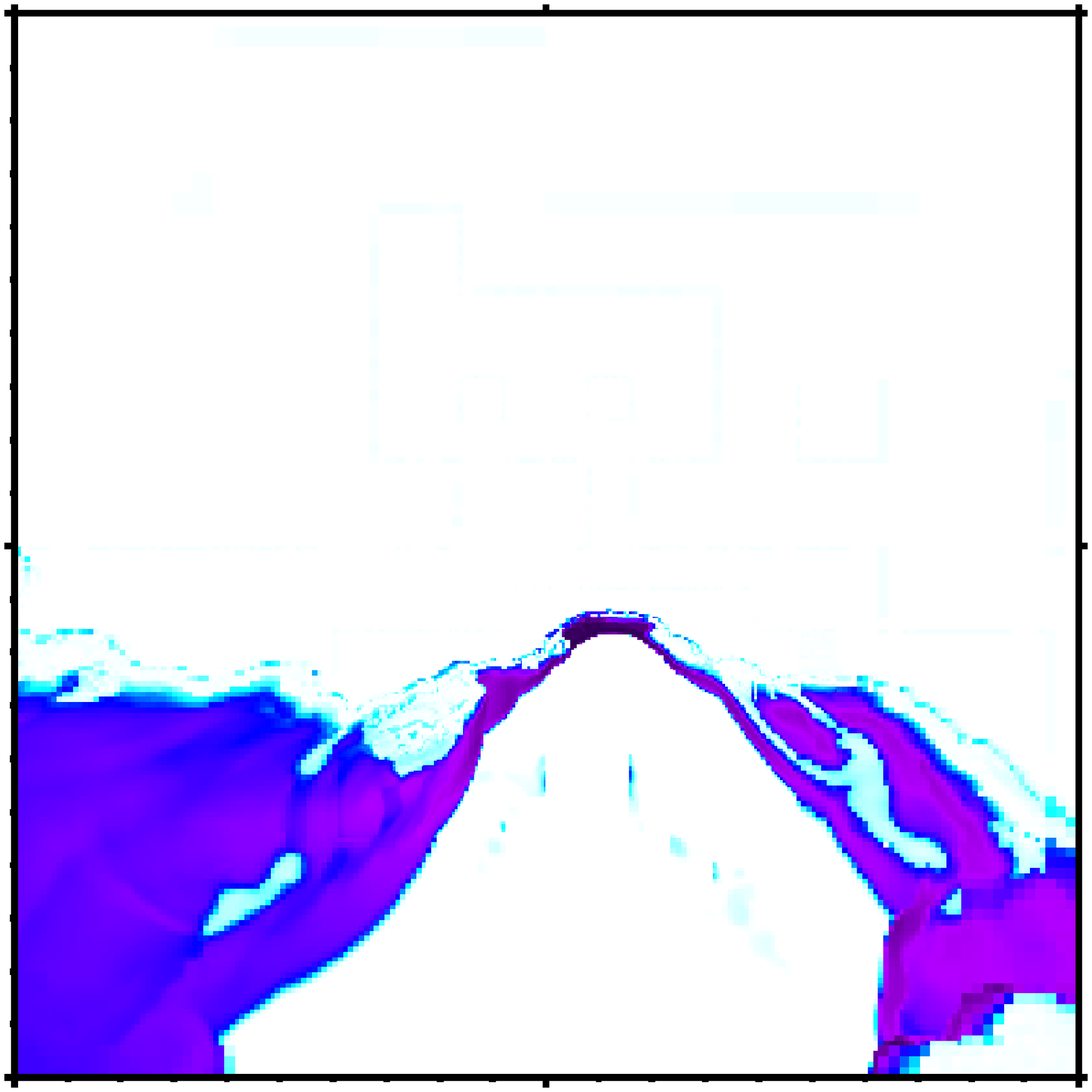}} & 
\resizebox{40mm}{!}{\includegraphics{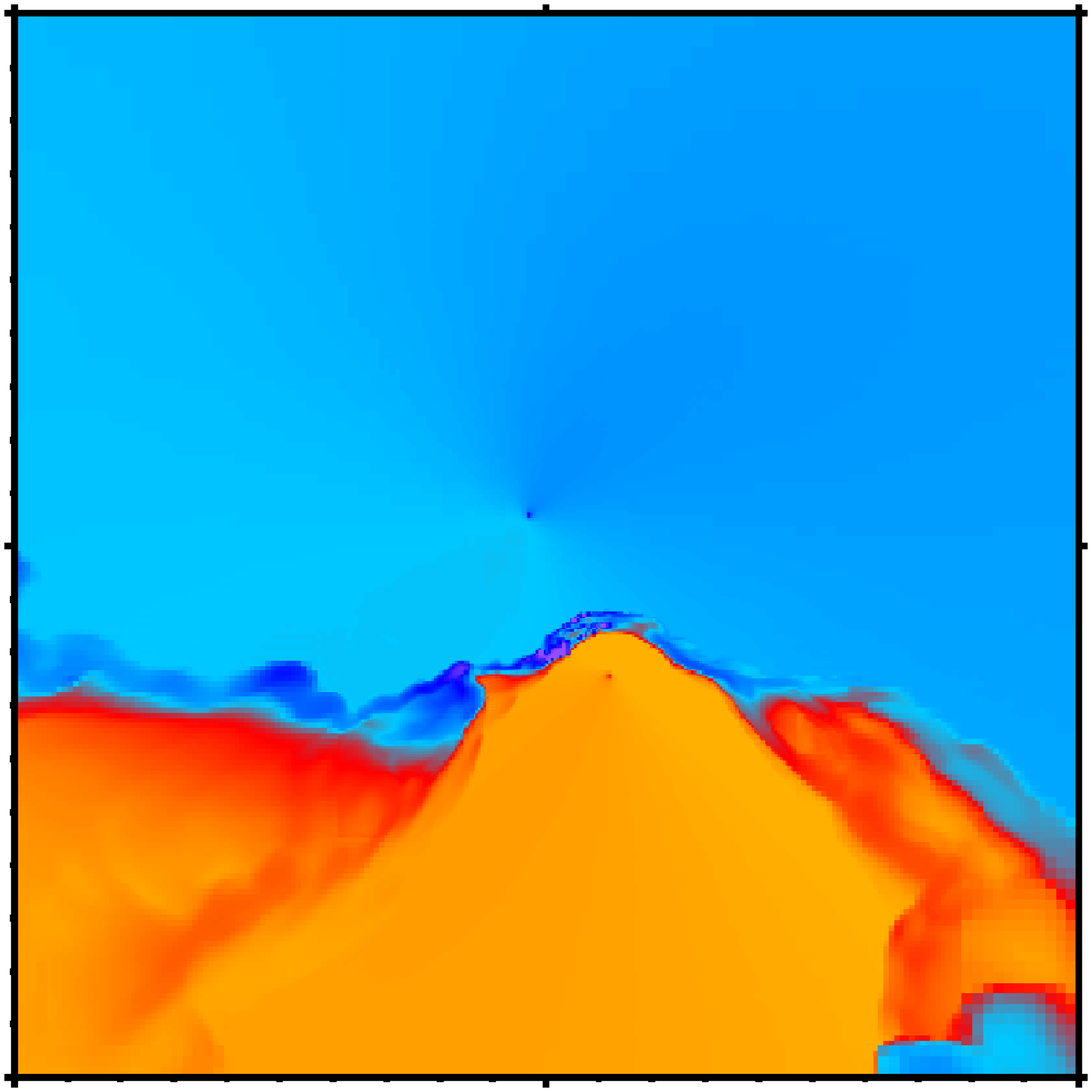}} \vspace{-1mm} \\ 

\resizebox{40mm}{!}{\includegraphics{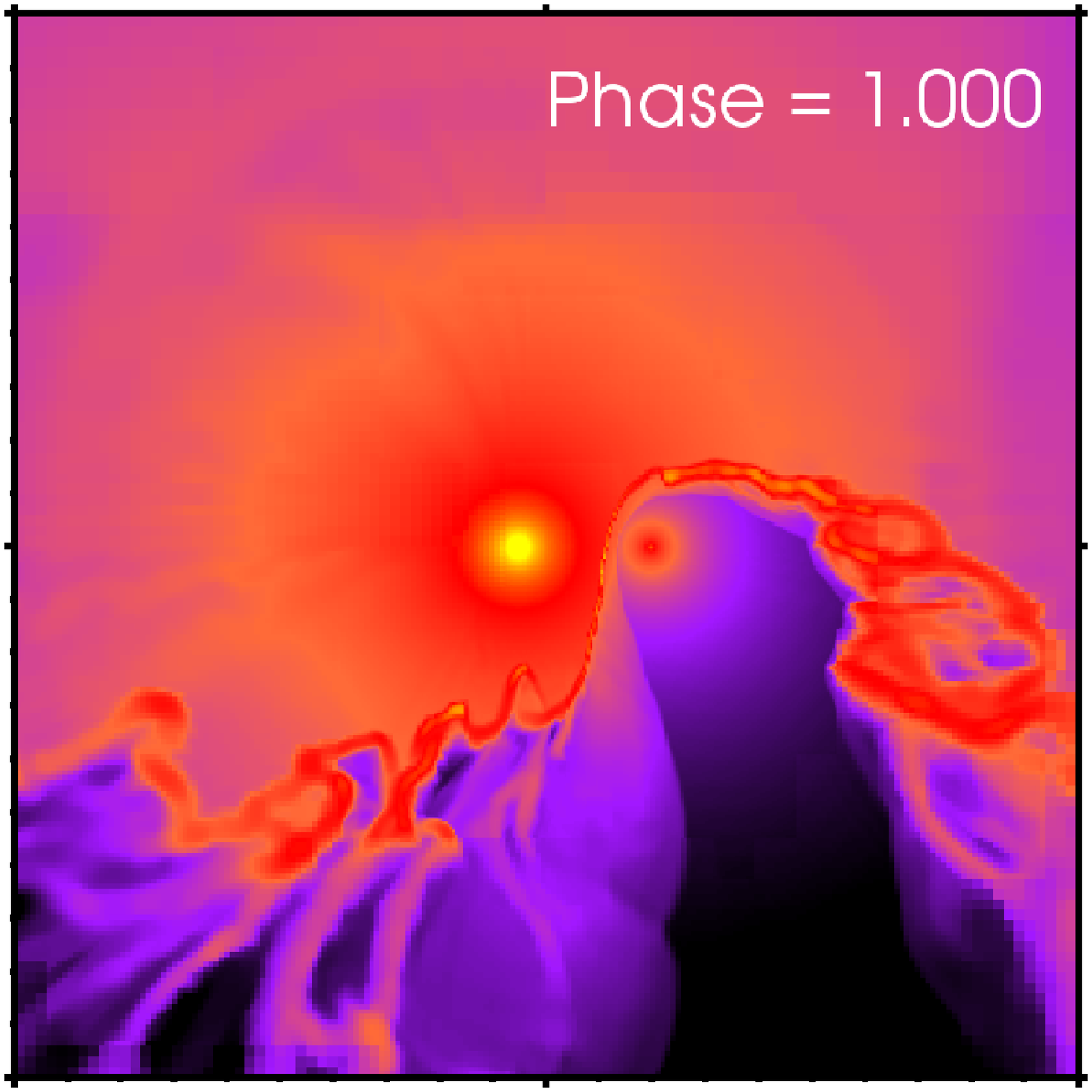}} & 
\resizebox{40mm}{!}{\includegraphics{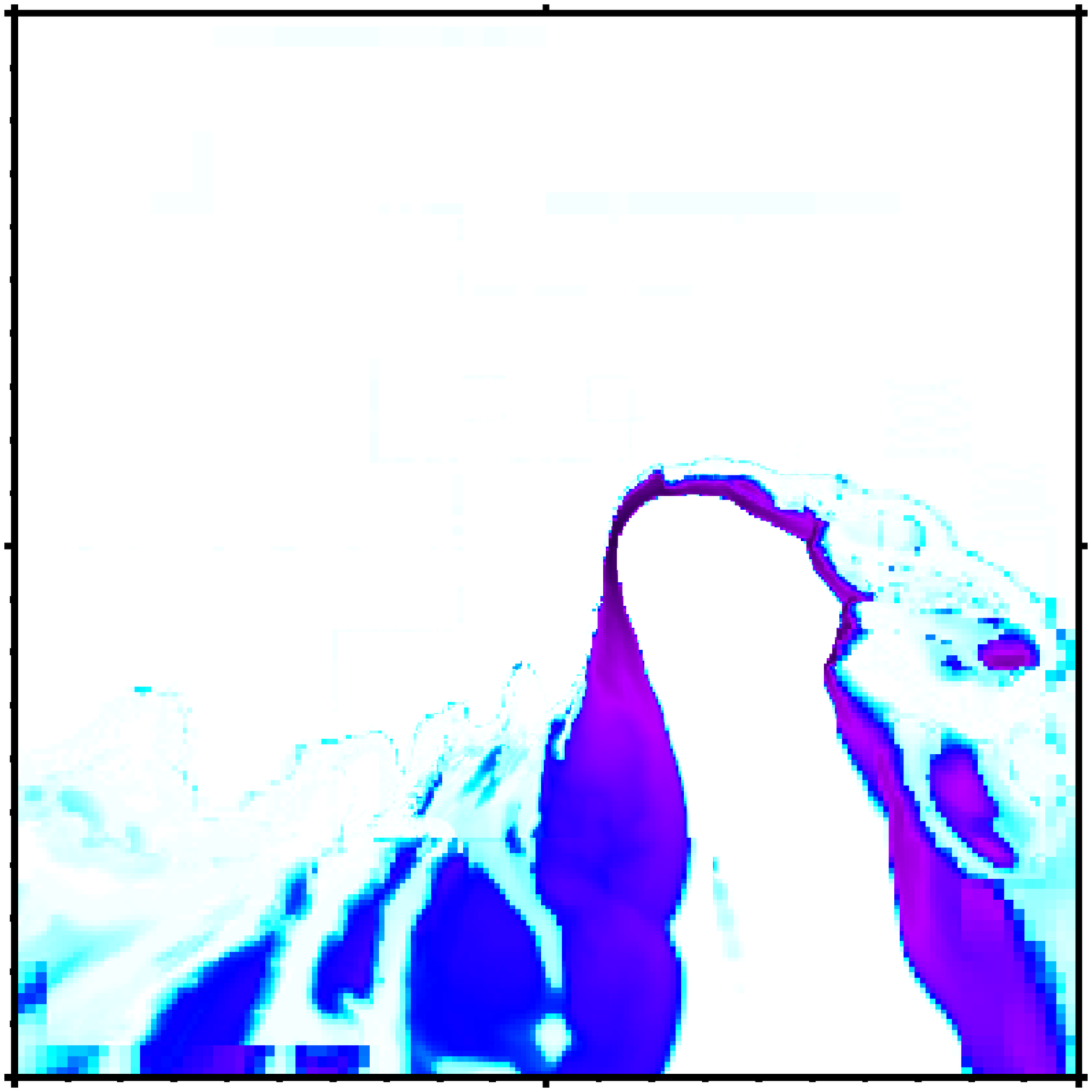}} & 
\resizebox{40mm}{!}{\includegraphics{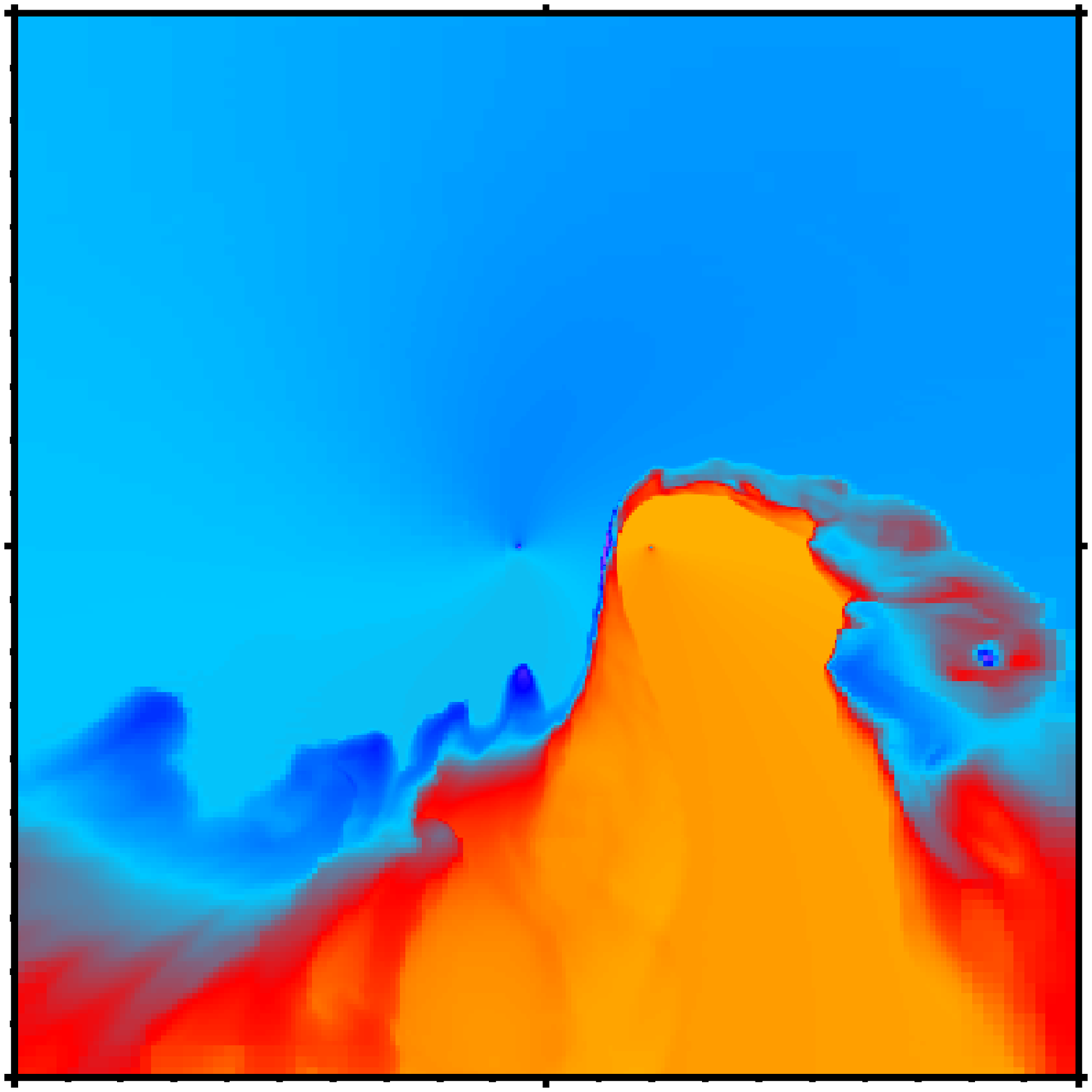}} \vspace{-1mm} \\ 

\resizebox{40mm}{!}{\includegraphics{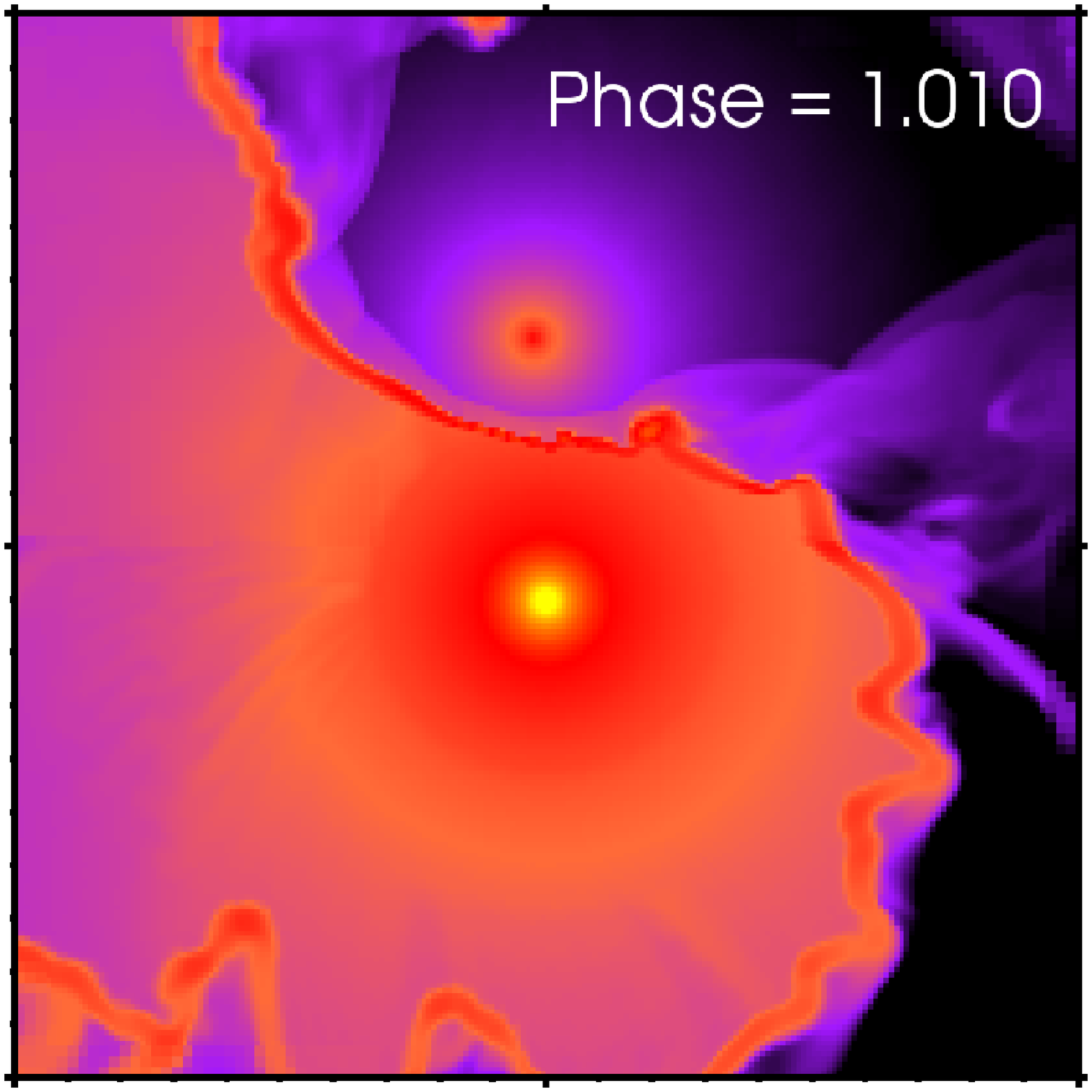}} & 
\resizebox{40mm}{!}{\includegraphics{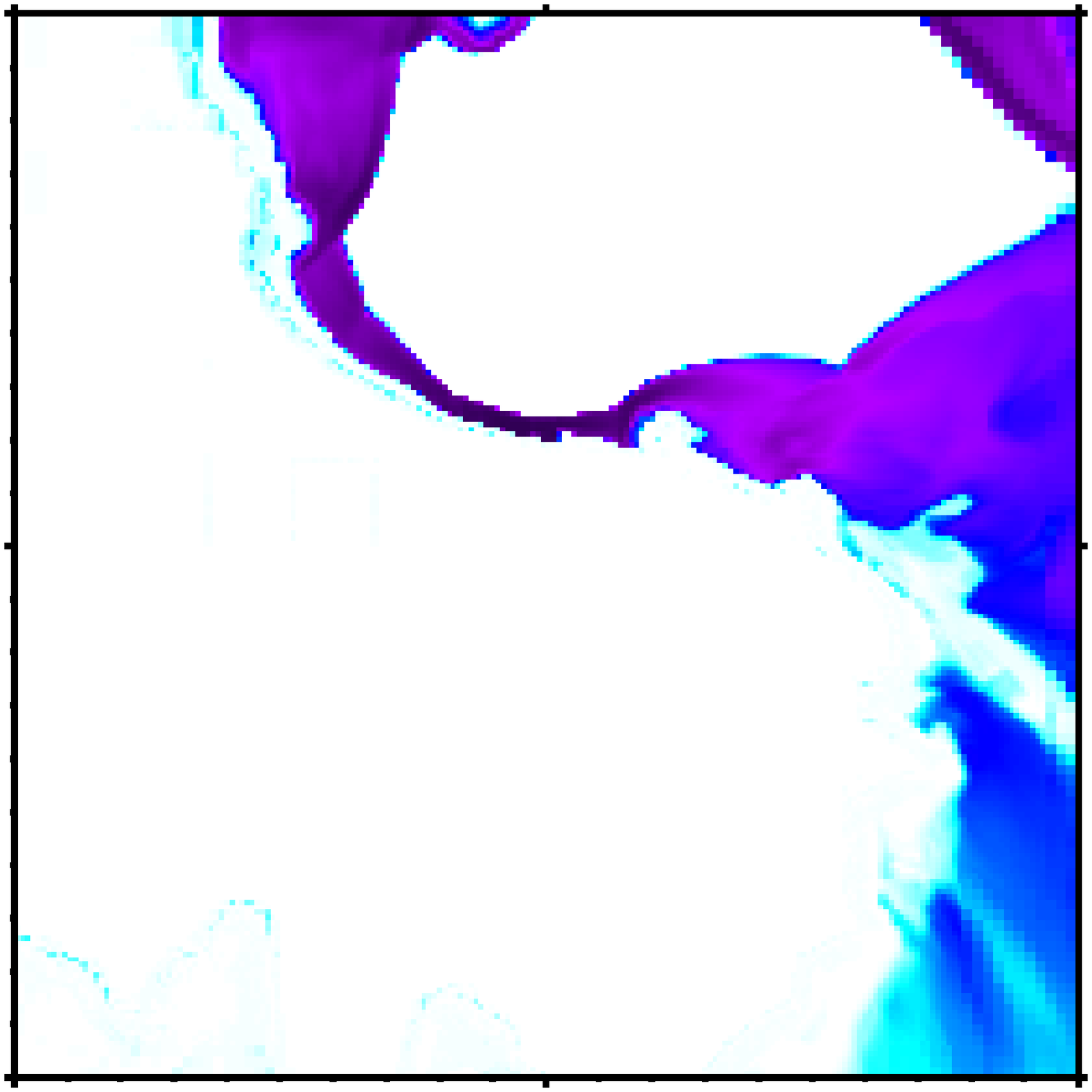}} & 
\resizebox{40mm}{!}{\includegraphics{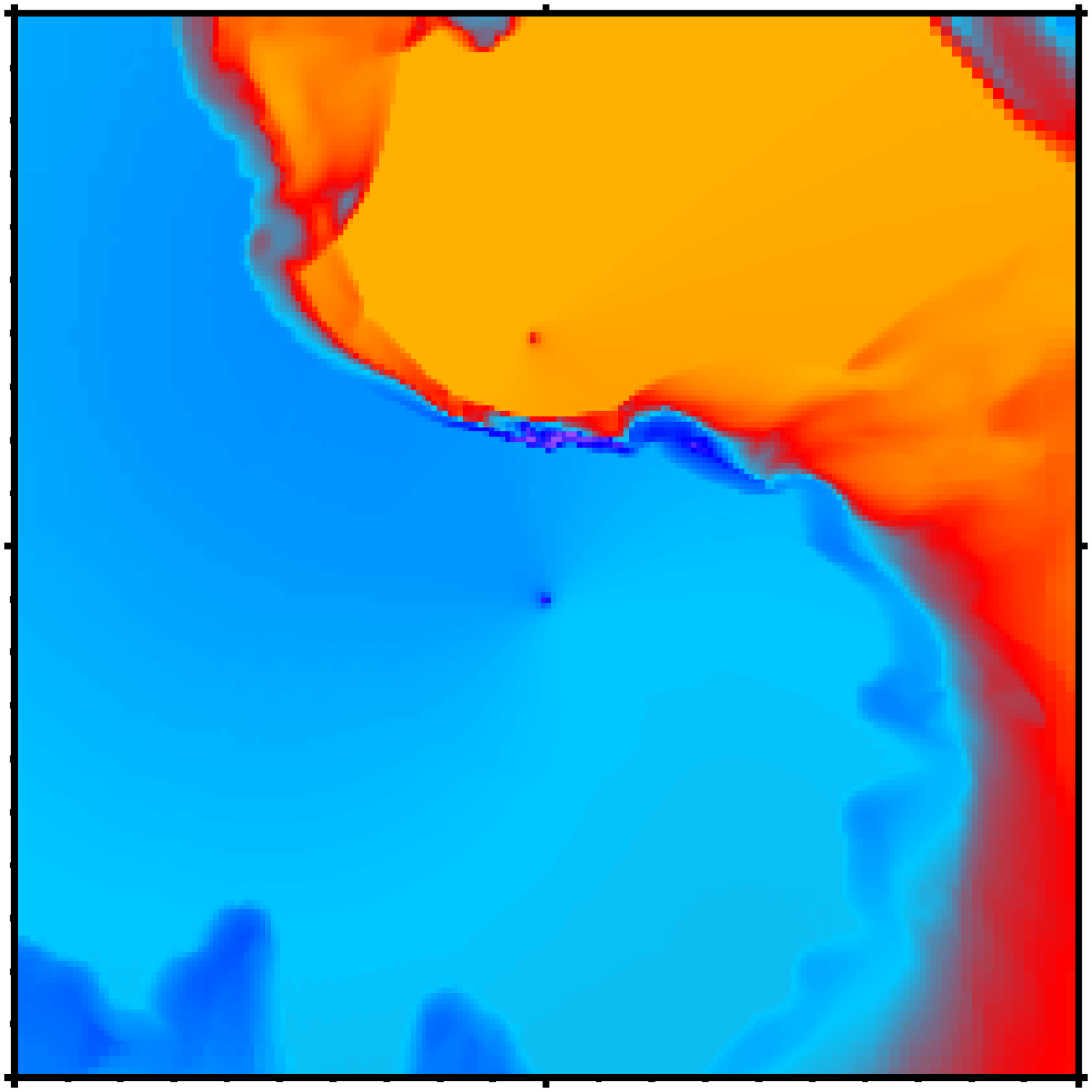}} \vspace{-1mm} \\ 

\resizebox{40mm}{!}{\includegraphics{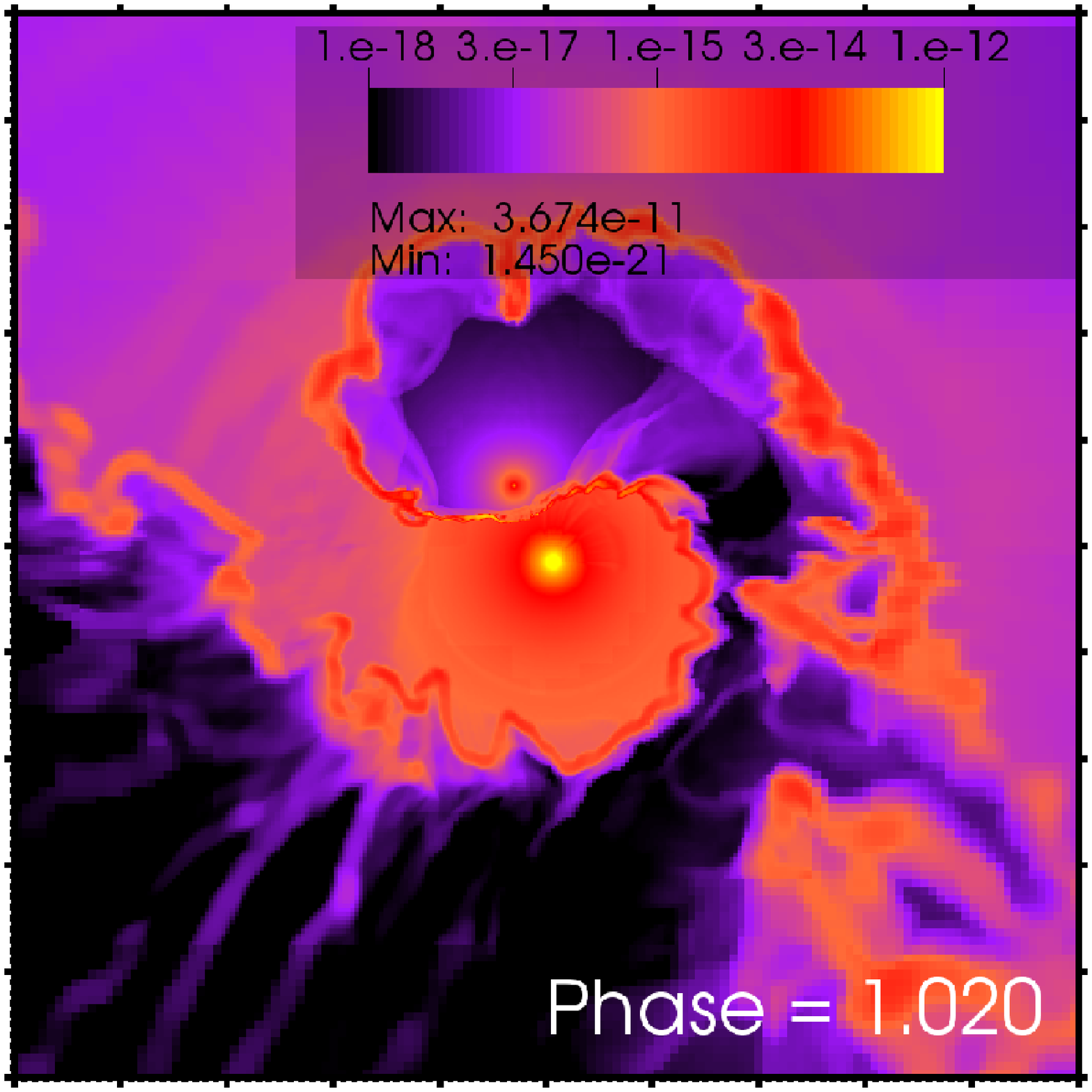}} & 
\resizebox{40mm}{!}{\includegraphics{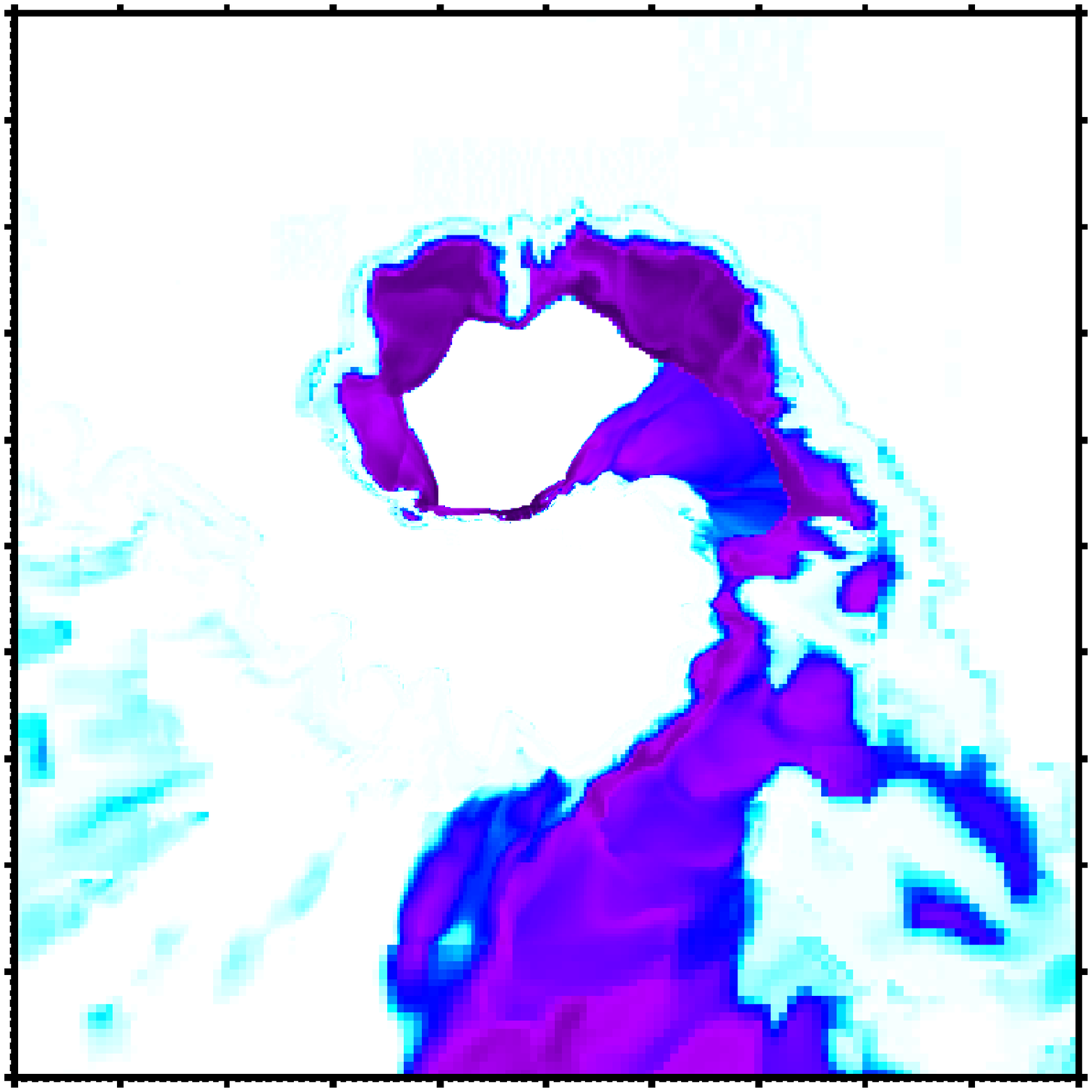}} & 
\resizebox{40mm}{!}{\includegraphics{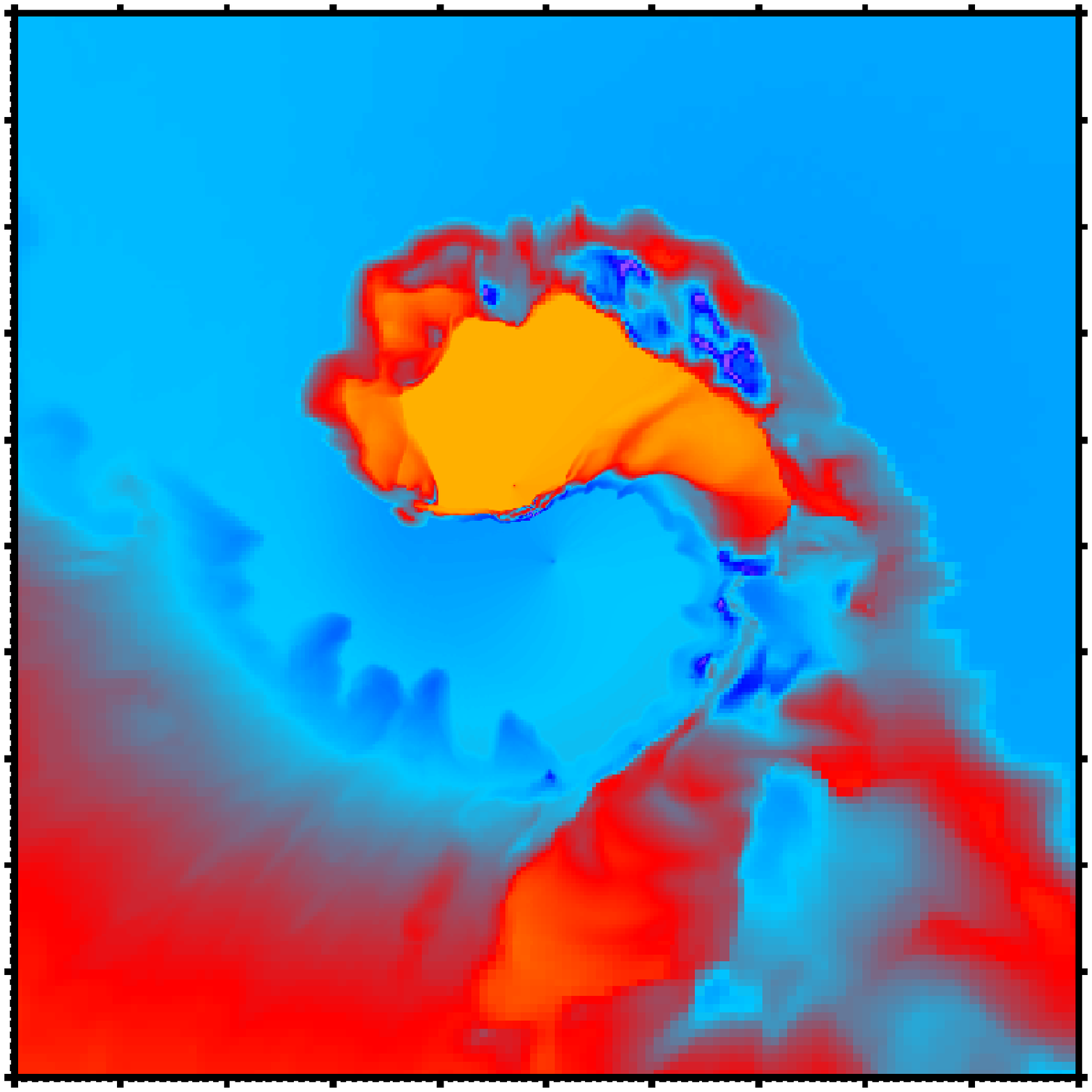}} \vspace{-1mm} \\ 

\resizebox{40mm}{!}{\includegraphics{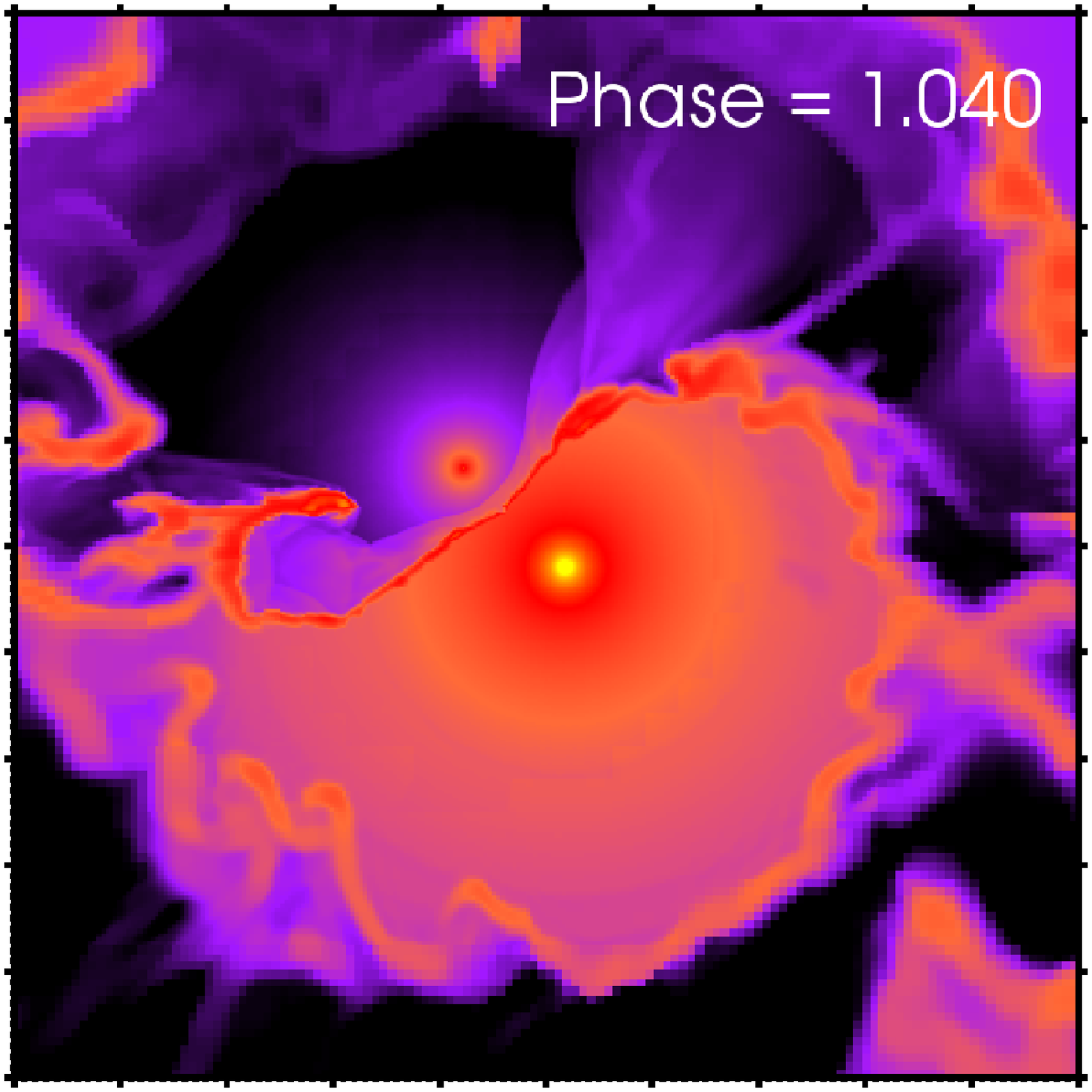}} & 
\resizebox{40mm}{!}{\includegraphics{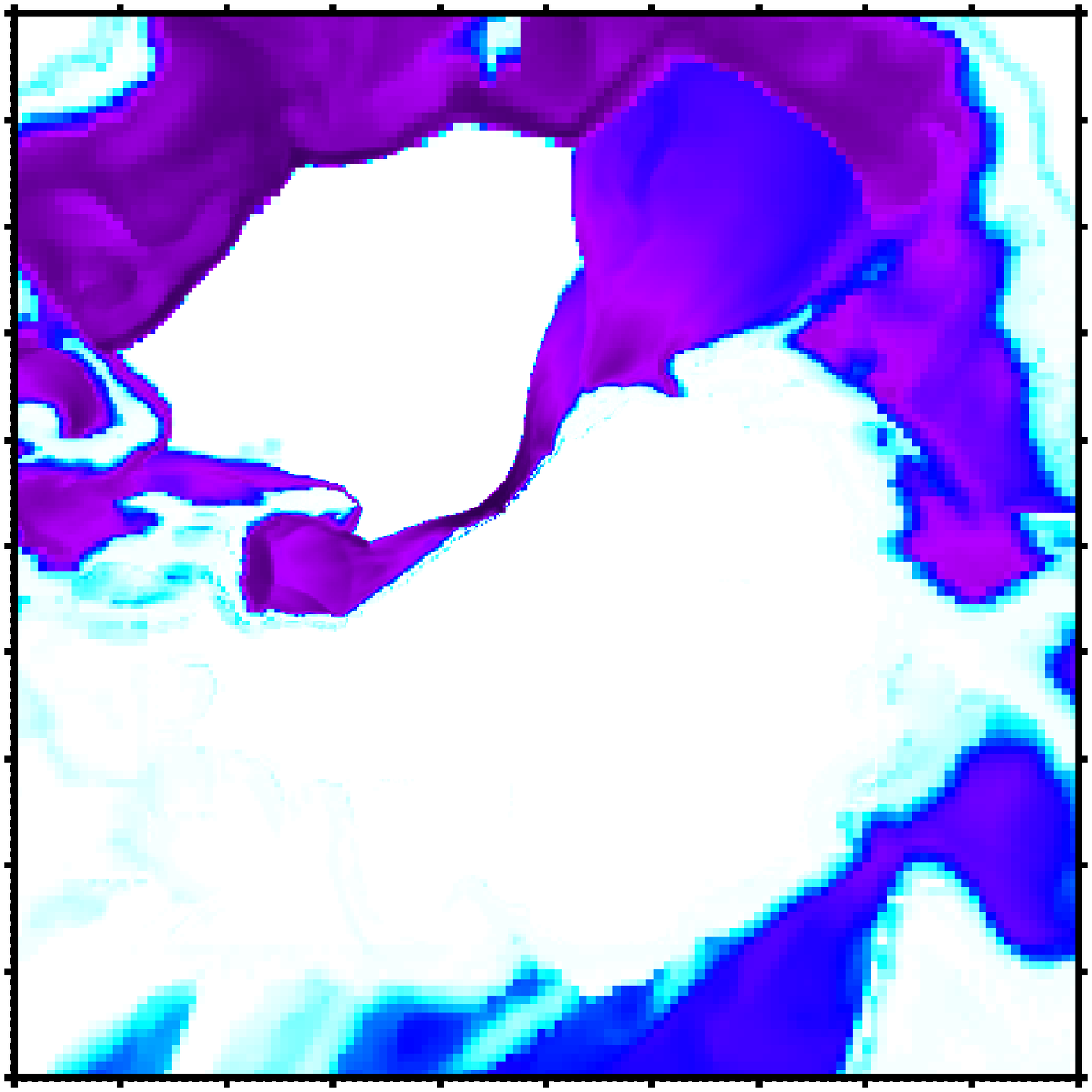}} & 
\resizebox{40mm}{!}{\includegraphics{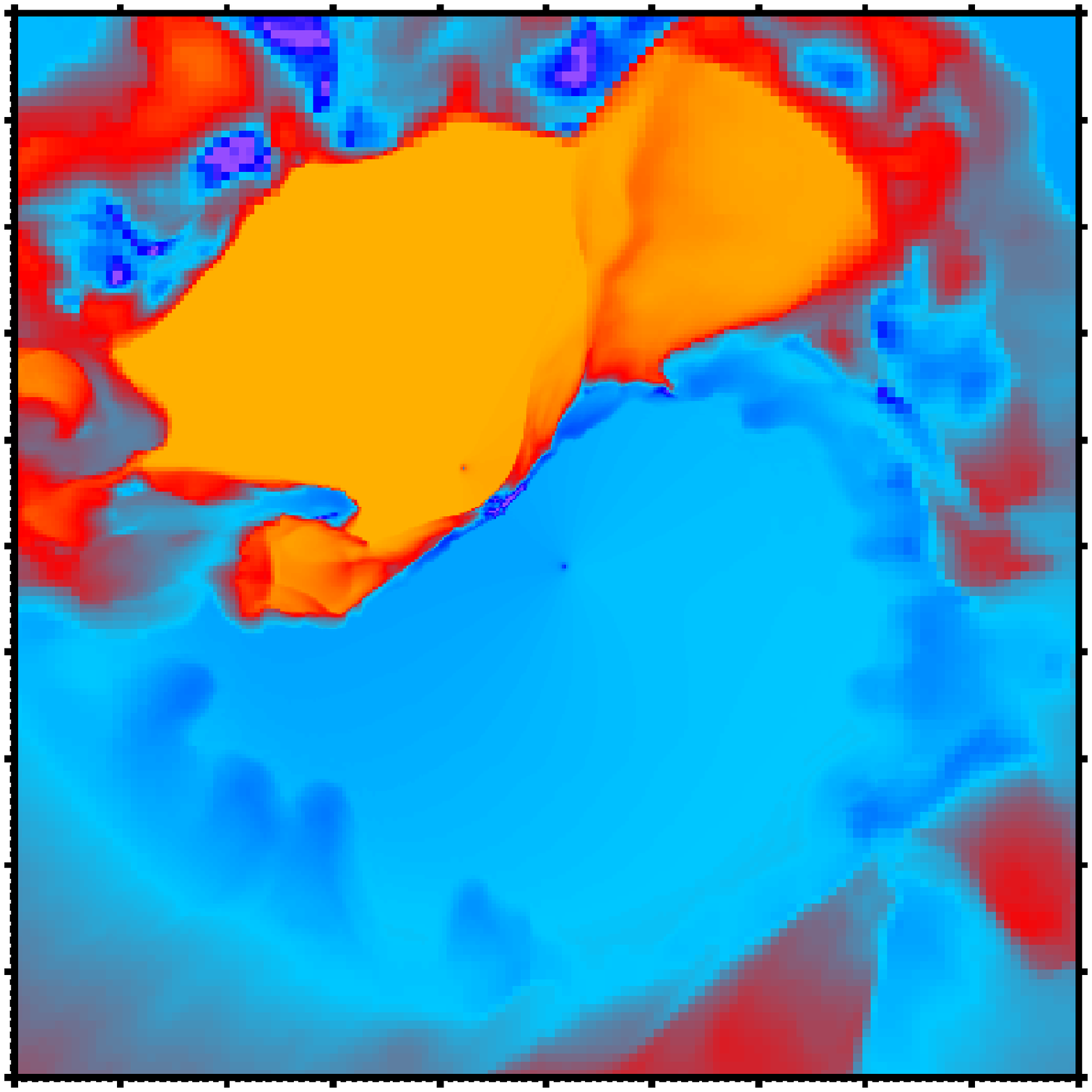}} \vspace{-1mm} \\ 

    \end{tabular}
    \caption{Snapshots of the orbital ($x-y$) plane from model
      Orbit-IA showing (from top to bottom) $\phi = 0.990$, 0.995,
      1.000, 1.010, 1.020, and 1.040, and (from left to right)
      density, temperature, and speed. At periastron ($\phi =1.000$)
      the primary star is to the left, and the companion star is to
      the right, of the image centre. The plots at $\phi = 0.990$,
      0.995, 1.000, and 1.010 show a region of
      $\pm1\times10^{14}\;$cm, whereas the plots at $\phi =$ 1.020 and
      1.040 show a region of $\pm5\times10^{14}\;$cm. In all plots
      large axis tick marks correspond to a distance of
      $\pm1\times10^{14}\;$cm. Note the difference in colour scale
      used for the density plots at $\phi=$1.020 and 1.040.}
    \label{fig:vterm_peri_images}
  \end{center}
\end{figure*}

\begin{figure*}
  \begin{center}
    \begin{tabular}{ccc}
\resizebox{40mm}{!}{\includegraphics{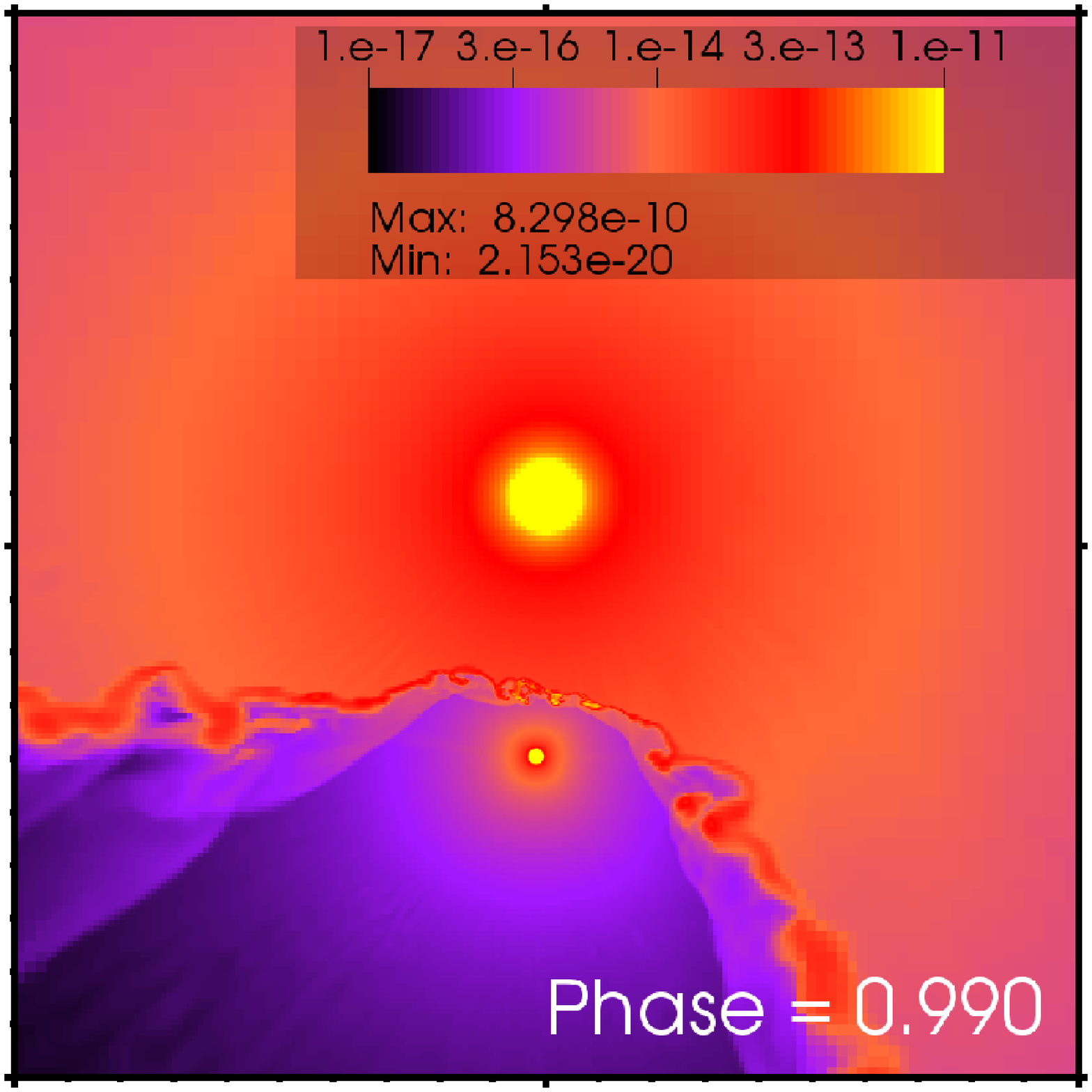}} & 
\resizebox{40mm}{!}{\includegraphics{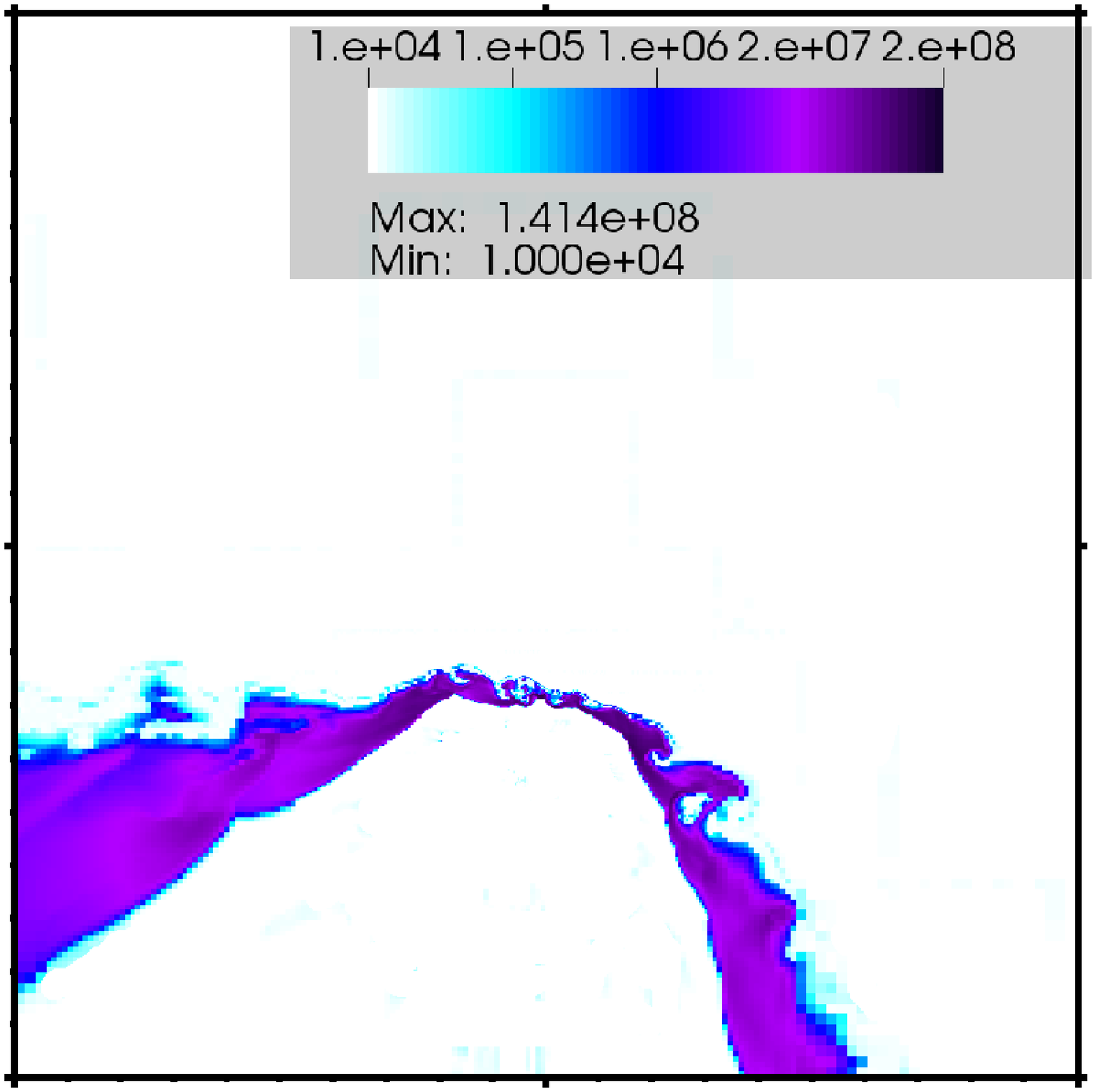}} & 
\resizebox{40mm}{!}{\includegraphics{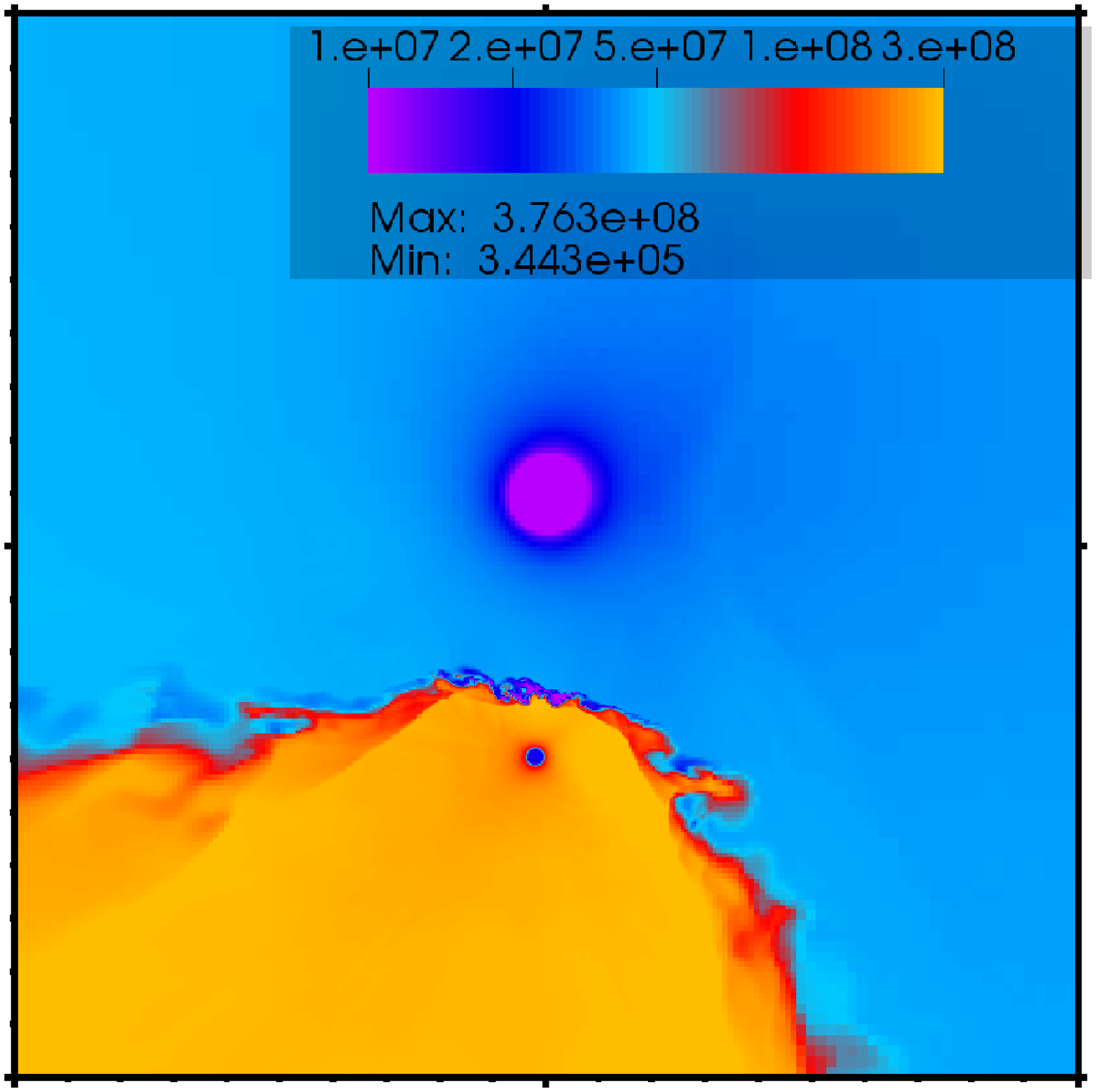}} \vspace{-1mm} \\ 

\resizebox{40mm}{!}{\includegraphics{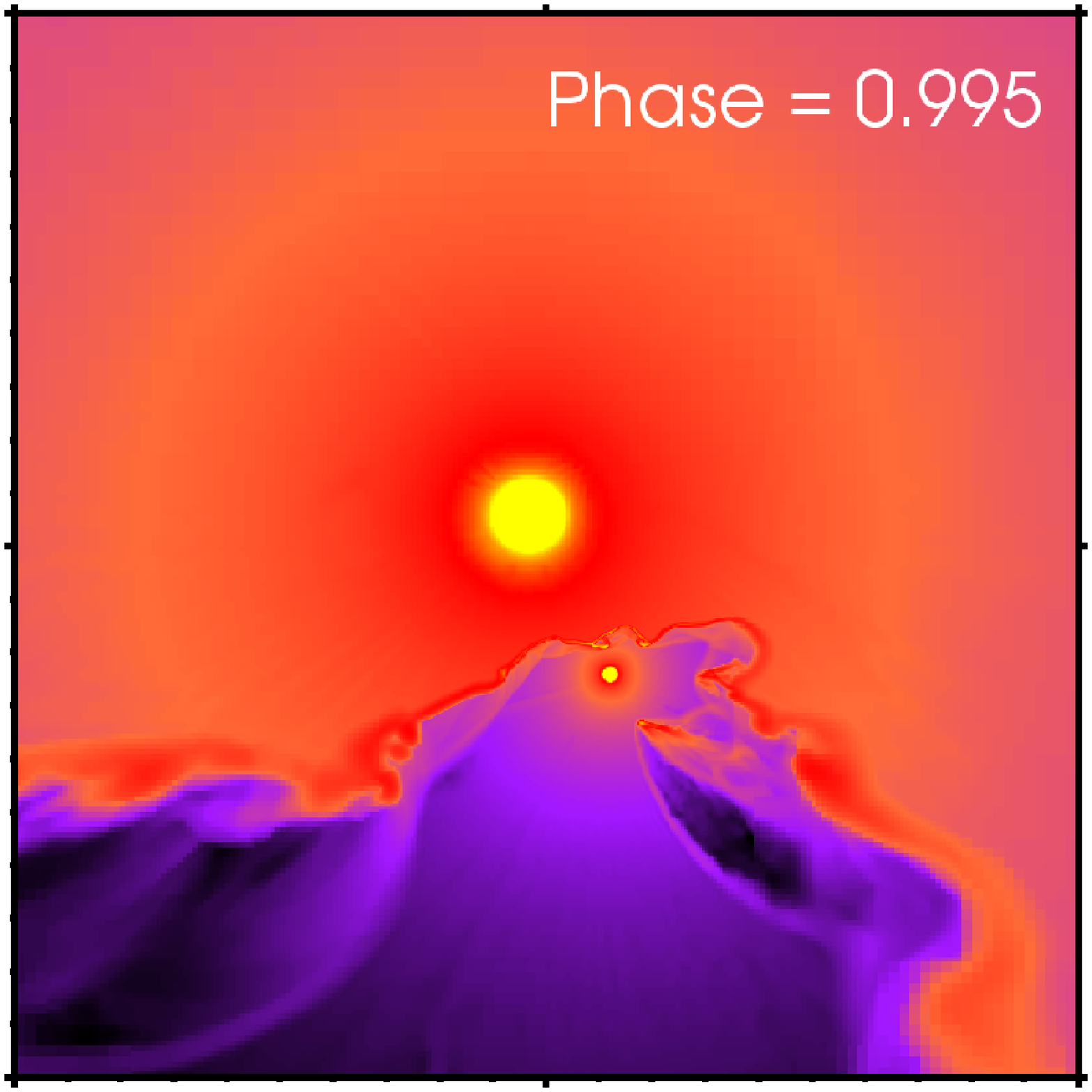}} & 
\resizebox{40mm}{!}{\includegraphics{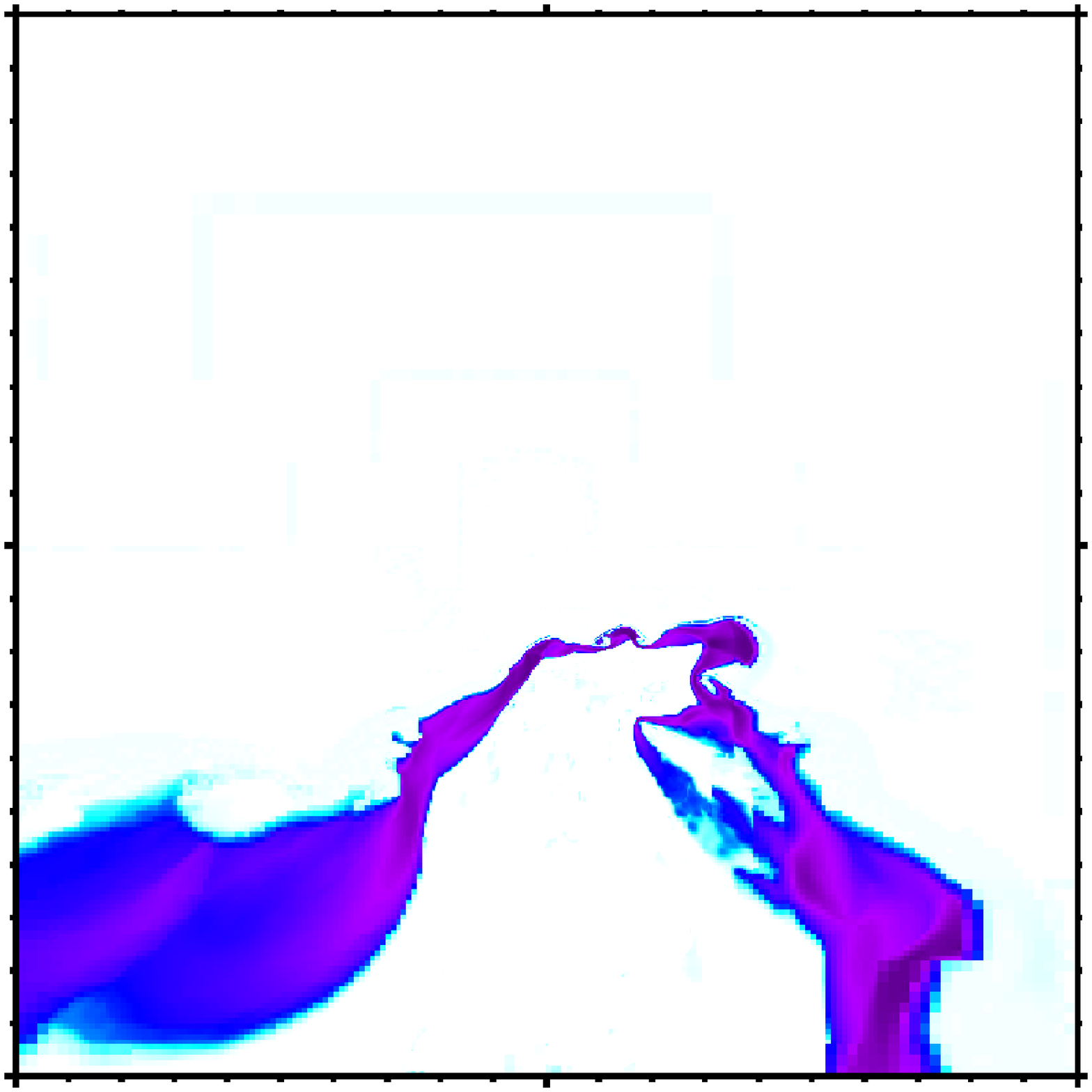}} & 
\resizebox{40mm}{!}{\includegraphics{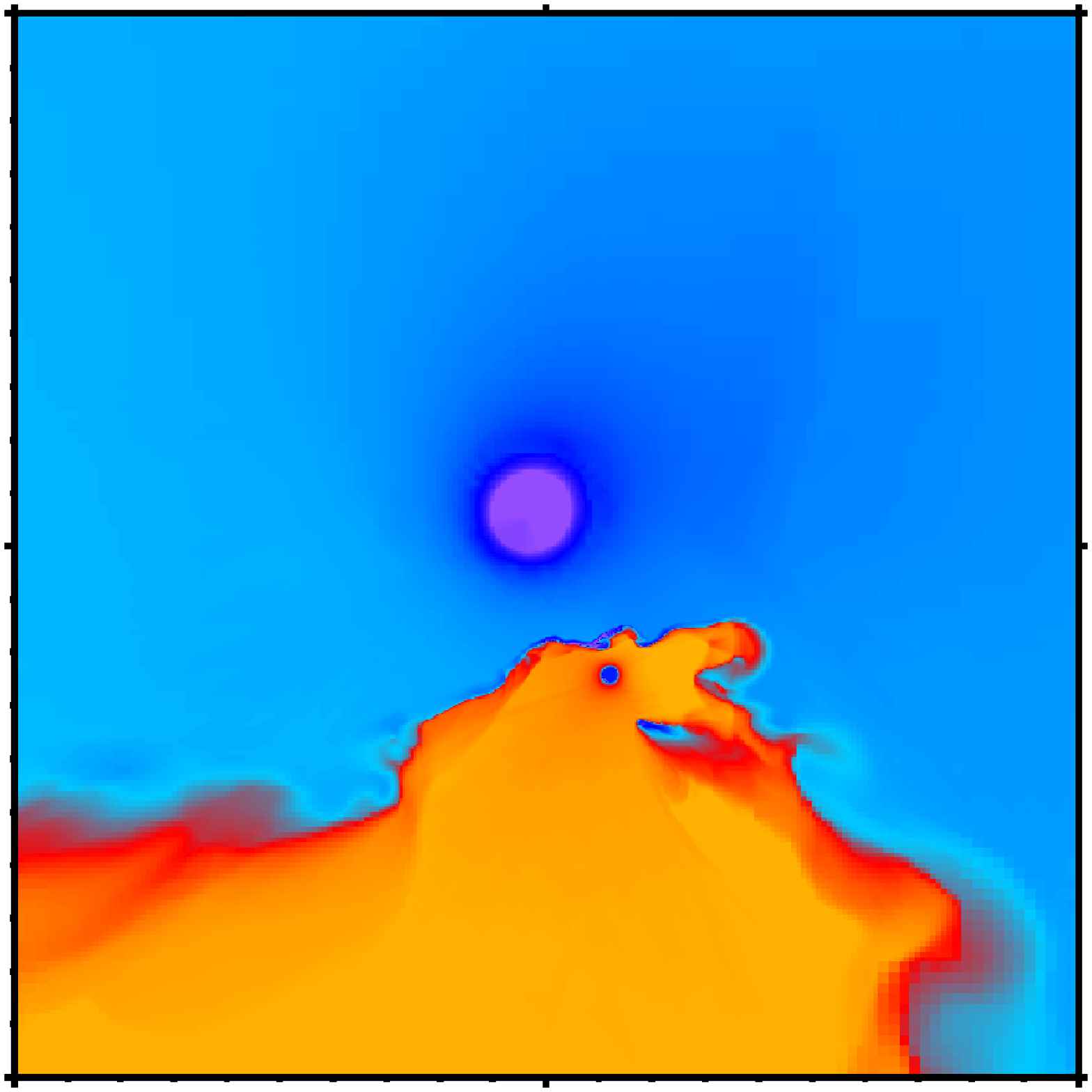}} \vspace{-1mm} \\ 

\resizebox{40mm}{!}{\includegraphics{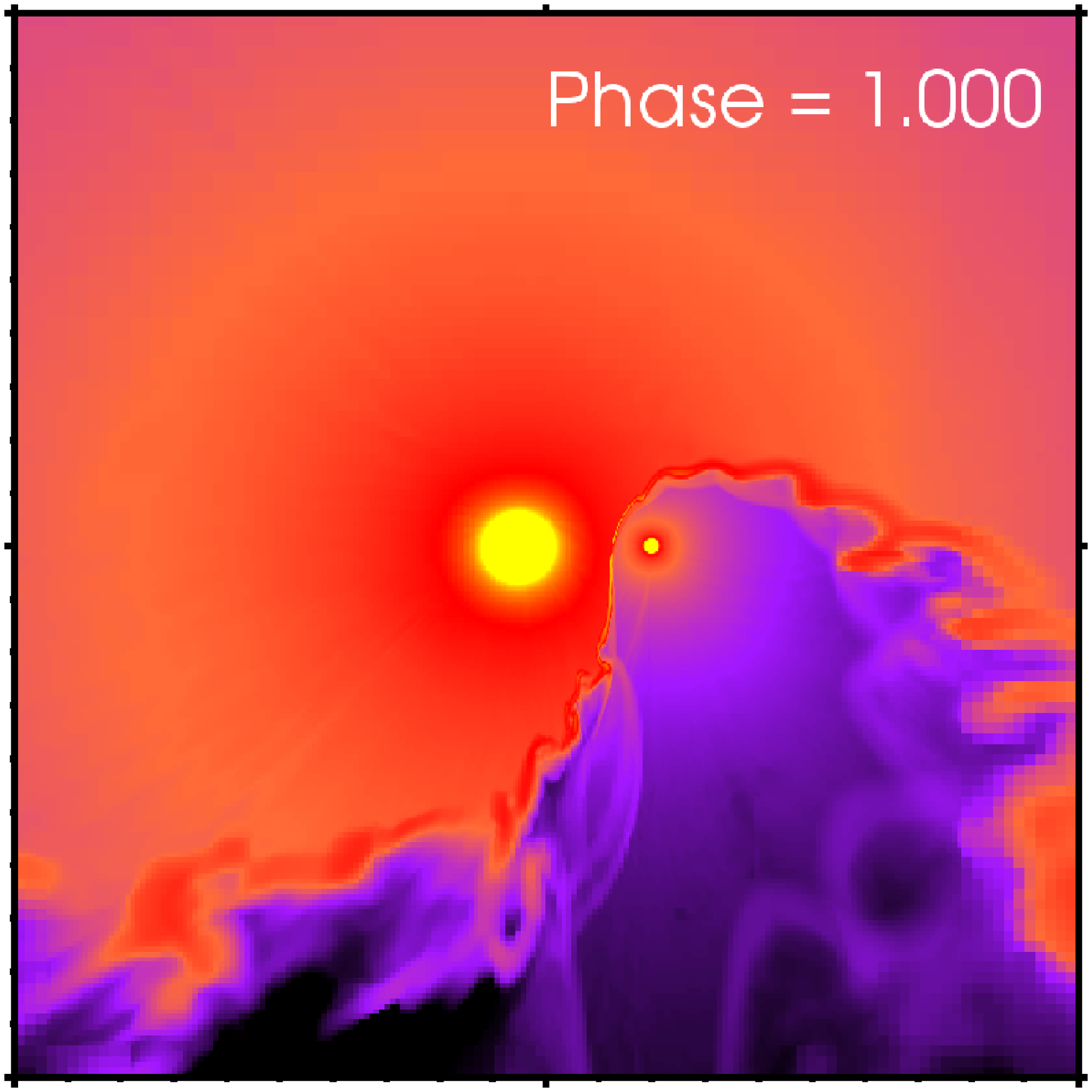}} & 
\resizebox{40mm}{!}{\includegraphics{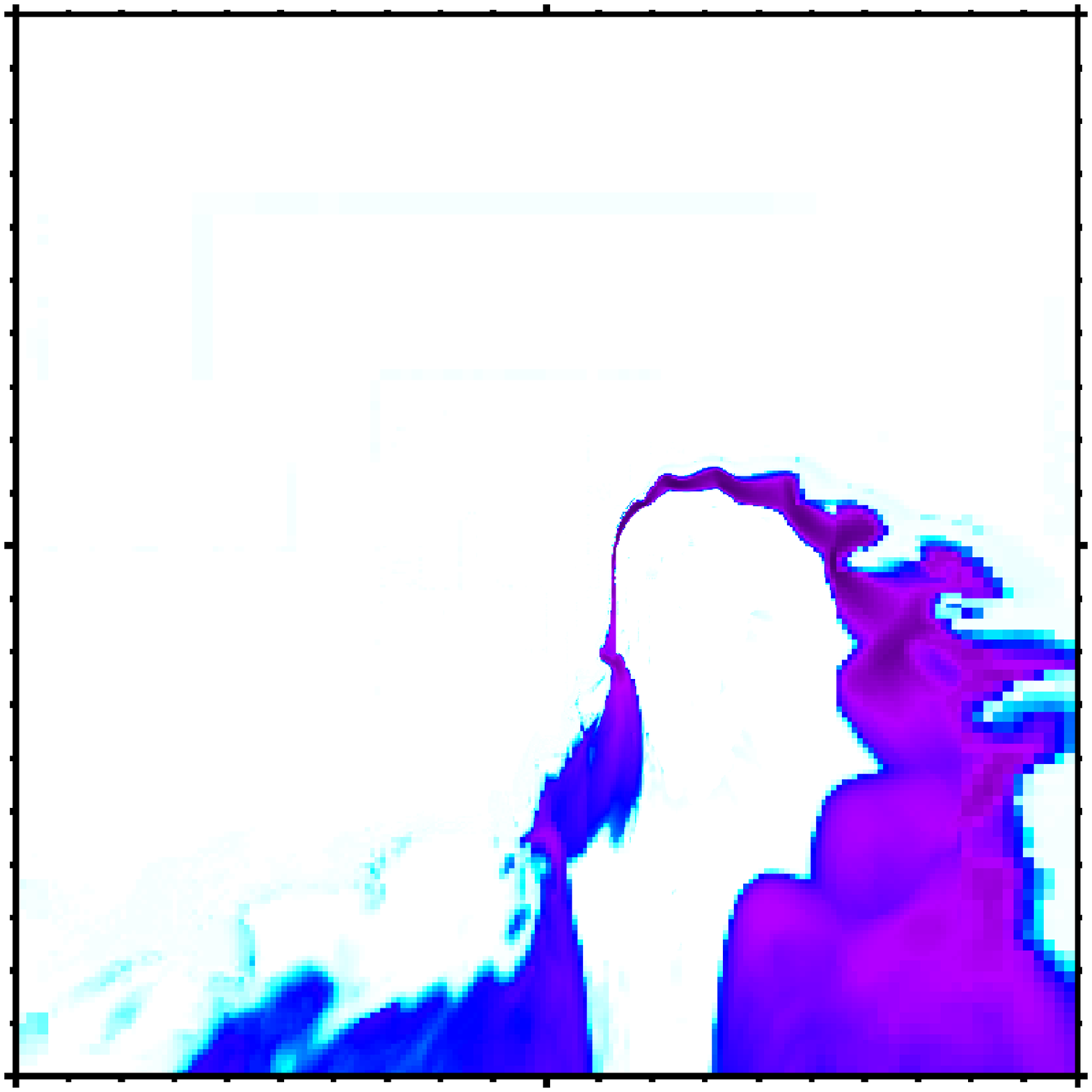}} & 
\resizebox{40mm}{!}{\includegraphics{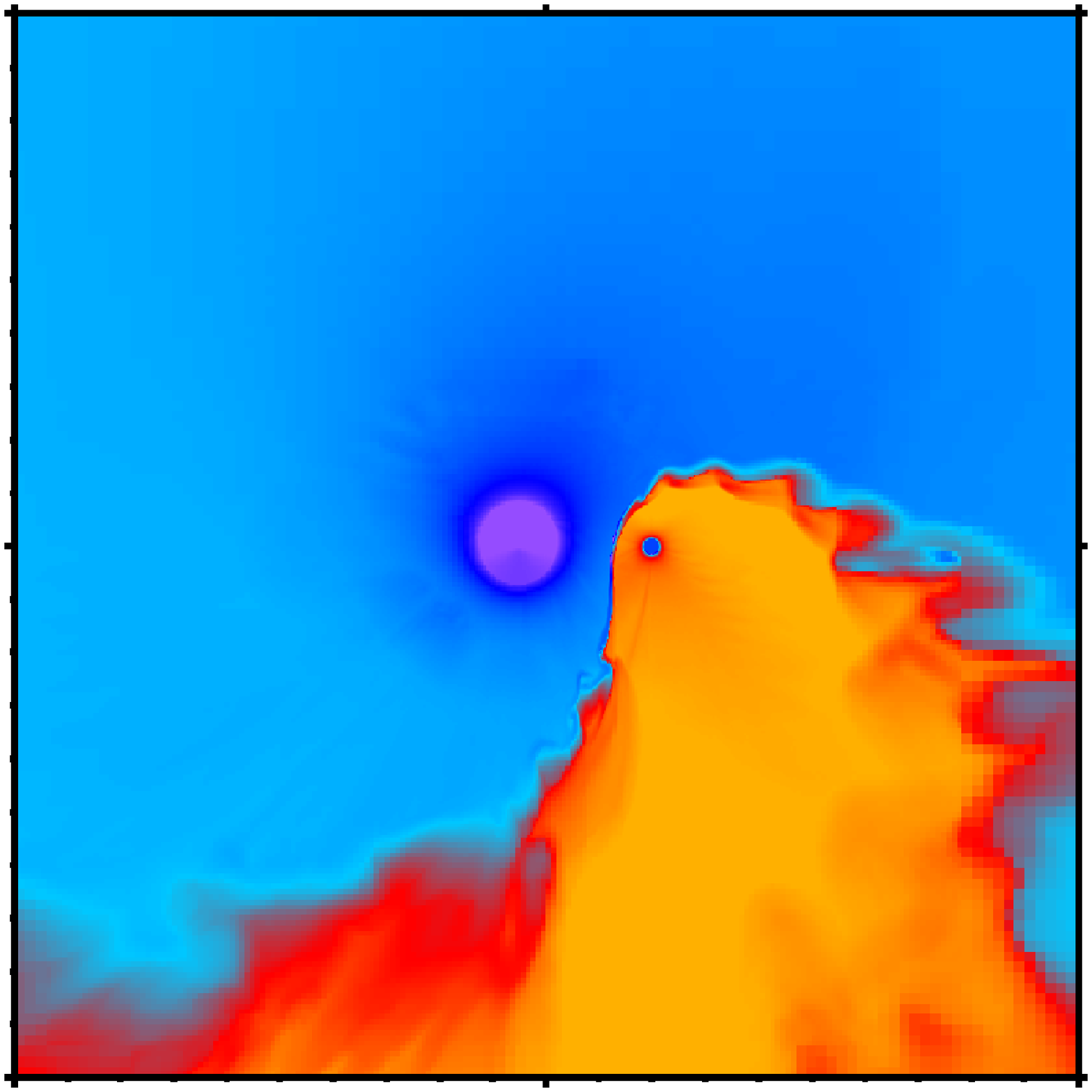}} \vspace{-1mm} \\ 

\resizebox{40mm}{!}{\includegraphics{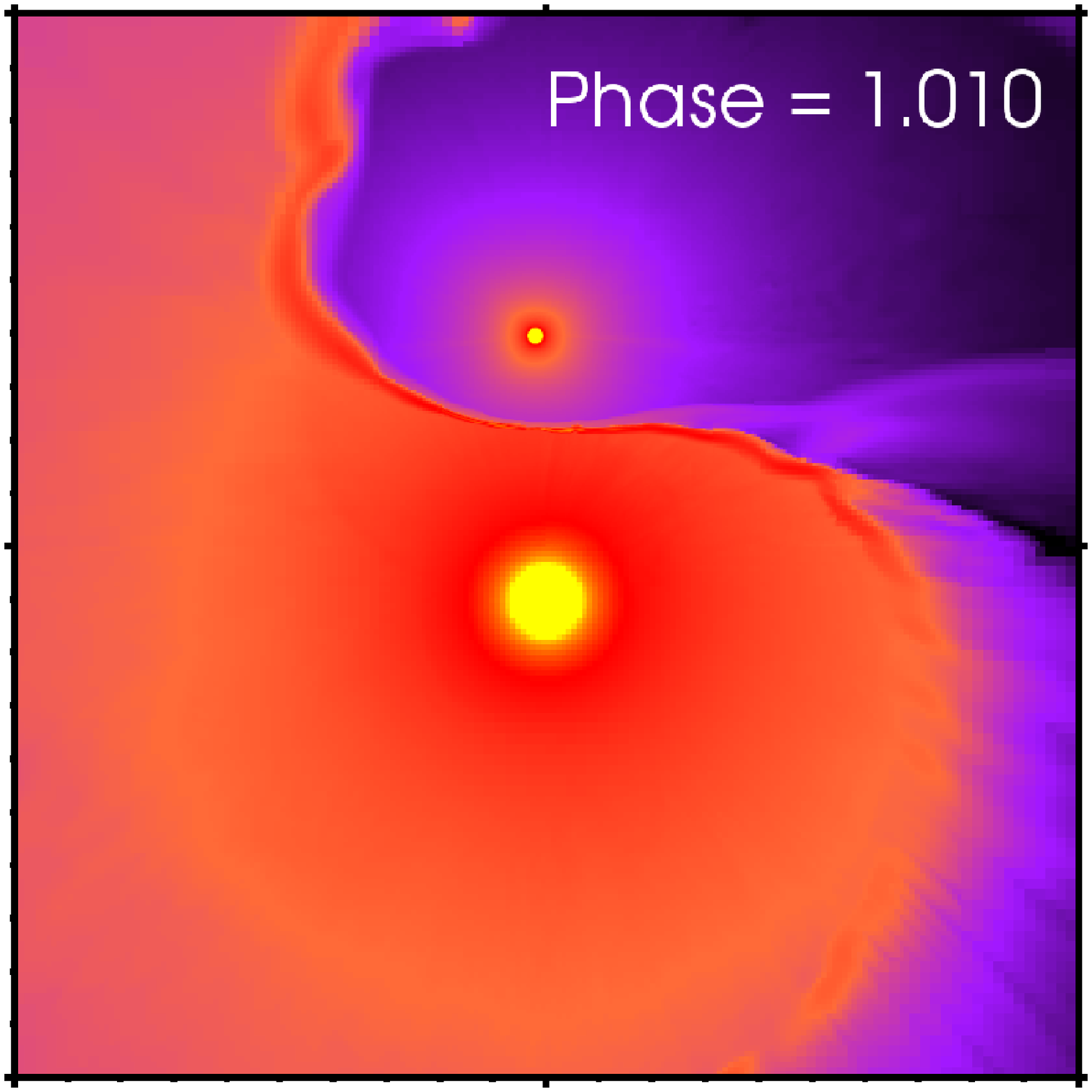}} & 
\resizebox{40mm}{!}{\includegraphics{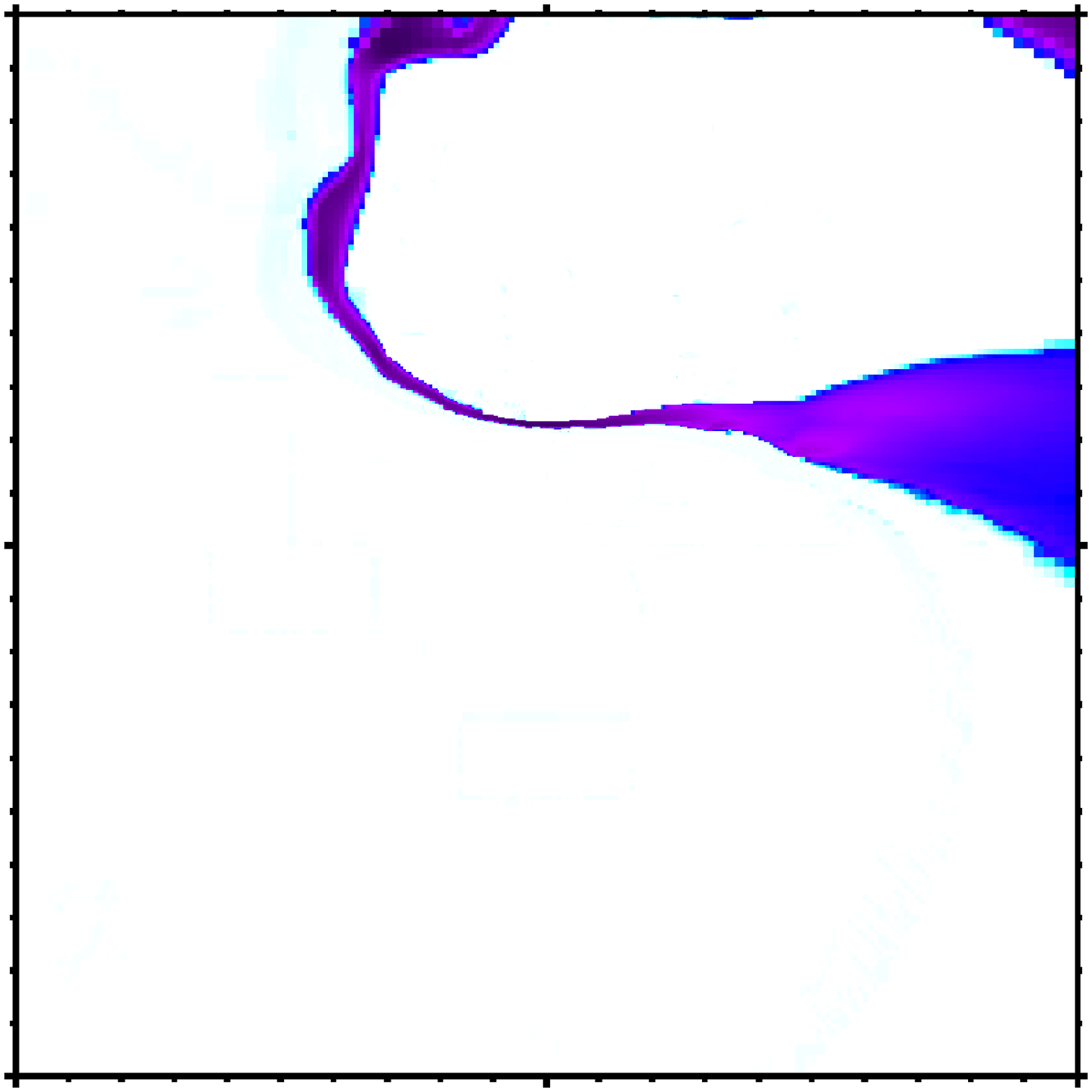}} & 
\resizebox{40mm}{!}{\includegraphics{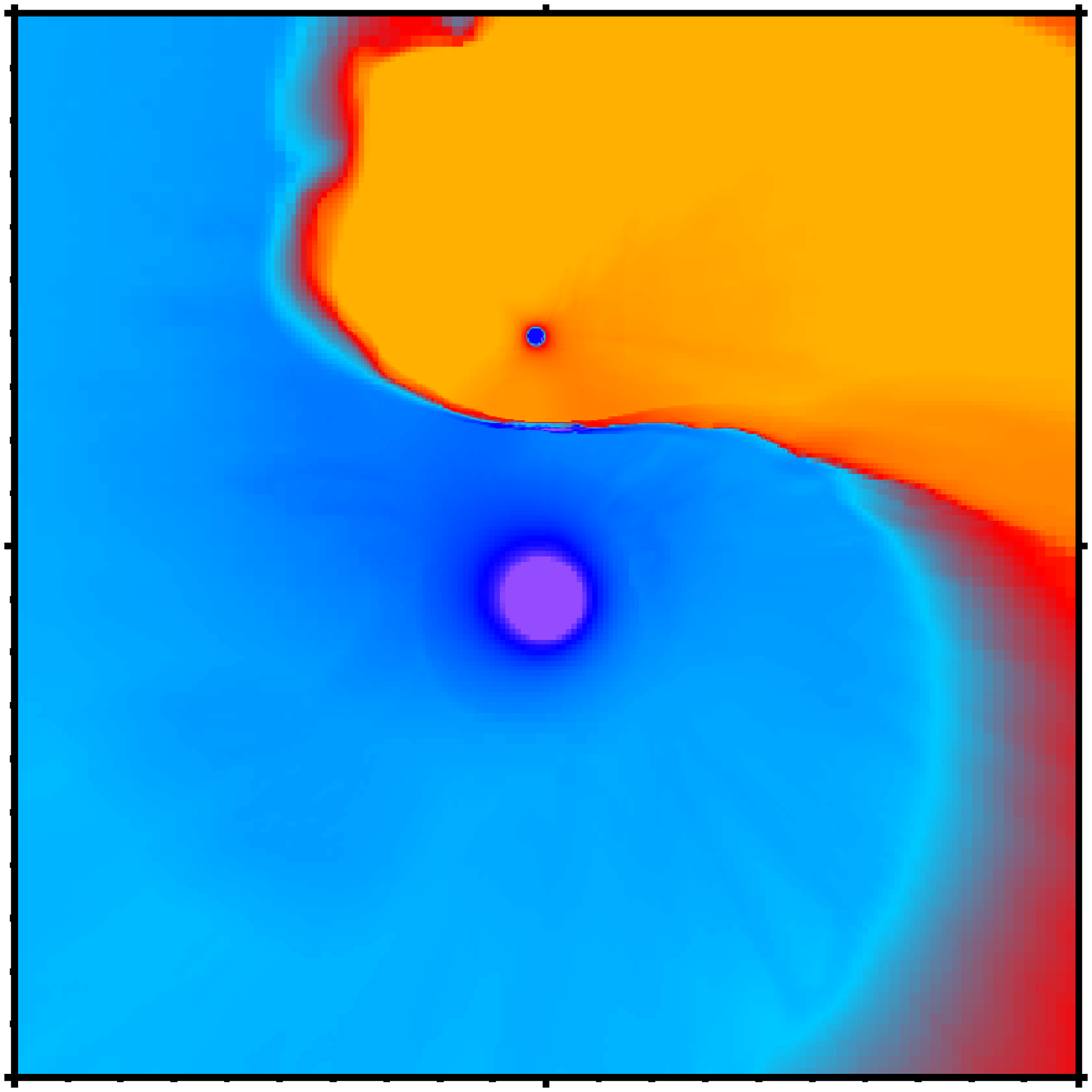}} \vspace{-1mm} \\ 

\resizebox{40mm}{!}{\includegraphics{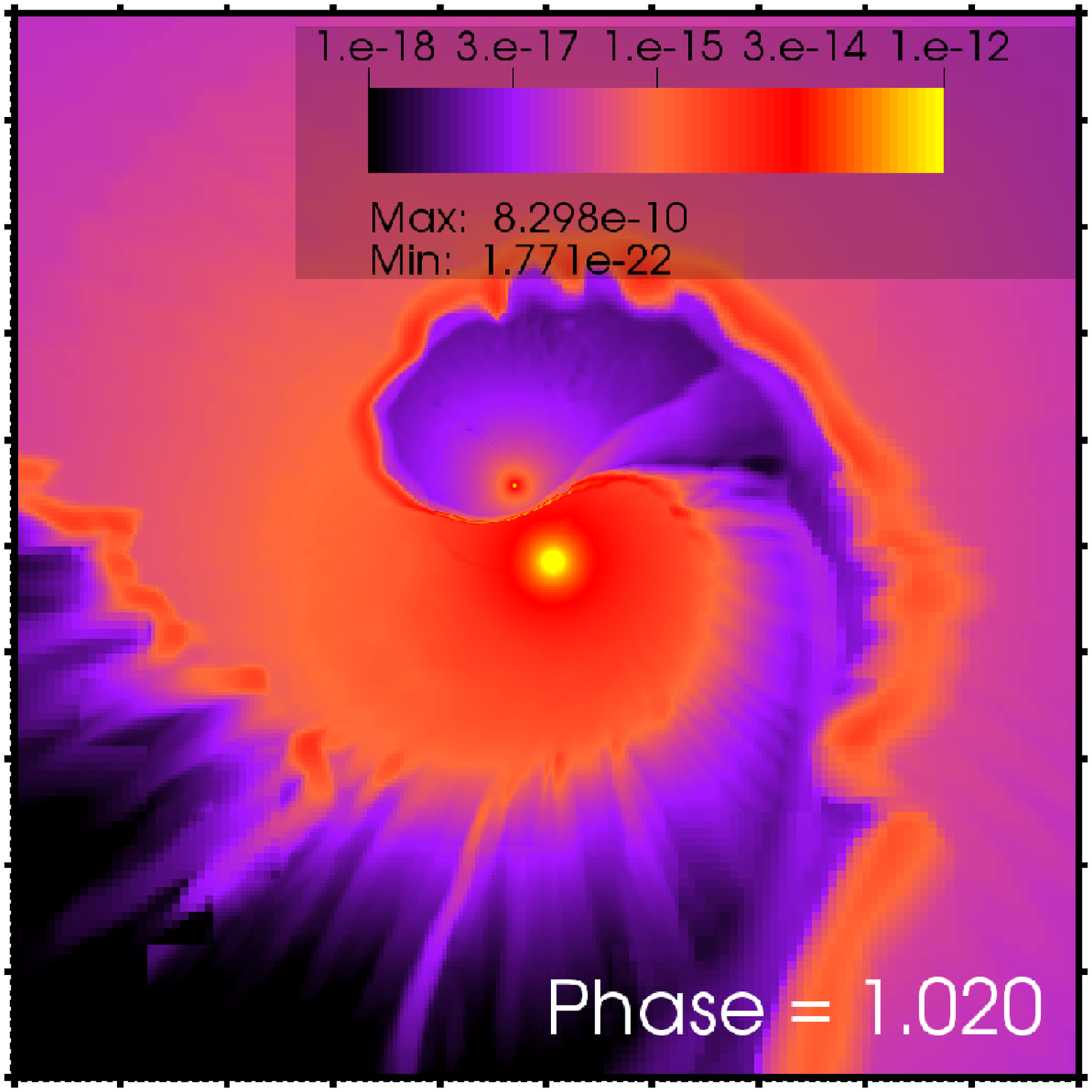}} & 
\resizebox{40mm}{!}{\includegraphics{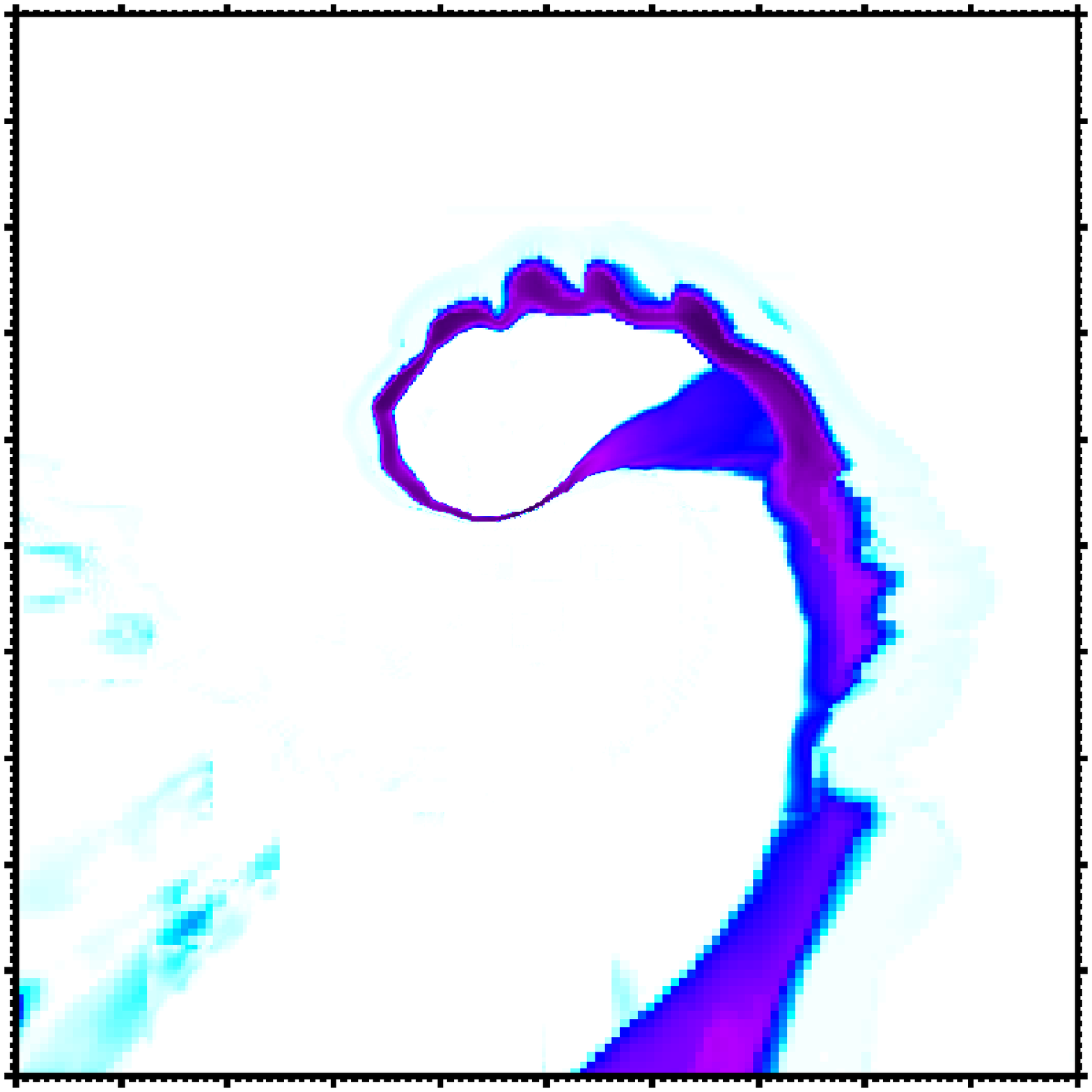}} & 
\resizebox{40mm}{!}{\includegraphics{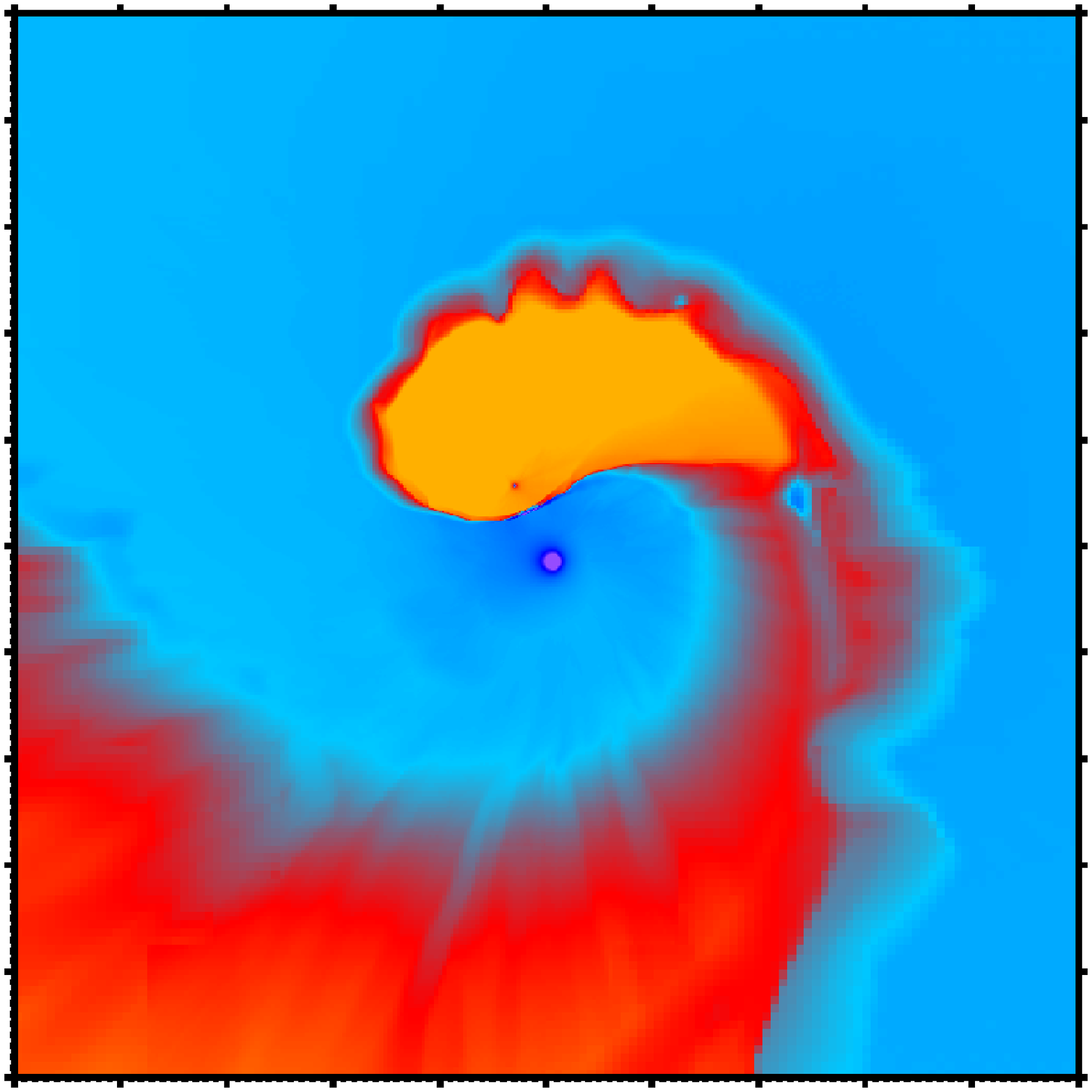}} \vspace{-1mm} \\ 

\resizebox{40mm}{!}{\includegraphics{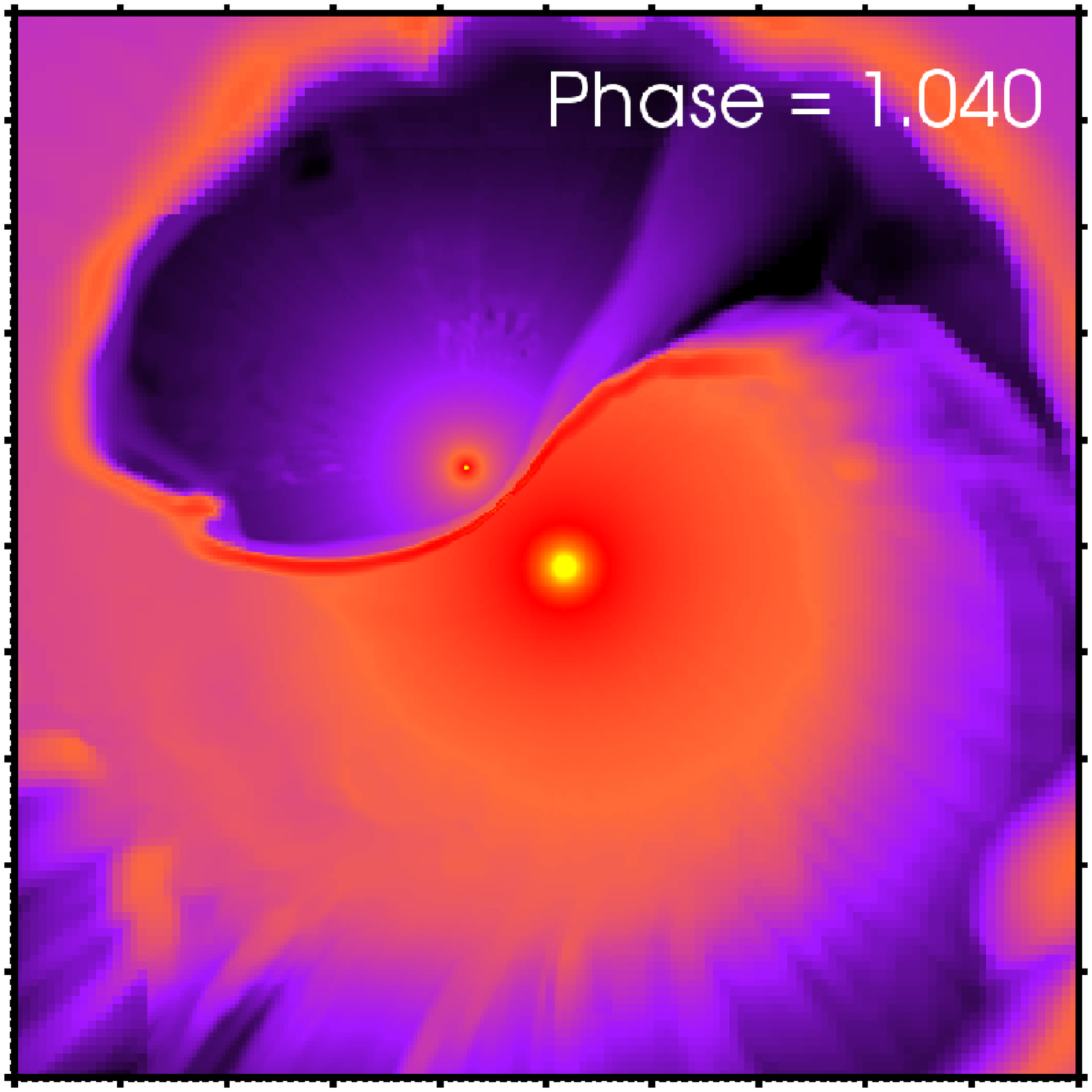}} & 
\resizebox{40mm}{!}{\includegraphics{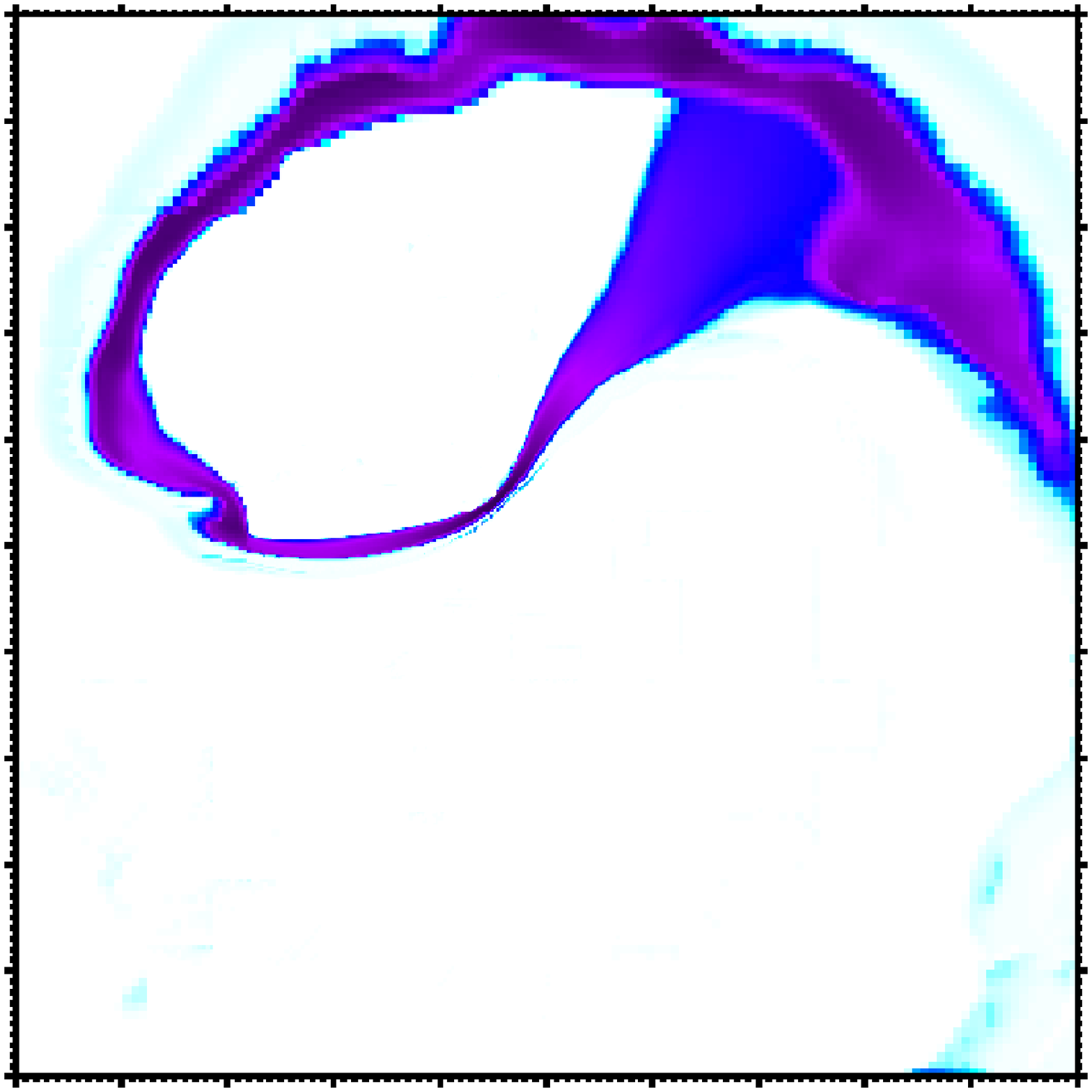}} & 
\resizebox{40mm}{!}{\includegraphics{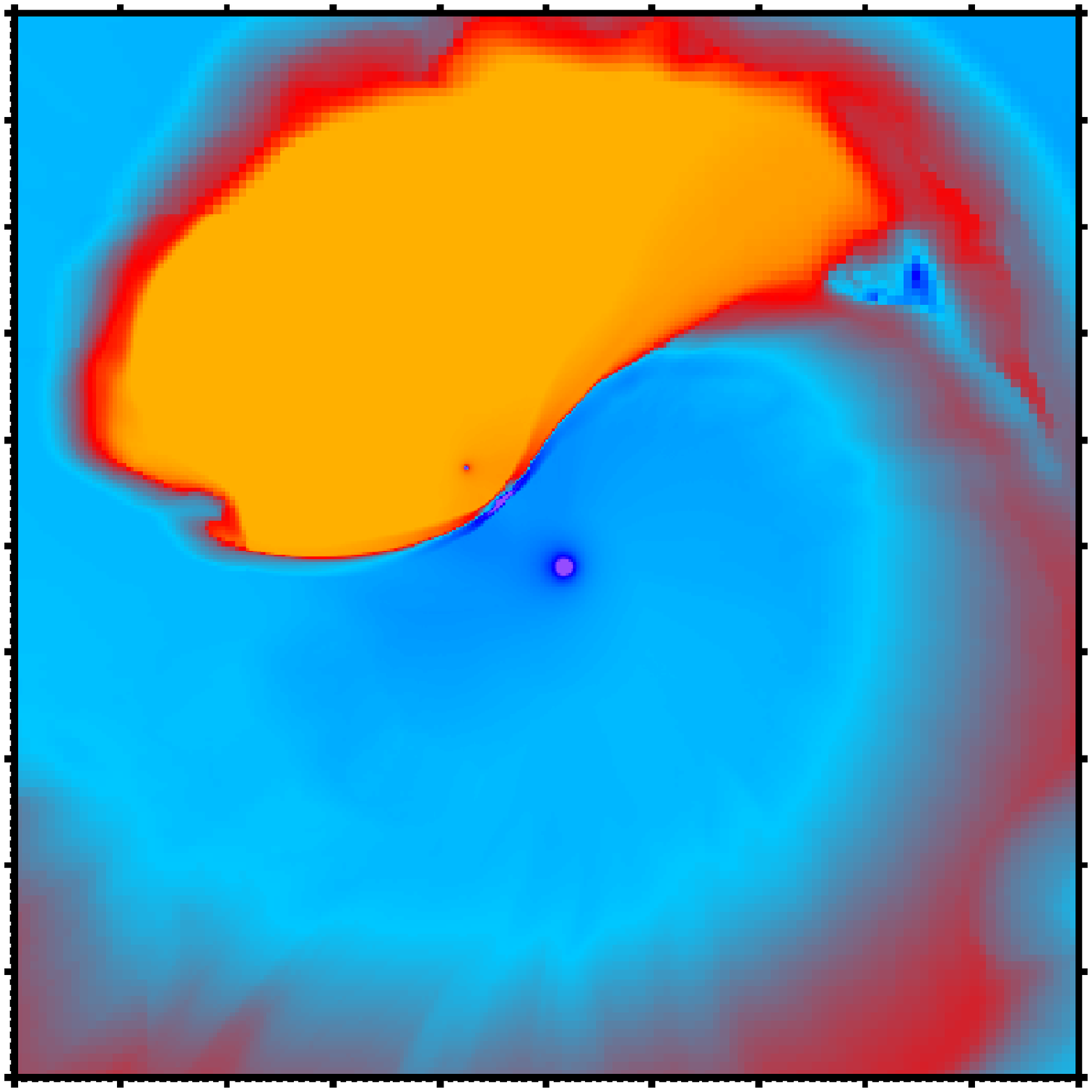}} \vspace{-1mm} \\ 

     \end{tabular}
    \caption{Same as Fig.~\ref{fig:vterm_peri_images} except model
      Orbit-RD is shown.}
    \label{fig:driven_peri_images}
  \end{center}
\end{figure*}

\subsubsection{Periastron passage}
\label{subsec:orbit_periastron}

Focusing now on the periastron passage of $\eta\;$Car, where there is
a considerable rate of change in the position of the stars and the
distribution of the winds, a number of interesting features can be
identified. Firstly, there appears to be a sudden change in the level
of instability in the shocks as periastron is approached. For
instance, prior to periastron the shocks seem to be unstable in both
model Orbit-IA and Orbit-RD (see $\phi=0.990$, and 0.995 snapshots in
Figs.~\ref{fig:vterm_peri_images} and \ref{fig:driven_peri_images}),
and one sees that, consistent with the predictions of the periastron
simulations in \S~\ref{sec:periastron}, the postshock gas is more
unstable when the wind acceleration regions are taken into
consideration (model Orbit-RD) than in the case where they are not
(model Orbit-IA). This fact is highlighted at $\phi=0.995$ where in
model Orbit-RD a shard of dense postshock primary wind gas is forced
through the shocks by instabilities in the leading arm of the WCR and
comes close to colliding against the companion star before being
rapidly ablated by its wind. However, as $\phi=1.000$ is approached,
and the orbital velocities of the stars increase, the shocks directly
between the stars appear much more stable and the catastrophic
disruption of the WCR seen in model Peri-RD is not reproduced in
model Orbit-RD. The reason for this is two-fold. Firstly, the level
of radiative inhibition of the preshock companion's wind is not as
severe when orbital motion is included - this point is discussed in
more detail below. And secondly, orbital motion, and the velocity
component that it provides perpendicular to the line of centres,
increases the preshock velocity. Consequently, the postshock gas is
less radiative and the prevailing thermal pressure stabilizes the WCR
against the growth of NTSIs. These results imply that a significant
disruption to the WCR must be afoot prior to the rapid rotation of the
stars (i.e. prior to $\phi=1.0$).

At orbital phases after periastron the contortion of the WCR by the
motion of the stars is most evident. The snapshots at $\phi=1.01$ and
1.02 in Figs.~\ref{fig:vterm_peri_images} and
\ref{fig:driven_peri_images} show how the unshocked companion wind is
completely enclosed by the WCR. Following periastron one can see that
the arms of the WCR become twisted to such an extent that they collide
downstream (see $\phi=1.02$ in Figs.~\ref{fig:vterm_peri_images} and
\ref{fig:driven_peri_images}), so that postshock gas in the trailing
arm is further heated as it passes through the shocks in the leading
arm. The wind of the companion star (on the side facing away from the
primary star) now collides against the inside of the leading arm of
the WCR and helps to heat this gas. In model Orbit-RD the preshock
companion wind in this direction is accelerated by the cumulative
radiative driving force from both stars, the result of which is that
immediately preshock this gas has reached a velocity of $3300\;{\rm
  km~s^{-1}}$, noteably higher than the (terminal) velocity of the
companion wind in model Orbit-IA. Consequently, the ram pressure of
the preshock gas is higher and it attains a pressure balance with the
postshock gas at a larger distance from the companion star, causing
the region of postshock companion wind to be thinner in model
Orbit-RD compared to model Orbit-IA. However, despite the higher ram
pressure of the companion's wind in this direction, the expansion of
the spiral at this time occurs at roughly the same speed because the
companion's wind is effectively running into a wall of dense postshock
primary wind with a high inertia. Therefore, it is the postshock
primary wind which largely dictates the rate of expansion. Examining
the tail of the WCR in the density snapshots at $\phi=1.01$, 1.02, and
1.04 one can see a striking example of photo-ablation by the radiation
fields of the stars (see Fig.~\ref{fig:driven_peri_images}).

\subsubsection{Preshock velocities}
\label{subsec:orbit_preshock_vels}

\begin{figure}
  \begin{center}
    \begin{tabular}{c}
\resizebox{80mm}{!}{\includegraphics{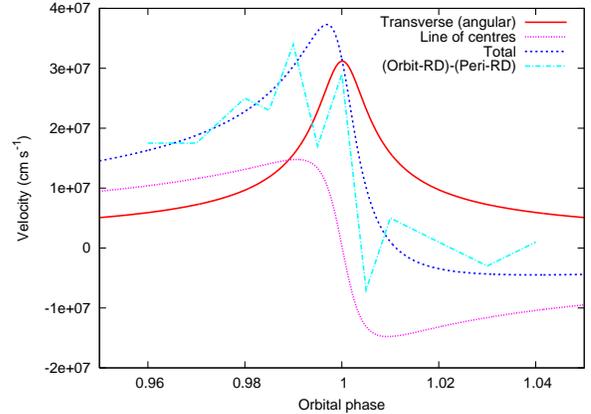}} \\
    \end{tabular}
    \caption{The orbital speed of the companion star around periastron
      passage. The components for the transverse and line-of-centres
      velocity, as well as the sum (total) are shown. For comparison,
      we also plot the difference between the preshock companion star
      wind velocities from models Orbit-RD and Peri-RD. Note that
      these values are calculated in the frame of reference of the
      centre of mass.}
    \label{fig:vels_peri}
  \end{center}
\end{figure}

\begin{figure}
  \begin{center}
    \begin{tabular}{c}
\resizebox{70mm}{!}{\includegraphics{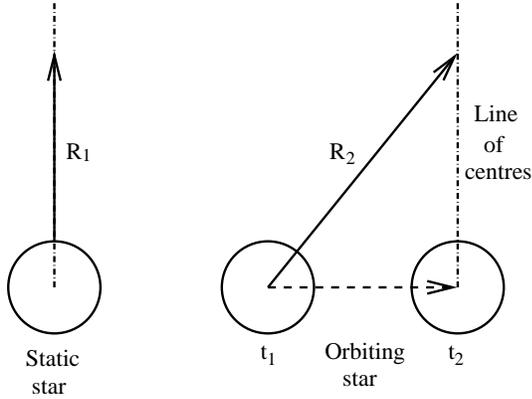}} \\
    \end{tabular}
    \caption{An illustrative example of the effect of orbital motion
      on the resultant line of centres wind acceleration. When the
      stars are static the line of centres (dot-dashed line) is
      aligned with radial flow (left). However, when a star undergoes
      orbital motion (right) the flow along the line of centres at
      time $t_{2}$ consists of wind material emitted at earlier times,
      $t_{1}$. Consequently, the path length for flow to reach the
      line of centres is larger (i.e. $R_2 > R_1$) and as $\rho
      \propto r^{-2}$ and $g_{\rm rad} \propto \rho^{-\alpha}$, this
      leads to an increase in the radiative acceleration.}
    \label{fig:streamline_effect}
  \end{center}
\end{figure}

The simulations presented in \S~\ref{sec:periastron}, in which the
stars were static with a separation corresponding to periastron,
predicted that the winds shock at such speeds that the WCR becomes
very unstable, so that dense clumps of gas may collide with the
companion star. Yet, when we add orbital motion to the mix we do not
see a disruption of such magnitude. If anything, the rapid motion of
the stars around periastron passage appears to have a stabilizing
effect on the apex of the WCR. We can begin to dissentangle this
puzzle by examining the preshock velocities from models Orbit-RD and
Peri-RD (Fig.~\ref{fig:preshock_velocity}). Clearly, at orbital
phases leading up to periastron the companion's wind is not reduced to
the same preshock velocity in model Orbit-RD as in model Peri-RD. In
fact, in model Orbit-RD the preshock companion's wind velocity does
not decrease to that in model Peri-RD until $\phi\simeq 1.005$, at
which point the lowest preshock velocity is attained. Following this
phase the agreement between model Orbit-RD and Peri-RD improves and
there appears to be a good correlation between the level of radiative
inhibition occuring. This behaviour is due to the combination of two
effects introduced by orbital motion (see
Fig.~\ref{fig:vels_peri}). In the frame of the WCR, the preshock gas
velocity is equal to the wind velocity (in the frame of the star) plus
the relative velocity of the star along the line of centres,
$\dot{d}_{\rm sep}$. Therefore, prior to $\phi \simeq 0.99$,
$\dot{d}_{\rm sep}$ is positive and the contraction of the stellar
separation will increase the preshock wind velocity. However,
$\ddot{d}_{\rm sep}$ is positive until $\phi \simeq 0.99$ and then
becomes negative (Fig.~\ref{fig:vels_peri}), so at $\phi > 0.99$ the
radial motion will decrease the preshock wind speed. The wind material
also has a considerable velocity in the transverse direction which
increases (decreases) as the stars approach (depart from)
periastron. This motion reduces the wind density along the line of
centres, and as $g_{\rm rad}\propto \rho^{-\alpha}$ the wind
acceleration increases, and the impact of radiative inhibition
decreases (see Fig.~\ref{fig:streamline_effect}). Note that the
alteration to $g_{\rm rad}$ will apply for both radiation fields, but
since $\alpha_{2} > \alpha_{1}$ the reduction in $\rho$ benefits the
secondary's radiative driving more than inhibition by the primary.

Prior to $\phi \simeq 1.01$ the combination of these two effects
increases the companion's preshock wind speed along the line of
centres compared to the static-star case (see
Figs.~\ref{fig:vel_comparison} and \ref{fig:vels_peri}). However,
following periastron the enhanced preshock velocity introduced by
transverse motion is countered by the receding relative motion of the
companion star along the line of centres. Examining the sum of the
orbital velocity components around periastron one sees good
qualitative agreement with the difference between models Orbit-RD and
Peri-RD (Fig.~\ref{fig:vels_peri}). We note that although the
critical point radius of the wind acceleration is not resolved in the
simulations, the influence of orbital acceleration on the preshock
winds will be negligible (Appendix~\ref{sec:orbital_motion}).

In simulation Orbit-IA a maximum wind speed of $\sim 3200\;{\rm
  km~s}^{-1}$ is attained. When the acceleration of the winds by the
radiation fields of both stars is considered the maximum speed for gas
grows to $\sim 4500 \;{\rm km~s}^{-1}$, with the highest velocities
possessed by unshocked companion wind gas which has been subjected to
the cumulative driving force from both stars. Although velocities of
this order will produce extremely hot gas when it shocks it is
unlikely that there will be a significant observable signature as the
total gas mass, its postshock density, and thus the free-free
emission, is very low.

\subsubsection{Resolution dependence}

Resolution tests were performed for the large-scale simulations using
finest cell sizes (for the postshock gas) in the range
$0.66-2.44\times10^{11}\;$cm ($0.88-3.5\;{\rm R_{\odot}}$). As the
simulation resolution is lowered the growth of KH instabilities at the
contact discontinuity is less pronounced, and vice-versa for higher
resolution. Importantly, there was a consistent finding that a
substantial disruption/collapse of the WCR does not occur in
simulations which include orbital motion (for our adopted stellar,
system, and radiation-wind coupling parameters).

We note again that the key reason that a catastrophic
disruption/collapse of the WCR does not occur at periastron in model
Orbit-RD is that the companion's preshock wind velocity does not reach
sufficiently low values to trigger radiative cooling in the postshock
gas. Therefore, although the simulations are only progressed through a
single orbit, continuing the simulation for subsequent orbits would
not result in a disruption/collapse.

\subsection{X-ray emission}
\label{subsec:xrays}

\etacar is an exceptionally bright source of X-ray emission. This
X-ray emission provides a direct probe of the WCR, and detailed
comparisons of model predictions against observations can determine
key parameters of the stars and their winds. To date, considerable
focus has been placed on constraining the orientation of the orbit on
the plane of the sky. There remains a lack of consensus on this issue,
with some models preferring the companion star to pass behind the
primary star at periastron \citep{Damineli:1996, Pittard:1998,
  Corcoran:2001, Pittard:2002, Corcoran:2005, Akashi:2006,
  Hamaguchi:2007, Nielsen:2007, Henley:2008, Okazaki:2008,
  Parkin:2009, Gull:2009, Richardson:2010, Groh:2010a}, whereas others
prefer the orientation the opposite way around \citep{Falceta:2005,
  Abraham:2005a, Abraham:2007, Kashi:2008b, Falceta:2009,
  Abraham:2010}, or the system at quadrature \citep{Ishibashi:2001,
  Smith:2004}. The inclination of the orbital plane, and whether or
not it is aligned with the equatorial skirt of the Homunculus nebula,
is still debated \citep[see][and references
  there-in]{Parkin:2009}. Our new models can also be used to test the
orientation. We define our line of sight geometry as follows: the
inclination angle, $i$, is measured against the $z$-axis
($i=0^{\circ}$ would view the system from directly above the orbital
plane), and the angle $\theta$ is measured against the {\it negative}
$x$-axis (companion star in front at apastron) such that $\theta$
increases in the prograde direction ($\theta =90^{\circ}$ would align
the line of sight with the negative $y$-axis). We adopt viewing angles
of $i=42^{\circ}$ and $\theta = 20^{\circ}$, in agreement with values
determined from the most recent modelling of the X-ray emission from
$\eta\;$Car \citep{Okazaki:2008, Parkin:2009}. In the following
analysis we define $\phi=1.0$, 2.0, and 3.0 as periastron passage in
1998, 2003.5, and 2009, respectively.

\subsubsection{X-ray lightcurves}
\label{subsec:lightcurves}

\begin{figure*}
  \begin{center}
    \begin{tabular}{ccc}
\multicolumn{3}{l}{\resizebox{180mm}{!}{\includegraphics{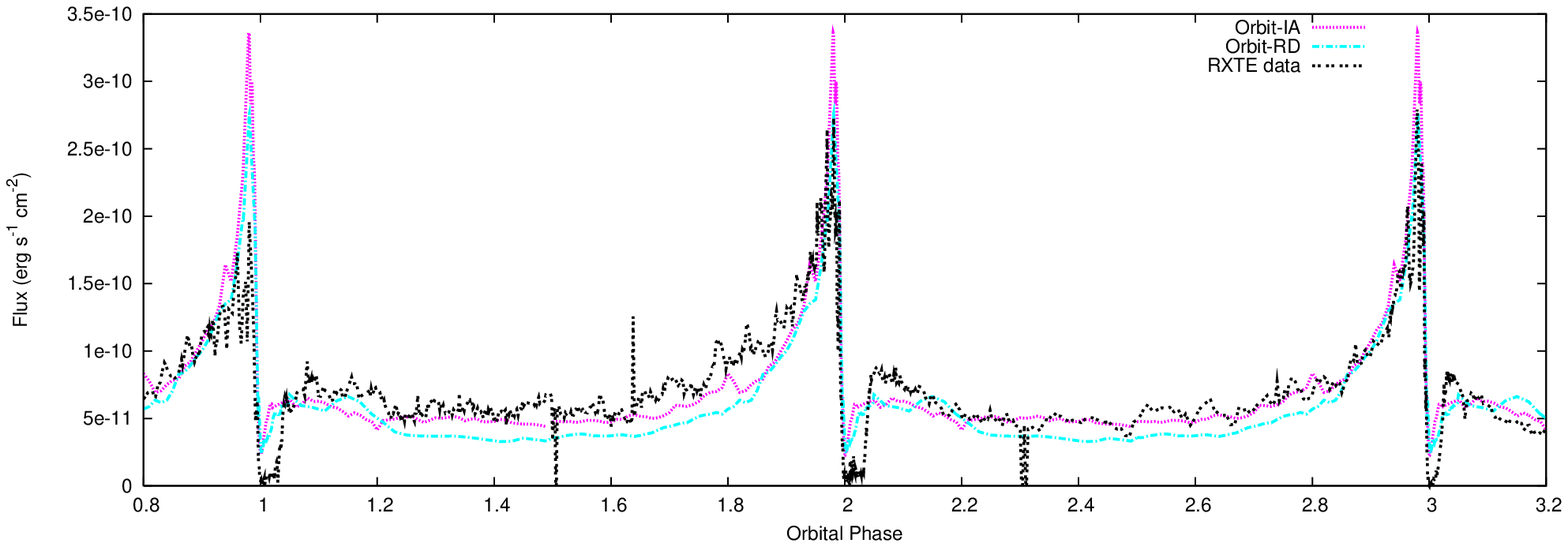}}} \\
\resizebox{60mm}{!}{\includegraphics{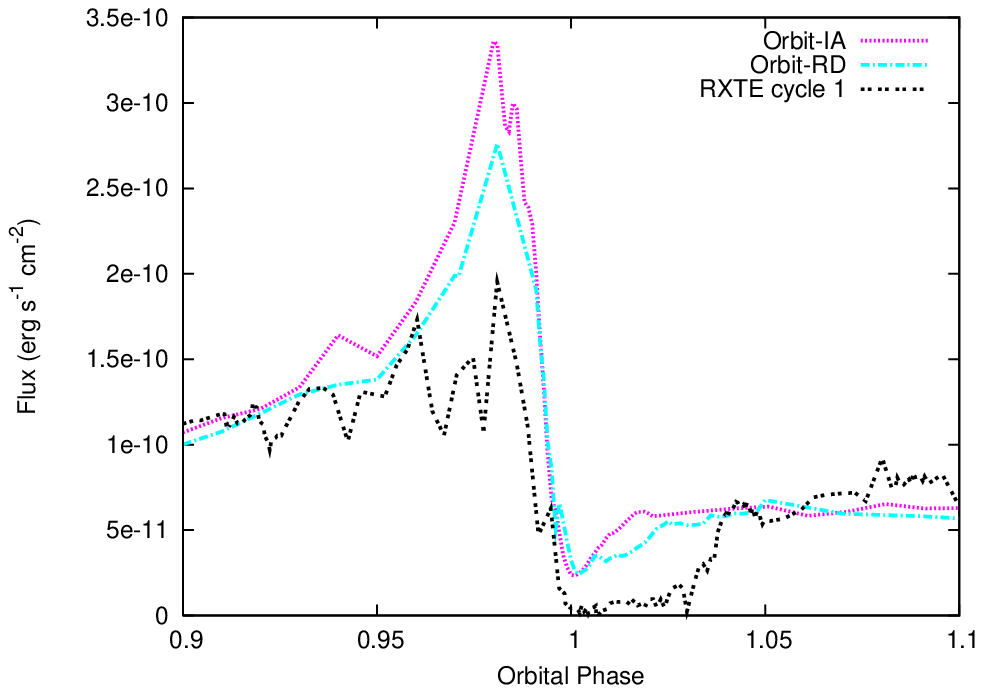}} &
\resizebox{60mm}{!}{\includegraphics{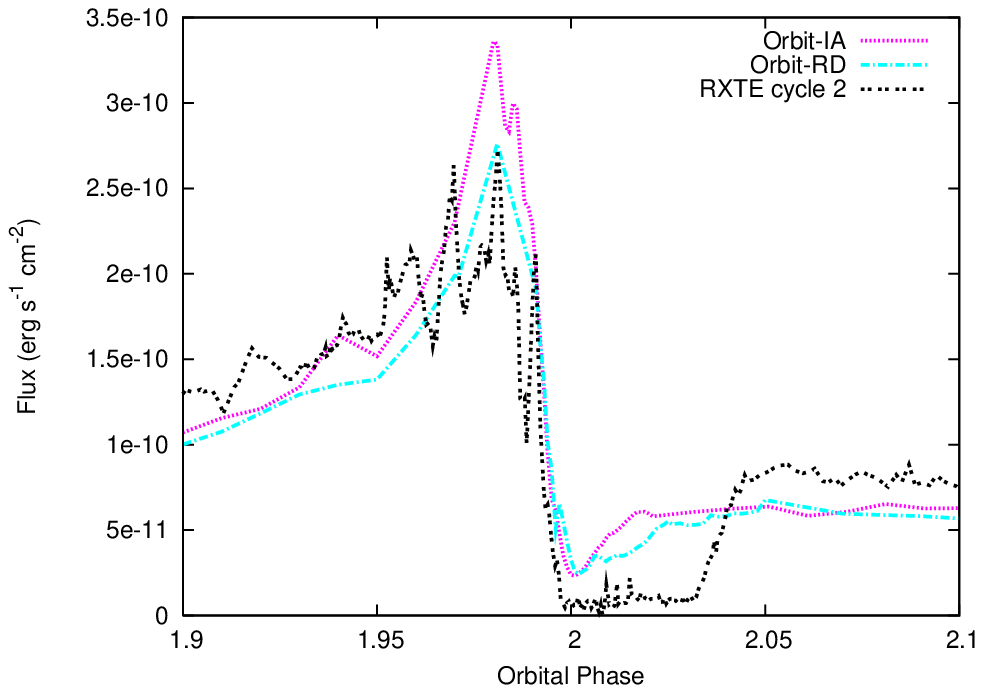}} &
\resizebox{60mm}{!}{\includegraphics{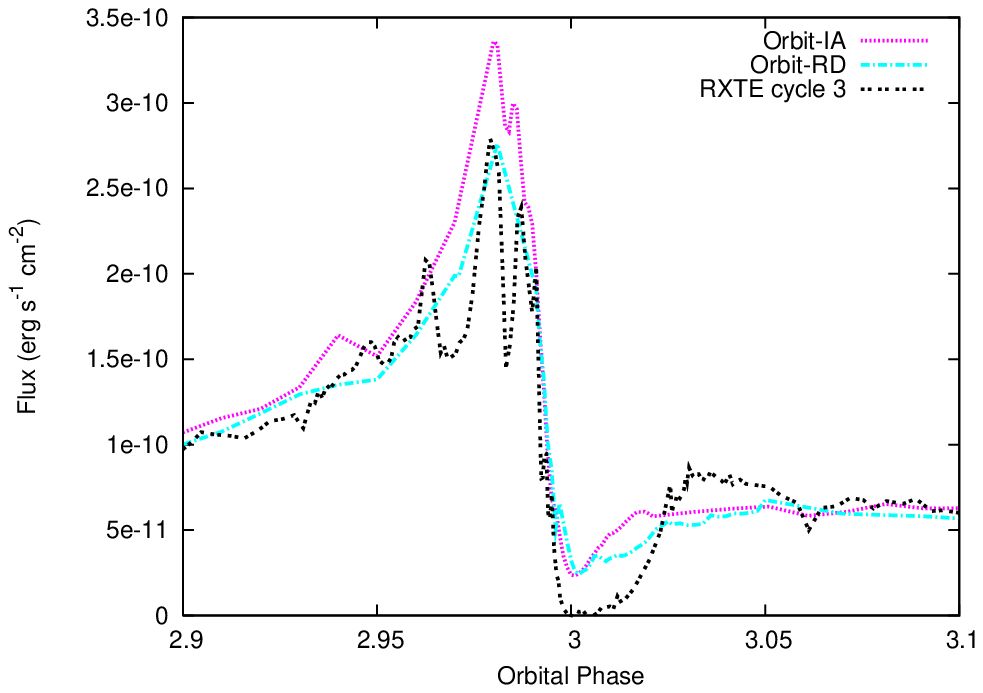}} \\
    \end{tabular}
    \caption{2-10 keV X-ray lightcurves for models Orbit-IA and
      Orbit-RD plotted against the {\it RXTE} lightcurve over three
      cycles (top row) and against the individual minima (bottom row -
      left to right: 1998, 2003.5, and 2009). The estimated cosmic
      background has been removed from the {\it RXTE} data
      \citep{Corcoran:2001, Corcoran:2005, Corcoran:2010}.}
    \label{fig:lc}
  \end{center}
\end{figure*}

The X-ray lightcurves for models Orbit-IA and Orbit-RD are shown in
Fig.~\ref{fig:lc}. The characteristic shape of $\eta~$Car's X-ray
lightcurve \citep{Corcoran:2001, Corcoran:2005, Corcoran:2010} is
reproduced by the models; the X-ray flux remains relatively flat for
the majority of the orbit ($\phi \simeq 0.05-0.7$) and then as the
stars approach periastron there is a sharp increase in X-ray flux
($\phi\simeq 0.7-0.99$) followed by an abrupt decline to a minimum at
periastron, and a recovery of the X-ray flux as the stars move away
from periastron. The difference in apastron flux between models
Orbit-RD and Orbit-IA is due to the WCR being closer to the primary
star in the former (see \S~\ref{subsec:dynamics}). Due to the slightly
different wind densities at equivalent distances from the stars in
models Orbit-IA and Orbit-RD we do not attempt to infer suggested
alterations to model parameters which would improve the model
fits. However, we note that on the whole the adopted wind momentum
ratio provides good agreement between the model and the observed
lightcurve.

\begin{figure*}
  \begin{center}
    \begin{tabular}{ccc}
\resizebox{55mm}{!}{\includegraphics{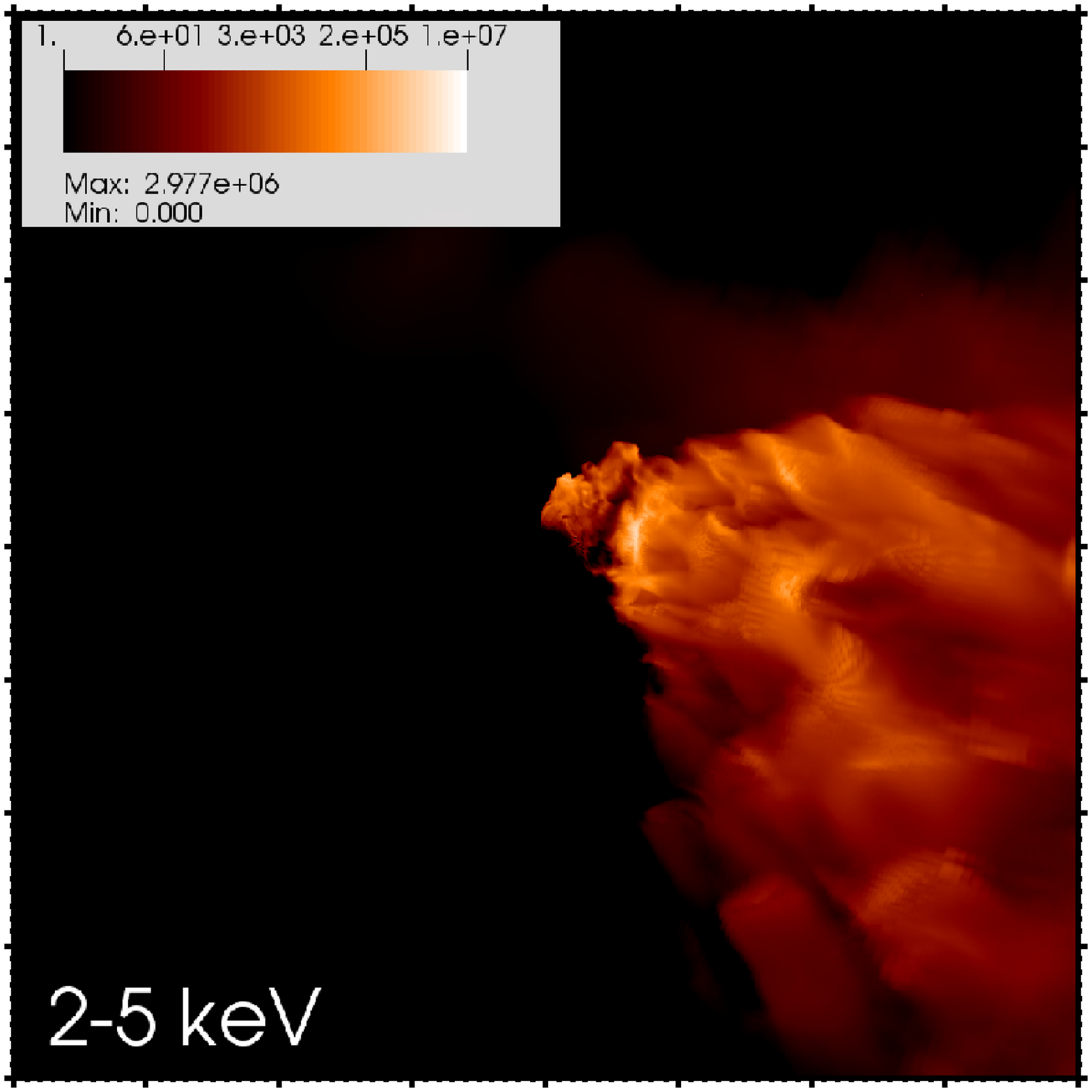}} &
\resizebox{55mm}{!}{\includegraphics{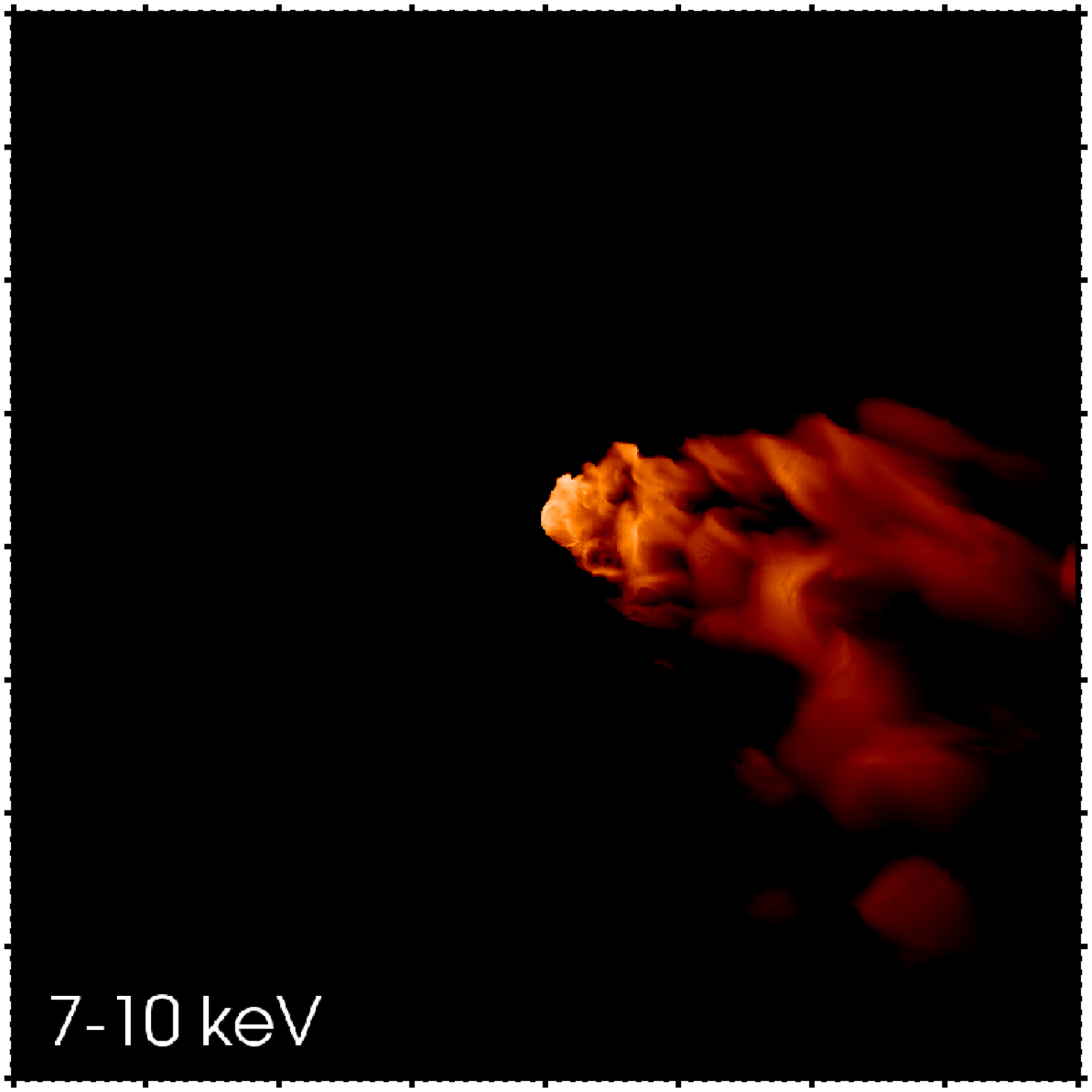}} &
\resizebox{55mm}{!}{\includegraphics{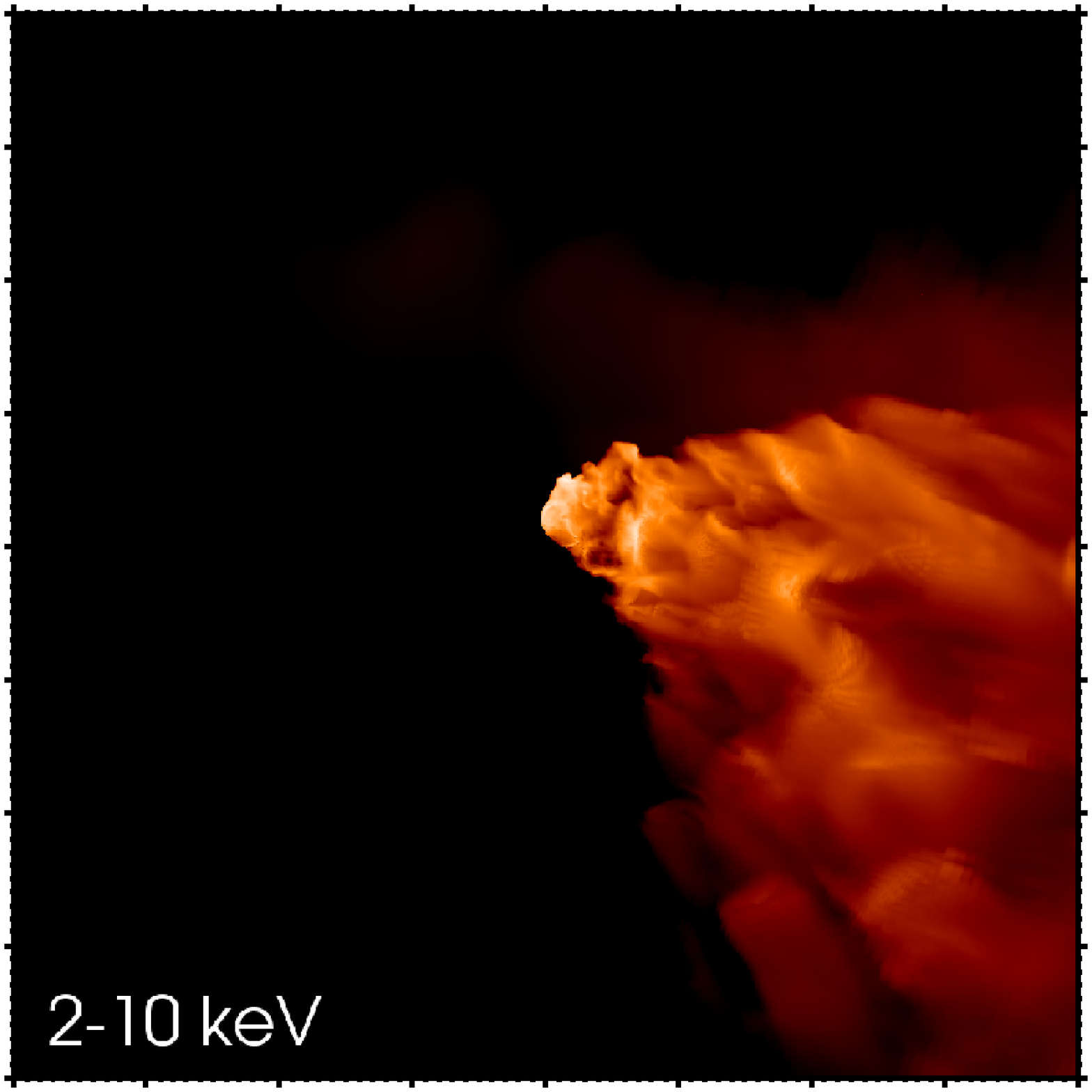}} \\
    \end{tabular}
    \caption{Ray-traced images showing X-ray emission
      (erg~s$^{-1}$~cm$^{-2}$~keV$^{-1}$~ster) from model Orbit-RD at
      $\phi = 1.0$ in the 2-5 (left), 7-10 (middle), and 2-10~keV
      (right) energy bands. The centre of the image is aligned with
      the centre of the simulation box and the viewing angles are
      $i=42^{\circ}$ and $\theta= 20^{\circ}$
      (\S~\ref{subsec:xrays}). The plots show a region of $\pm
      4\times10^{14}\;$cm - large axis tick marks correspond to a
      distance of $1 \times10^{14}\;$cm.}
    \label{fig:bbimages}
  \end{center}
\end{figure*}

Examining the orbital phase range $\phi=0.9-1.1$ reveals some noteable
differences between the observations and the models. Both model
Orbit-IA and, to a lesser extent, model Orbit-RD overestimate the
peak in {\it average} flux observed by {\it RXTE} prior to
periastron. At $\phi = 1.0$ the models do not reach the same minimum
in flux as the {\it RXTE} data. This is, at first, a somewhat puzzling
result as previous models were able to reproduce the low flux level at
$\phi=1.0$ via an eclipse of the X-ray emission region
\citep{Okazaki:2008, Parkin:2009}. However, inspection of the 2-5 keV
broadband image at this phase reveals considerable emission from
downstream gas in the tail of the WCR (Fig.~\ref{fig:bbimages}). This
emission predominantly comes from the trailing arm of the WCR with a
weaker contribution from the downtream gas in the leading arm (see
Fig.~\ref{fig:peri_emiss} - note that due to our adopted viewing angle
the emission close to the apex in the {\it leading} arm is strongly
absorbed by the intervening primary wind). This lack of agreement
provides an important clue as to the nature of the X-ray minimum of
\etacar - to match the low flux level at $\phi=1.0$ the apex of the
WCR must be disrupted at a slightly earlier phase such that the
disruption propagates out and removes the X-ray emitting plasma in the
trailing arm. The residual flux observed by {\it
  XMM-Newton}\footnote{During the deep minimum there will be some
  contamination from the cosmic background in the {\it RXTE}
  data.} at $\phi = 1.0$ could then be contributed by the X-ray
emitting plasma in the leading arm of the WCR \citep[i.e. in the
  downstream gas {\it behind} the companion star
  -][]{Parkin:2009}. The flow time for the postshock companion star
wind to be accelerated from the shock apex to point ``A'' in
Fig.~\ref{fig:peri_emiss} is $\simeq12\;$days ($\delta\phi\simeq
0.006$), which suggests that the disruption of the WCR must be afoot
by $\phi=0.994$, in agreement with the 0-40 days proposed by
\cite{Soker:2005} for the commencement of an accretion event
\citep[see also][]{Akashi:2006}.

A key failing is that neither model reproduces the observed extended
X-ray minimum. The X-ray flux from model Orbit-IA recovers by $\phi =
1.017$, whereas model Orbit-RD shows a slightly extended minimum with
a recovery by $\phi = 1.025$. It would therefore seem that the width
of the extended X-ray minimum cannot be reproduced when the spatial
extent of the emitting region is taken into consideration, which poses
difficulties for models based on an eclipse of the X-ray emission
region \citep[][ see also
  \citeauthor{Parkin:2009}~\citeyear{Parkin:2009}]{Okazaki:2008}. A
step is seen in the minimum of model Orbit-RD, which corresponds to
lines of sight passing through the dense shell of postshock primary
wind, although it is not as pronounced as in the model of
\cite{Parkin:2009}. Encouragingly, the width of the X-ray minimum in
model Orbit-RD agrees well with the less extended cycle 3 minimum
observed by {\it RXTE} in 2009 (see lower right panel of
Fig.~\ref{fig:lc}). This result implies that in the previous two
cycles the extended X-ray minimum resulted from a catastrophic
disruption of the WCR, whereas the shorter duration of the recent
minimum indicates that the WCR was not so severely disrupted the last
time around.

\subsubsection{X-ray line profiles}
\label{subsec:line_profiles}

\begin{figure}
  \begin{center}
    \begin{tabular}{c}
\resizebox{65mm}{!}{\includegraphics{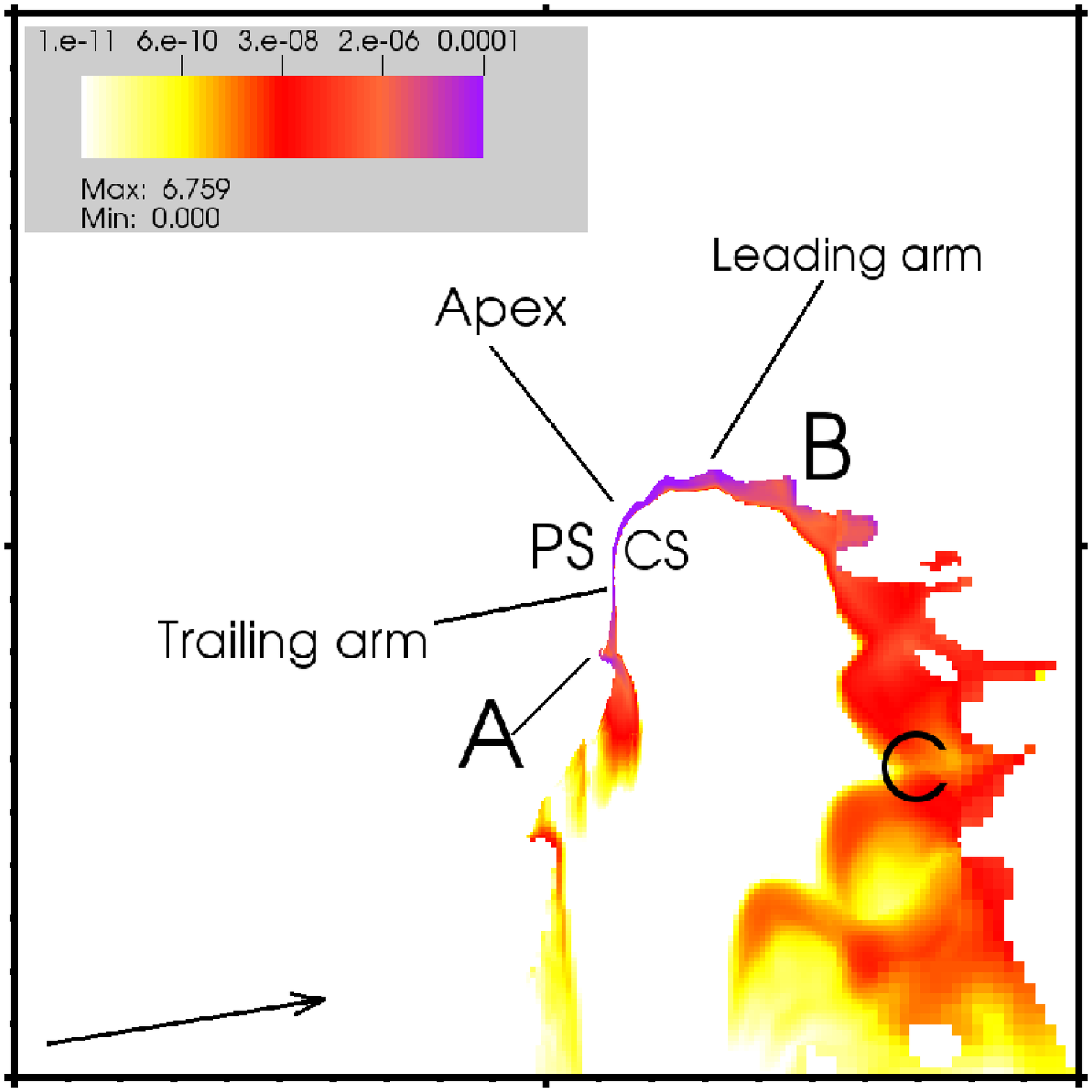}} \\
    \end{tabular}
    \caption{Snapshot of the orbital ($x-y$) plane from model
      Orbit-RD at $\phi = 1.0$ showing the 2-5~keV X-ray emissivity
      (erg~s$^{-1}$~cm$^{-3}$~keV$^{-1}$~ster). The region of
      postshock gas in the trailing arm of the WCR responsible for the
      excess of 2-5~keV X-ray emission is labelled ``A''. The labels
      ``B'' and ``C'' refer to suggested sites for S and Si X-ray line
      emission. The position of the primary and companion star are
      indicated by the ``PS'' and ``CS'' labels, respectively. The
      arrow in the lower left of the image indicates the direction
      vector (in the orbital plane) for our adopted line of sight. The
      plots show a region of $\pm1\times10^{14}\;$cm - large axis tick
      marks correspond to a distance of $1\times10^{14}\;$cm. For the
      corresponding density, temperature, and speed images see
      Fig.~\ref{fig:driven_peri_images}.}
    \label{fig:peri_emiss}
  \end{center}
\end{figure}

If the apex of the WCR is heavily disrupted the projected velocity
along the lines-of-sight from high temperature, X-ray emitting
postshock gas will be different. A high spectral resolution
observation with {\it Chandra} found that at $\phi=1.009$ there was a
positive line shift (i.e. away from the observer) for the S~XVI line,
whereas S~XV, Si~XIV, and Si~XIII had negative line shifts
\citep[][see also
  \citeauthor{Behar:2007}~\citeyear{Behar:2007}]{Henley:2008}. In the
absence of a stable WCR between the stars the higher excitation S~XVI
line could still have been emitted by hot postshock companion star
wind {\it close to} the WCR disruption which (for our adopted viewing
angle) is travelling away from the observer (labelled ``B'' in
Fig.~\ref{fig:peri_emiss}). The S~XV, Si~XIV, and Si~XIII lines could
then originate from lower temperature gas further downstream in the
leading arm of the WCR (labelled ``C'' in Fig.~\ref{fig:peri_emiss}),
which is moving towards the observer, albeit at a steep angle, such
that there will be a negative projected line shift.

\subsubsection{Column density}
\label{subsec:column}

\begin{figure*}
  \begin{center}
    \begin{tabular}{cc}
\resizebox{80mm}{!}{\includegraphics{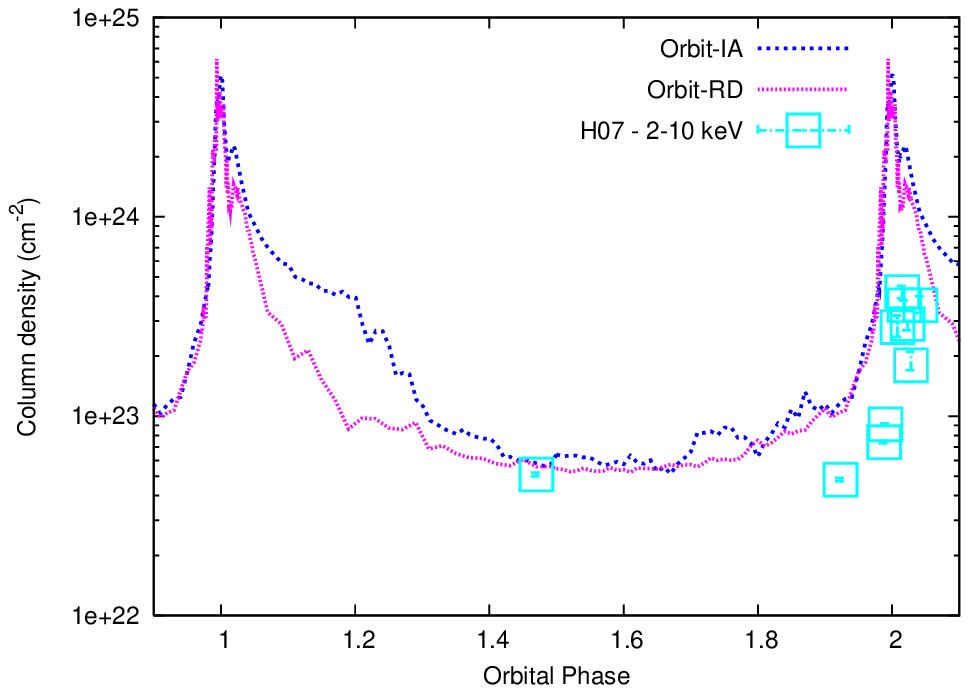}} &
\resizebox{80mm}{!}{\includegraphics{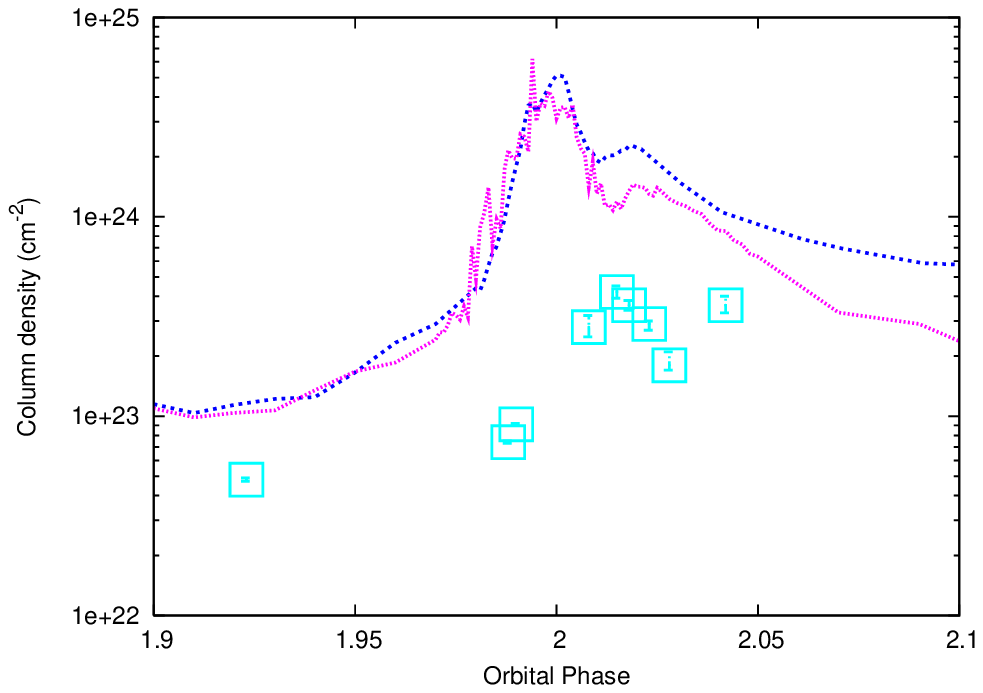}} \\
    \end{tabular}
    \caption{Emission weighted column density (EWC) for models
      Orbit-IA and Orbit-RD for orbital phase range $\phi = 0.9 -
      2.1\;$(left panel) and 1.9-2.1 (right panel). For comparison,
      the column densities derived by \cite{Hamaguchi:2007} for the
      2-10~keV X-ray emission are shown (H07 - 2-10keV), with the
      formal error bars obtained from spectral fitting. The
      interstellar and nebula column density ($\sim
      5\times10^{22}\;$cm$^{-2}$) are additional to these plots.}
    \label{fig:EWC}
  \end{center}
\end{figure*}

The emission-weighted column densities\footnote{The emission-weighted
  column density (EWC) is calculated as $N_{\rm EWC}= \Sigma N_{\rm H}
  L_{\rm int X} / \Sigma L_{\rm int X} $, where $N_{\rm H}$ and
  $L_{\rm int X}$ are the column density and 1-10~keV intrinsic
  luminosity from a given line of sight, and the summation is over all
  sight lines (pixels) in the X-ray image.} from models Orbit-IA and
Orbit-RD appear to be largely similar for the majority of the orbit
(Fig.~\ref{fig:EWC}), though model Orbit-IA shows a slower decline in
column density in the orbital phase range 1.0-1.4. This difference is
caused by the distribution of the dense layer of postshock primary
wind in relation to the lines of sight to the X-ray emitting plasma
(see Figs.~\ref{fig:vterm_big_images}, \ref{fig:driven_big_images},
\ref{fig:vterm_peri_images}, and \ref{fig:driven_peri_images}). In
model Orbit-IA the growth of KH instabilities at the contact
discontinuity in the leading arm of the WCR distort the dense layer of
postshock primary wind and push it into the path of X-rays as they
exit the system, consequently increasing the column density. In
contrast, the lack of KH instabilities in model Orbit-RD results in a
more rapid decline in column density as the stars move away from
periastron. Interestingly, the column density calculated from model
Orbit-RD agrees very well with the results from the model used by
\cite{Parkin:2009}, which did not allow dynamical instabilities.

The column densities derived from fits to \XMM~spectra in the 2-10~keV
energy range \citep{Hamaguchi:2007} are also plotted in
Fig.~\ref{fig:EWC}. At $\phi\simeq1.47$ the models and the
\cite{Hamaguchi:2007} values are in good agreement, with a rather
modest reduction of 1.15 required to align the models and
observation. Our adopted viewing angle places the companion star in
front of the primary at this orbital phase. Therefore, $N_{\rm
  H}\propto \dot{M}_{2}$, and the aforementioned reduction factor
would bring the companion star mass-loss rate to a value of
$1.2\times10^{-5}\Msolpyr$. However, at phases close to periastron the
column densities calculated from models Orbit-IA and Orbit-RD are
noticeably higher than the \cite{Hamaguchi:2007} values. During this
part of the orbit there is a more complicated distribution of both the
shocked and unshocked winds which lines of sight to the X-ray emitting
plasma must pass through. Therefore, we cannot infer such a clear-cut
reduction in mass-loss rates from a comparison between the models and
the \cite{Hamaguchi:2007} column densities. At face value,
Fig.~\ref{fig:EWC} implies that a reduction in the mass-loss rates of
at least a factor of 2 is necessary, which gives lower limits to the
mass-loss rates of $\dot{M}_1=2.4\times10^{-4}\Msolpyr$ and
$\dot{M}_2=7\times10^{-6}\Msolpyr$. However, this seems unlikely,
given other evidence for mass-loss rates greater than those assumed
here \citep{Hillier:2001, Hillier:2006, vanBoekel:2003, Groh:2010b},
unless these estimates have in turn been ``contaminated'' by emission
from the WCR. Another, perhaps more palatable way to reconcile the
difference would be to invoke the presence of (X-ray) ionized gas
close to the stars, which would reduce the opacity and thus allow for
higher mass-loss rates. Alternatively, perhaps observationally derived
column densities around periastron are suspect, due to the
complexities of disentangling different spectral components, and/or
the vagaries of fitting simpler plasma models to complicated systems
\citep[][]{Antokhin:2004, Pittard_Parkin:2010}. For instance, any
residual soft X-ray emission near $\eta~$Car (unresolved to {\it
  XMM-Newton}) will reduce the apparent column density.

Despite the model column densities being higher than the observed
values, we can extract some useful information from a comparison of
the variation in column density as a function of orbital phase. For
instance, the models and the \cite{Hamaguchi:2007} values agree in
that there is a sharp rise immediately prior to periastron which is
followed by a dip, and then there is a second smaller peak. As shown
by \cite{Parkin:2009}, the first peak ($\phi=2.00$) corresponds to a
rapid increase in absorption as the emission region is obscured by the
dense unshocked primary's wind, and the second peak ($\phi=2.02$)
corresponds to sightlines to regions of high intrinsic luminosity
becoming closely tangential with the leading arm of the WCR (see the
$\phi=1.02$~snapshots in Figs.~\ref{fig:vterm_peri_images} and
\ref{fig:driven_peri_images}). The differences between the time at
which the EWC in the models and the data begin to increase as
periastron is approached, and also the time of the peak in column,
suggest that our adopted azimuthal angle $\theta=20^{\circ}$ is too
small - if the value were increased slightly then the peak in column
density would occur at a later orbital phase.

We note that in the model of \cite{Okazaki:2008}, which assumed the
emission to originate from a point source situated close to the apex
of the WCR, it was this secondary peak in column density at
$\phi=2.02$ which allowed the extended X-ray minimum to be
reproduced\footnote{There was an additional source of absorption from
  a pile-up of unshocked primary wind gas in the SPH model of
  \cite{Okazaki:2008}. In our grid-based hydrodynamical models we do
  not see such a build-up as this gas is processed through the shocks
  and thereafter flows downstream within the WCR.}. This is evident if
one compares the position of the apex of the WCR relative to the
leading arm of the WCR in the $\phi=1.02\;$and 1.04 snapshots in
Figs.~\ref{fig:vterm_peri_images} and \ref{fig:driven_peri_images}.

\subsubsection{Spectra}

\begin{figure*}
  \begin{center}
    \begin{tabular}{cc}
\resizebox{65mm}{!}{\includegraphics{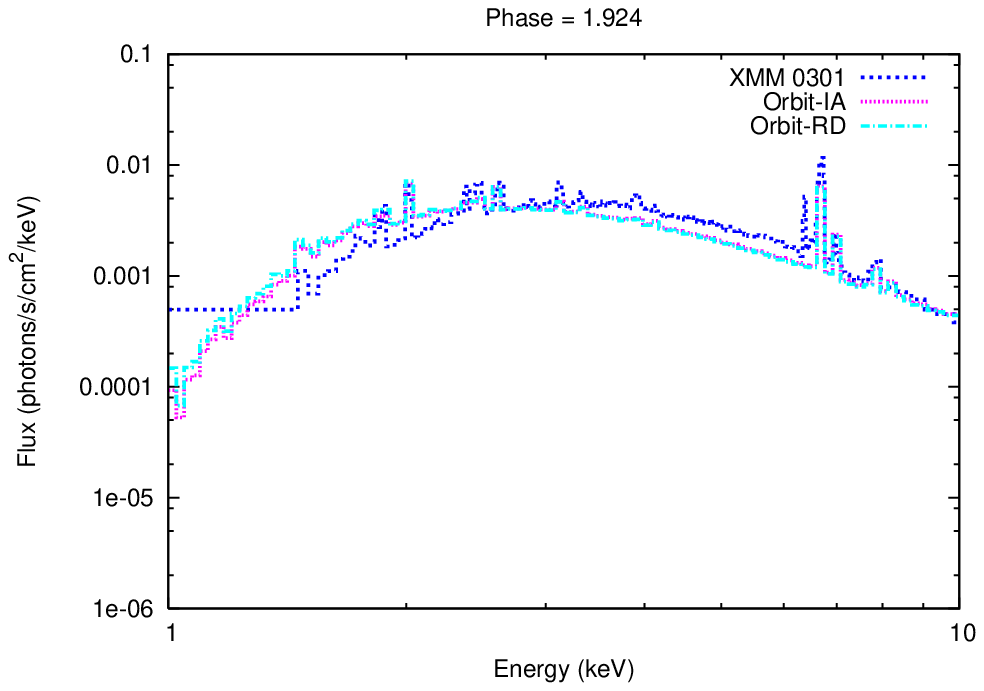}} &
\resizebox{65mm}{!}{\includegraphics{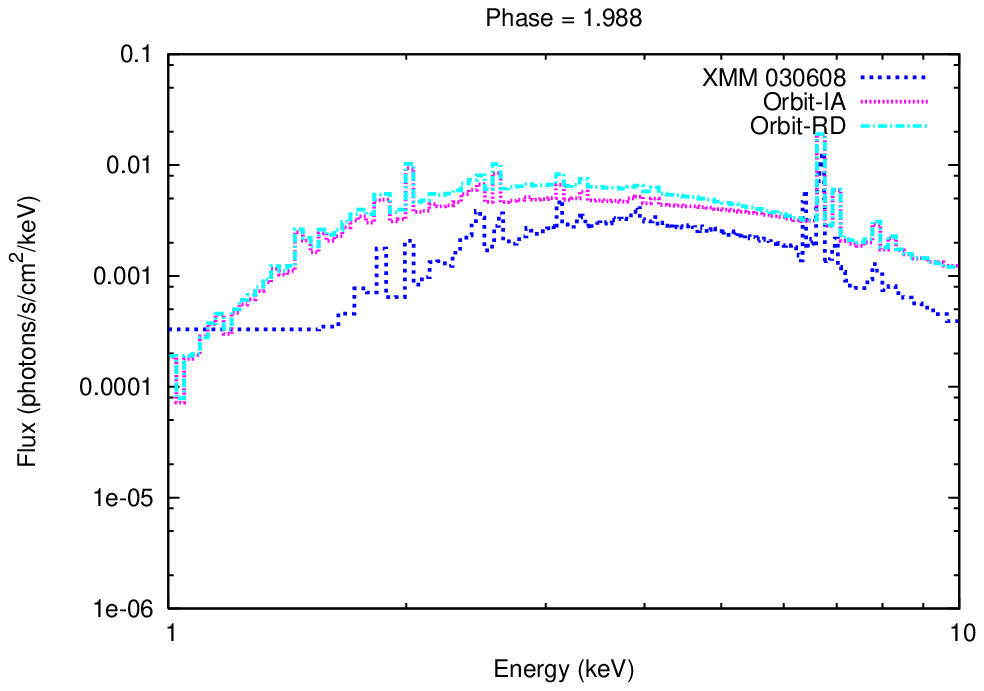}} \\
\resizebox{65mm}{!}{\includegraphics{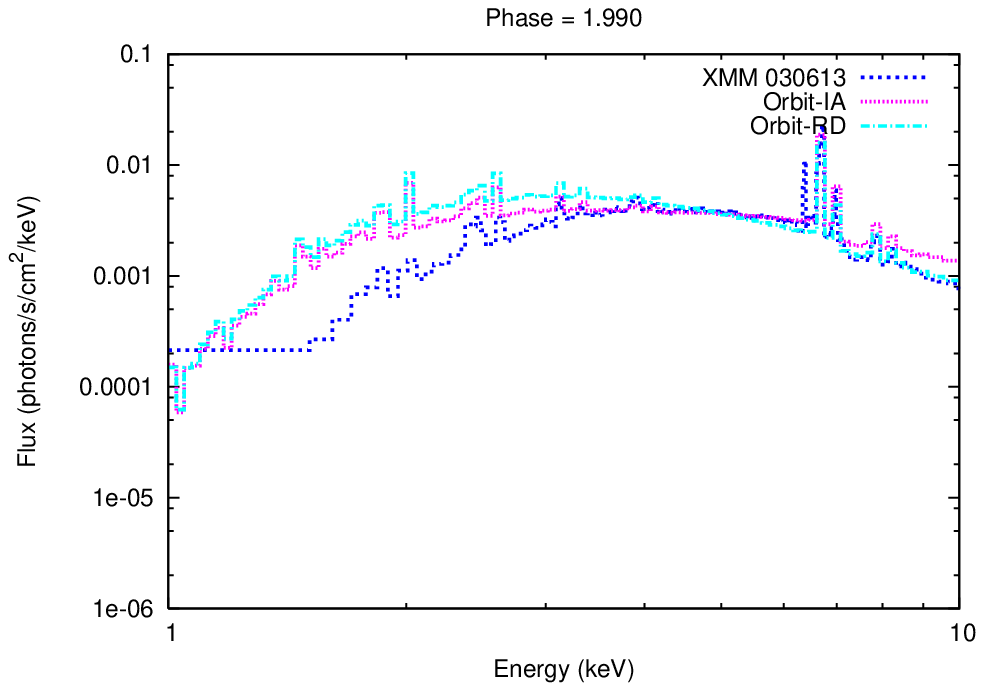}} &
\resizebox{65mm}{!}{\includegraphics{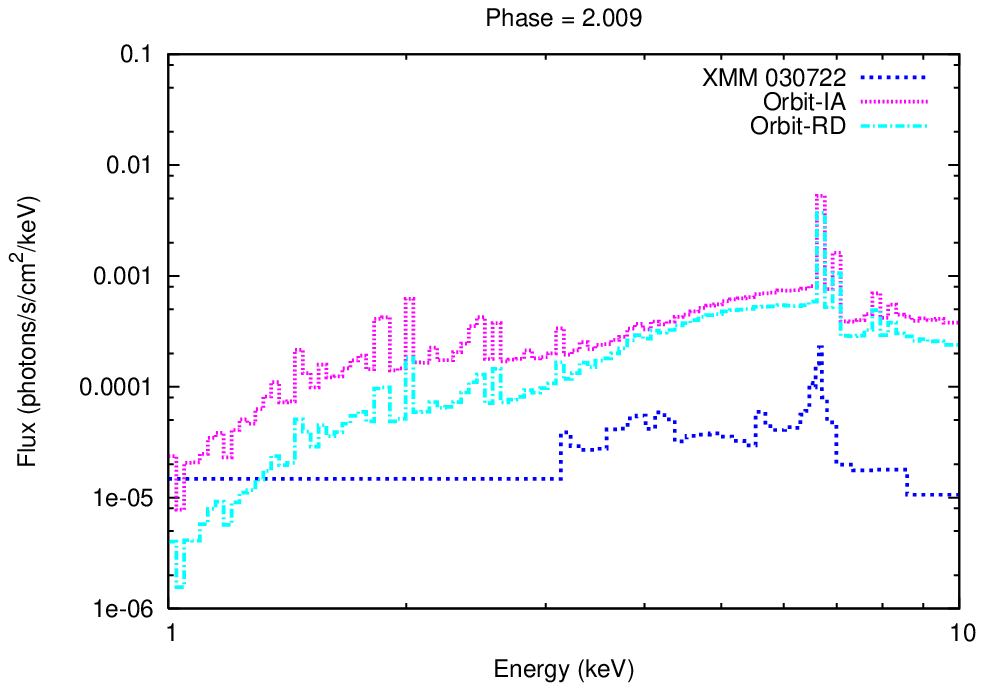}} \\
\resizebox{65mm}{!}{\includegraphics{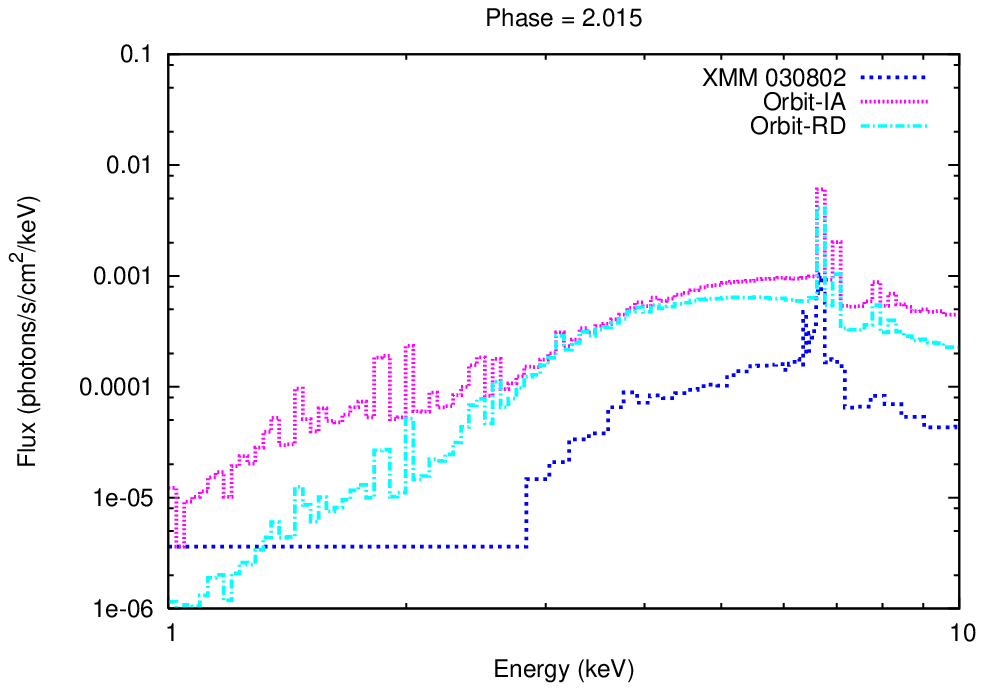}} &
\resizebox{65mm}{!}{\includegraphics{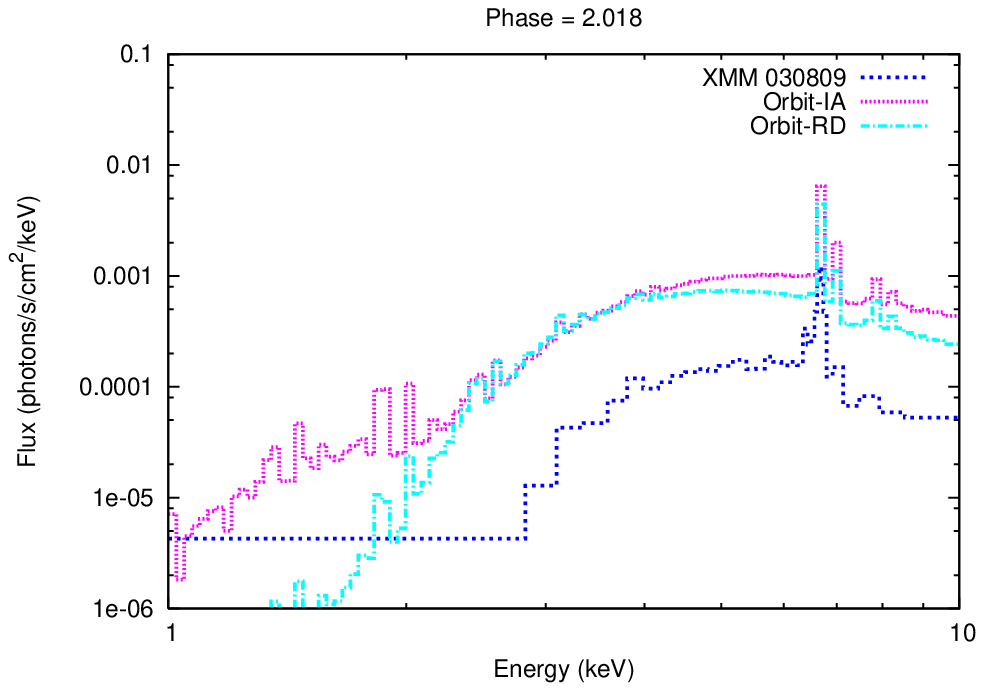}} \\
\resizebox{65mm}{!}{\includegraphics{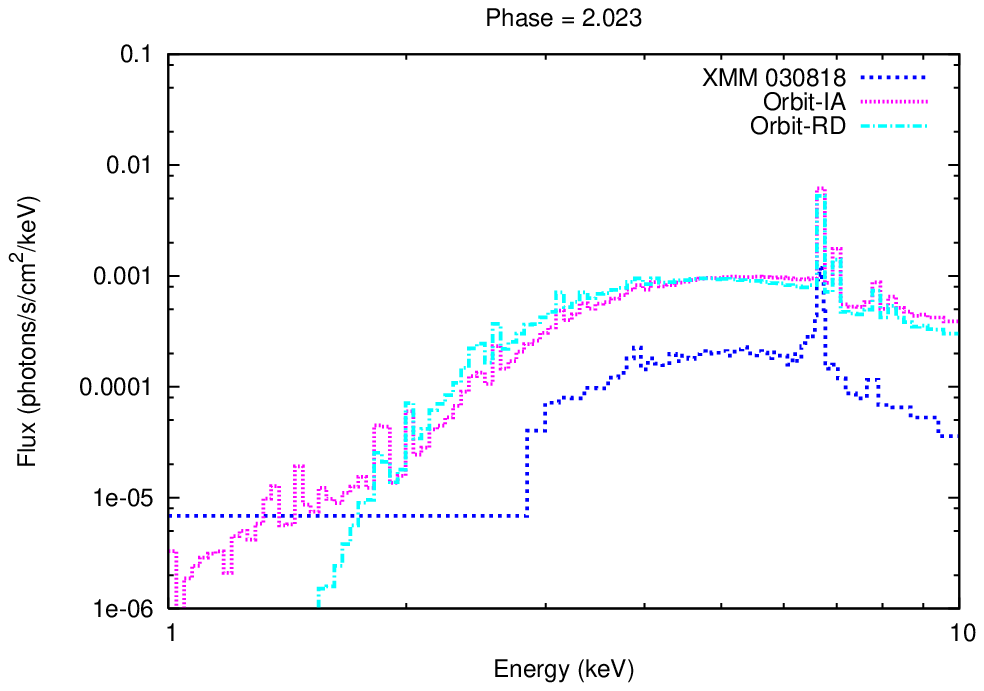}} & \\
    \end{tabular}
    \caption{1-10 keV X-ray spectra from models Orbit-IA and Orbit-RD
      at orbital phase (from top left to bottm right): $\phi =1.924$,
      1.988, 1.990, 2.009, 2.015, 2.018, and 2.023.}
    \label{fig:specs}
  \end{center}
\end{figure*}

X-ray spectra of $\eta\;$Car are particularly useful as they put tight
constraints on the energy dependence of the X-ray emission from the
postshock gas, which constrains the postshock gas temperature and the
preshock velocity. Therefore, the spectra provide valuable clues as to
the variation of the preshock companion's wind speed as a function of
orbital phase.

In Fig.~\ref{fig:specs} we compare the X-ray spectra from models
Orbit-IA and Orbit-RD to \XMM~spectra attained around periastron
passage by \cite{Hamaguchi:2007}. To facilitate our comparison the
non-variable emission components from the outer-ejecta, the X-ray
Homunculus nebula, and the central constant emission component
identified by \citeauthor{Hamaguchi:2007} have been removed, so that
the remaining emission is almost entirely contributed by the WCR. At
$\phi=1.924$ the match between models Orbit-IA and Orbit-RD and the
\XMM~spectrum appears quite good with only a slight overestimation
(underestimation) of the flux at $E\ltsimm 2\;$keV
($E\simeq3-7\;$keV). The relative normalization of the models and the
\XMM~spectra suggests that our adopted companion wind mass-loss rate
is reasonable, whereas the slope of the 7-10 keV flux indicates that a
slightly lower preshock companion wind speed is required at this
phase.

The \XMM~spectra at $\phi=1.988$ and 1.990 lie at the bottom and top
of a flare-like feature, respectively. Therefore, the fact that both
model Orbit-IA and Orbit-RD overestimate the flux level at
$\phi=1.988$ is unsurprising. Ideally, we would then like the models
to slightly underpredict the observed flux level at $\phi=1.990$ so
that the {\it average} flux is approximately correct (i.e. neglecting
the mechanism that produces the flare-like variability in the
lightcurve). However, as the models are unable to reproduce the
flare-like features in the observed lightcurve (Fig.~\ref{fig:lc}), we
take the level of agreement as reasonably good within the scope of the
current investigation, and defer a detailed study of the flare-like
features to future work. 

Comparing the goodness of fit between the \XMM~spectra and the models
presented in this work with the results of \cite{Parkin:2009} we see a
number of interesting details. Firstly, at orbital phases prior to
periastron in cycle 2 ($\phi=1.924, 1.988$, and 1.990)
\citeauthor{Parkin:2009} showed that a reduced preshock companion star
wind speed improved the agreement between the observed spectra and
their model. However, we now find that even without any inhibition of
the preshock companion's wind (model Orbit-IA) the fits to the spectra
are considerably better. This implies that a more accurate description
of the hydrodynamic structure of the WCR around periastron (i.e. the
highly asymmetric temperature distribution in the leading and trailing
arms - see Figs.~\ref{fig:vterm_peri_images} and
\ref{fig:driven_peri_images}) helps to account for the behaviour of
the observed X-ray emission. In addition, our new model with modest
radiative inhibition matches the spectra about as well as model
Orbit-IA. This is because prior to $\phi=0.990$ radiative inhibition
of the preshock companion's wind is actually relatively minor.

The spectra taken at orbital phases during the extended X-ray minimum
of cycle 2 ($\phi=2.009$, 2.015, 2.018, and 2.023) are not well
matched by the models, which overpredict the observed 7-10 keV flux by
roughly an order of magnitude. Also apparent is the fact that the
spectra do not drop off sharply as one tends towards lower
energies. Examining the broadband images at these orbital phases shows
that a significant amount of soft ($E\ltsimm3\;$keV) X-ray emission
originates from the tail of the WCR (see
e.g. Fig.~\ref{fig:bbimages}). In essence, there are two spatially
discrete emission regions: the postshock gas near the apex of the WCR
from which we only observe a significant flux of hard
($E\gtsimm5\;$keV) X-rays due to the obscuration by the primary's
dense wind, and the downstream gas in the tail of the WCR which lies
in the wake behind the companion star. The additional softer X-rays in
model Orbit-IA at $\phi=2.009,$ 2.015, 2.018, and 2.023 compared to
model Orbit-RD is therefore due to the higher level of instabilities
in the tail of the WCR in the case of the former. Such instabilities
perturb the dense layer of postshock primary wind into the path of the
postshock companion's wind and in so doing they help to reheat the gas
through a series of shocks - this is well illustrated by a comparison
of the $\phi=1.020$ snapshots in Figs.~\ref{fig:vterm_peri_images} and
\ref{fig:driven_peri_images}.

\cite{Parkin:2009} found that an improved agreement between the model
and data could be attained at $\phi=2.009$ by either strongly
inhibiting the preshock companion's wind to a velocity of 1600
km~s$^{-1}\;$ (not tested in this work), or via a disruption/collapse
of the WCR. The net effect of these alterations was to reduce the 7-10
keV flux by a factor of roughly 10. Interestingly, a
collapse/disruption of the shock apex may also help to resolve the
problem with the excess emission at $E\ltsimm3\;$keV by later
modifying the conditions in the WCR further downstream. We note again
that although a disruption/collapse of the WCR does not occur in model
Orbit-RD, our adopted formalism for the coupling of the radiation
fields to the opposing wind (Eq.~\ref{eqn:RI2}) provides a lower limit
to the degree of radiative inhibition of the companion's wind by the
primary's radiation field. For instance, if we were instead to adopt
$k_2$ and $\alpha_2$ for the coupling of the companion's wind and the
primary's radiation field the coupling would be stronger, and likewise
the decelerative force would be greater. Therefore, the preshock
companion's wind would be more strongly inhibited \citep[see figure 21
  of ][]{Parkin:2009} and postshock gas would cool more rapidly making
conditions more favourable for instability growth, and thus a
disruption/collapse of the WCR.

\cite{Corcoran:2010} compared {\it Chandra} X-ray spectra taken at
roughly the same point during the 2003.5 and 2009 periastron passages,
i.e. $\phi=2.03\;$and 3.03. Interestingly, the $\phi=3.03$ spectrum
showed an approximately identical shape to that at $\phi=2.03$, albeit
with a factor of $\sim12$ higher flux level. Encouragingly, the models
agree much better with the $\phi=3.03$ spectra and support our
conjecture that the lack of an extended WCR disruption/collapse is
consistent with the recent periastron passage.

\section{Discussion}
\label{sec:discussion}

The simulations presented in \S~\ref{sec:periastron} of static stars
at a separation corresponding to periastron (model Peri-RD) show that
if the companion wind is suppressed sufficiently then the WCR displays
catastrophic instabilities where dense clumps may hit the companion
star. However, when orbital motion is included (\S~\ref{sec:orbit},
model Orbit-RD) the reduction in the preshock companion wind speed
caused by radiative inhibition is smaller and higher preshock
velocities stabilize the WCR. If one compares the results of
simulations Peri-RD and Orbit-RD there appears to be a smoking gun
for explaining the observed extended X-ray minima. The key lies in the
radiative cooling of the postshock companion wind. We have argued that
this can be brought about by the suppression of the preshock companion
wind by radiative inhibition which, given our adopted relatively weak
coupling between the primary's radiation field and the companion's
wind, may not actually be such a hard thing to achieve. For example,
the coupling between the radiation fields and the winds which we have
adopted essentially provides a lower limit to the level of radiative
inhibition which will be felt by the companion's wind (see
\S~\ref{sec:theory} and Fig.~\ref{fig:vel_comparison}). In reality,
the coupling between the radiation fields and the winds may lie
somewhere between the two extremes of each star, and in that case a
greater inhibiting force would be imparted on the preshock companion's
wind by the primary's radiation field \citep{Parkin:2009}. In
addition, observations of massive star winds show them to be
inhomogeneous \citep[e.g.][]{Fullerton:1996, Markova:2005,
  Davies:2005, Moffat:2008}, and the processing of clumps in the WCR
could provide a sufficient density enhancement to trigger radiative
cooling of the postshock companion's wind. Therefore, a useful
direction for future investigations would be to perform simulations
focused on periastron passage which explore a broader range of
radiation-wind couplings and also the influence of inhomogeneous winds
on the postshock gas conditions.

The factors which we mention in relation to triggering a
disruption/collapse of the WCR may also help to explain the observed
flare-like rapid variations in X-ray brightness. \cite{Moffat:2009}
presented separate models based on clumps from the primary's wind
entering the WCR and on an unstable WCR and found that each model had
its fair share of difficulties reproducing the observations. Based on
the results of models Peri-IA and Peri-RD, i.e. oscillations in the
WCR causing rapid variations in the X-ray flux
(Fig.~\ref{fig:peri_lx}), re-examining the combined r\^{o}le of
instabilities and wind clumping using hydrodynamical models may be
worthwhile. Furthermore, although relatively small clumps can be
efficiently destroyed in largely adiabatic WCRs \citep{Pittard:2007},
this will not be the case for a radiative WCR. Large clumps traversing
a highly unstable WCR consisting of radiative shocks may cause much
greater variability, and could therefore account for the onset of
large amplitude flare-like features as periastron is approached.

In model Peri-RD the clumps which collide with the companion star are
removed from the computational domain when the wind is initiated after
every time step. In reality, these clumps might disrupt the generation
of the wind. While aspects of this scenario are reminiscent of the
model proposed by \cite{Soker:2005}, there are some crucial
differences. In the \citeauthor{Soker:2005} model the clumps originate
in the preshock primary wind, whereas in model Peri-RD the clumps are
formed by the fragmentation of the dense layer of postshock primary
wind by the NTSI. Furthermore, in our work clumps do not traverse the
shocks due to gravitational attraction, but instead are catapulted by
violent instabilities. Since in our simulations the preshock stellar
winds are modelled as homogeneous, we cannot directly test
\citeauthor{Soker:2005}'s model \citep[although see][for a discussion
  of the ablation of clumps in postshock flow]{Walder:2002,
  Pittard:2007, Parkin_Pittard:2010}.

Ideally, one would like to resolve the region around the stars such
that the critical point radius of the wind acceleration region is well
sampled. It is, however, a computationally demanding task to resolve
the highly eccentric orbit of the stars and also the critical point
radius (for the companion star this was $\sim1.03\;R_{\ast 2}$). It
may therefore be advantageous in future models to adopt a sub-grid
treatment of wind accretion which modifies the driving of the
winds. For instance, ``switching off'' the initiation of the wind over
those parts of the stellar surface in which the ram pressure of
incident clumps exceeds that of the wind.

Photo-ablation by the stellar radiation fields clearly has important
consequences for the fate of cool ($\sim 10^{4}\;$K), dense postshock
primary wind. With the stars in \etacar being so immensely luminous,
photo-ablation seems inevitable. However, in this work we adopted the
\cite{Castor:1975} formalism for radiative driving, whereby the
radiation-wind coupling is calibrated by the stars ability to drive
its own wind. Therefore, the quantitative accuracy of future models
would benefit from the use of a more suitable treatment of
photo-ablation of circumstellar gas.

\section{Conclusions}
\label{sec:conclusions}

Three dimensional hydrodynamical simulations of the enigmatic
super-massive binary star system \etacar have been presented which
include the radiative driving of the stellar winds, radiative cooling,
gravity, and orbital motion. A suite of simulations were then
performed to explore the r\^{o}le of wind acceleration, interacting
radiation fields, and instabilities in the WCR on the gas dynamics and
resulting X-ray emission. We summarize our key conclusions as follows:
\begin{itemize}
\item When one separates the coupling between the radiation fields of
  the stars and the opposing wind the level of radiative inhibition of
  the companion's wind is reduced. Consequently, in static-star
  calculations with a periastron separation the companion's wind
  attains a preshock velocity of $\simeq 2200\;{\rm km~s^{-1}}$ -
  notably higher than the estimate of $\simeq 1500\;{\rm km~s^{-1}}$
  when the radiation-wind couplings are not separated
  \citep[][]{Parkin:2009}.
\item Despite a reduced level of radiative inhibition, when the wind
  acceleration regions are considered (in static-star calculations
  with a periastron separation) the WCR becomes massively disrupted by
  non-linear thin shell instabilities (NTSIs). This occurs because the
  lower preshock companion wind speed acquired due to wind
  acceleration and radiative inhibition leads to a lower postshock gas
  temperature. As such, radiative cooling becomes important in the
  postshock companion's wind and thermal pressure - which prevents the
  growth of the NTSI - is lost. The WCR is distorted to such an extent
  that dense fragments of cool postshock primary wind are repeatedly
  driven deep into the companion's wind acceleration region and in
  some instances collide against the star.
\item When orbital motion is included the catastrophic disruption of
  the WCR is not reproduced. The root cause of this difference lies in
  the rapid orbital motion of the stars around periastron which acts
  to increase the preshock velocity, and thus postshock pressure, of
  the companion's wind. In so doing the stability of the WCR against
  thin-shell oscillations is increased. The influence of orbital
  motion will depend on the adopted system parameters.
\item Large-scale, high resolution simulations show a number of
  interesting dynamical effects, including the influence of the
  stellar radiation fields on the growth of instabilities at the
  shocks, the photo-ablation of the cold dense layer of postshock
  primary wind, and the highly asymmetric temperature distribution in
  the arms of the WCR.
\item The models provide a reasonable match to the majority of the
  {\it RXTE} lightcurve and, compared to previous models, the X-ray
  spectra agree much better with {\it XMM-Newton} observations
  obtained just prior to periastron. However, the extended X-ray
  minima are not reproduced by model Orbit-RD (which uses a
  relatively weak coupling between the primary's radiation field and
  the companion's wind) and the 7-10 keV X-ray emission is
  overestimated by roughly an order of magnitude. Yet, when the shock
  is heavily disrupted (in the static-stars simulation), the 7-10~keV
  X-ray luminosity of the postshock gas is an order of magnitude lower
  than the undisrupted case. This shows that dynamical instabilities
  in the WCR could be a key mechanism in explaining the X-ray
  observations.
\item From a comparison between the model and {\it XMM-Newton} column
  densities around the 2003.5 periastron passage, a reduction of the
  stellar wind mass-loss rates by a factor of $\sim2\;$is implied. The
  revised mass-loss rates are $\simeq2.4\times10^{-4}\Msolpyr$ and
  $(7-12)\times10^{-6}\;\Msolpyr$ for the primary and companion wind,
  respectively. However, these estimates do not account for the
  possible presence of highly ionized, and thus lower opacity, gas
  close to the stars, and seem unlikely given other evidence for
  mass-loss rates greater than those assumed here \citep{Hillier:2001,
    Hillier:2006, vanBoekel:2003, Groh:2010b}. Alternatively, it may
  be that the observationally derived column densities around
  periastron are subject to uncertainty due to fitting simpler plasma
  models to complicated systems \citep[see][]{Pittard_Parkin:2010},
  and/or the process of disentangling different spectral components.
\item To reproduce the extended minima likely requires effective
  radiative cooling in the postshock companion's wind close to the
  apex of the WCR. This could be achieved by stronger radiative
  inhibition of the preshock companion's wind \citep{Parkin:2009},
  albeit with a stronger radiation-wind coupling than explored in this
  work. If the companion's wind is sufficiently suppressed, or
  oscillations in the WCR are especially vigorous, a catastrophic
  disruption/collapse of the WCR onto the companion star may occur.
\end{itemize}

We close with a note that the differences between the recent X-ray
minimum and those previously observed could be accounted for by the
stochastic nature of the growth of non-linear instabilities in the
WCR. For example, we can speculate that the growth of instabilities in
the 1998 and 2003 minima sufficiently disrupted the WCR to result in
an extended quenching of the 7-10~keV X-ray emission. In contrast,
during the shorter 2009 minimum the WCR was contorted by instabilities
to a lesser degree, perhaps due to less significant radiative cooling
of the postshock companion's wind, so that ultimately an extended
disruption and/or collapse of the WCR against the companion star did
not occur. This line of reasoning suggests that the differences
between the most recent and the previous X-ray minima may be divided
by a fine line related to the postshock gas conditions, and is an
intriguing possibility given that multiwavelength observations suggest
recent changes in the wind of \etacar~\citep{Martin:2010,
  Mehner:2010b, Corcoran:2010}.

\subsection*{Acknowledgements}
We thank Ian Stevens and Sven van Loo for helpful discussions, and the
referee for useful suggestions which improved the presentation of this
work. This work was supported in part by a Henry Ellison Scholarship
from the University of Leeds, and by a PRODEX XMM/Integral contract
(Belspo). JMP thanks the Royal Society for funding. This research has
made use of NASA's Astrophysics Data System. This research has made
use of data obtained from the High Energy Astrophysics Science Archive
Research Centre (HEASARC) provided by NASA's Goddard Space Flight
Center. Some of the software used in this work was in part developed
by the DOE-supported ASC/Alliance Center for Astrophysical
Thermonuclear Flashes at the University of Chicago. We thank the White
Rose Grid and the UK National Grid Service (NGS) for use of their
computer facilities.


\appendix
\section{Initiating the instantaneously accelerated winds}
\label{sec:remap}

In simulations where the winds are instaneously accelerated we do not
need to resolve the wind acceleration region. Therefore, the only
resolution requirement comes from having a sufficent number of cells
between the stars to accurately model the WCR dynamics. With this in
mind, computational resources can be saved by making the radial
distance into which the winds are initiated $R_{\rm map}$ a function
of the separation of the stars,
\begin{equation}
  R_{\rm map} = R_{\rm peri} \times 2^{l_{\rm peri} - l_{\phi}}
\end{equation}
\noindent where $R_{\rm peri}$ is a radial distance corresponding to
approximately 10\% of the distance between the stars at periastron
($\ge 6\;$cells), $l_{\phi}$ is the number of nested levels of
refinement required at a given orbital phase to accurately model the
flow dynamics between the stars, and $l_{\rm peri}=l(\phi = 0.)$.

\section{Refinement condition for postshock gas}
\label{sec:refinement}

To ensure that postshock gas is sufficiently well resolved to
accurately describe the WCR dynamics we have implemented an additional
criterion for grid refinement. Given the desired number of cells
between the stars, $n_{\rm sep}$, the number of nested levels of
refinement required at a given orbital phase
\begin{equation}
  l_{\phi} = \log_{2} \left(\frac{C_{\rm req}}{C_{\rm base}}\right),
\end{equation}
\noindent where $C_{\rm req}= d_{\rm sep}(\phi)/n_{\rm sep}$ is the
resolution required, $d_{\rm sep}(\phi)$ is the separation of the
stars, and $C_{\rm base}$ is the coarse grid resolution. For models
Orbit-RD and Orbit-IA, $n_{\rm sep}\simeq 200\;$cells.

\section{Estimating the influence of orbital acceleration on the wind}
\label{sec:orbital_motion}

\begin{figure}
  \begin{center}
    \begin{tabular}{cc}
\resizebox{80mm}{!}{\includegraphics{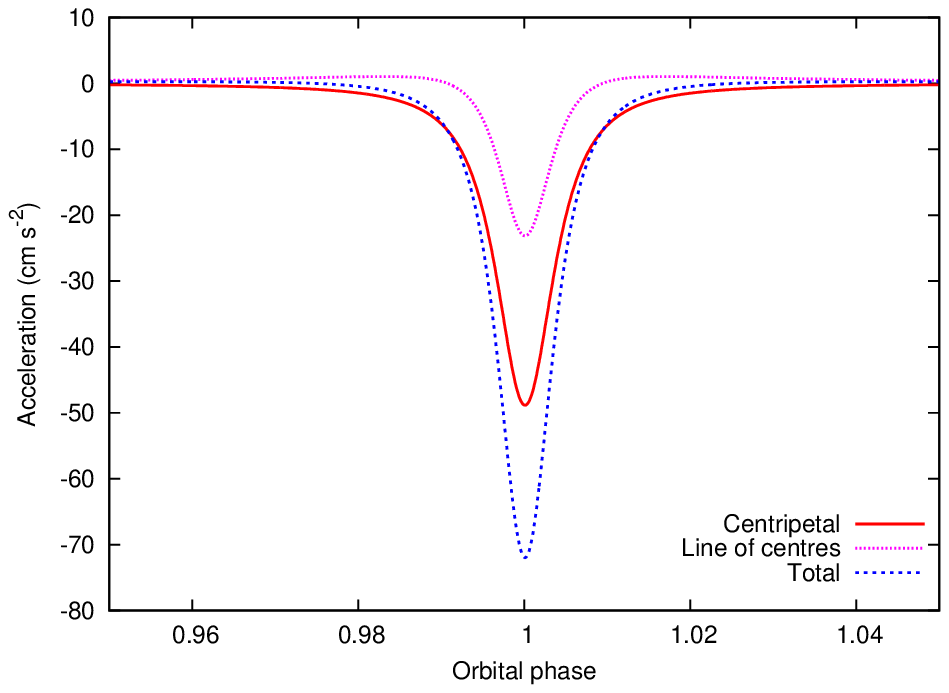}} &
\resizebox{80mm}{!}{\includegraphics{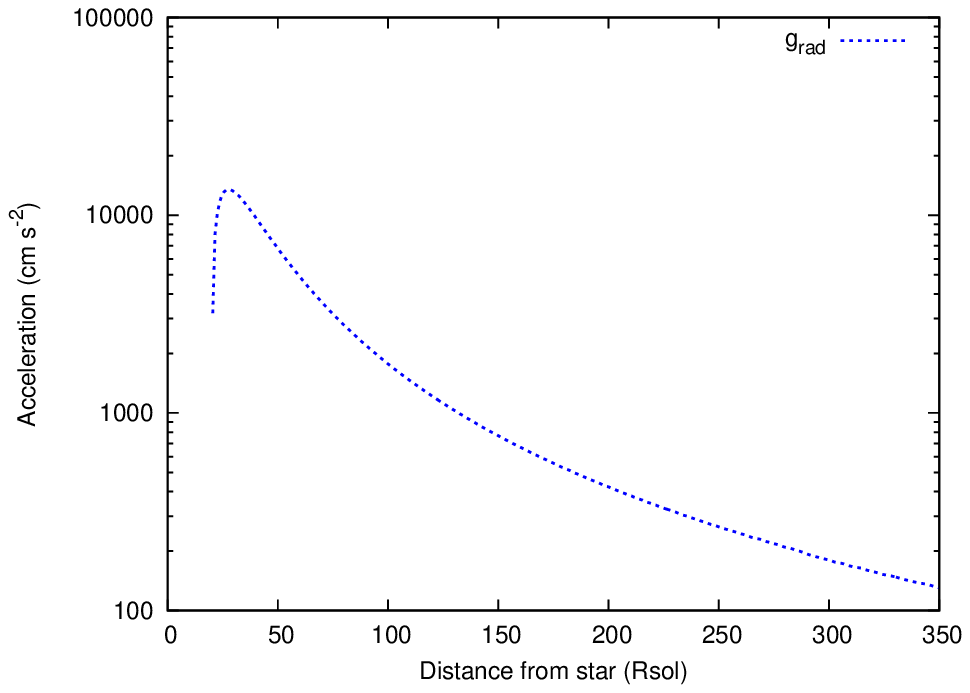}} \\
    \end{tabular}
    \caption{The orbital acceleration of the companion star around
      periastron passage (left panel) and the radiative line force
      calculated for the companion's single star wind profile (right
      panel). The components for the centripetal and line-of-centres,
      as well as the total are shown in the left panel. Note that
      these values are calculated in the frame of reference of the
      companion star (c.f. Fig~\ref{fig:vels_peri}).}
    \label{fig:gorbs_peri}
  \end{center}
\end{figure}

The motion of a star in an eccentric orbit introduces centripetal,
$g_{\rm cen}$, and line-of-centres, $g_{\rm loc}$,
accelerations. Considering the flow along the line of centres,
\begin{equation}
  g_{\rm cen} = \frac{v_{\rm orb}^2}{R} = (1 + e\cos \omega) \left(
  \frac{M_{\rm orb}}{a(1 -e^{2}) } \right)^{1/2}, \label{eqn:gcen}
\end{equation}
\noindent and,
\begin{equation}
  g_{\rm loc} = \frac{v_{\rm orb}}{R}\frac{2 \pi a e \cos \omega}{P \sqrt{1 - e^{2}}}, \label{eqn:gloc}
\end{equation}
\noindent where $v_{\rm orb}$ is the orbital velocity, $R$ is the
radial distance to the system centre of mass ($=a(1 -e^2)/(1 +
e\cos\omega)$), $\omega$ is the true anomaly ($= 2\pi \phi$), $M_{\rm
  orb}$ is the mass term relevant to the particular orbit under
consideration (e.g. for the barycentric orbit of the companion star
$M_{\rm orb} = G M_1^3/(M_1 + M_2)^2$), and $P$ is the orbital
period. The left panel of Fig.~\ref{fig:gorbs_peri} shows $g_{\rm
  cen}$ and $g_{\rm loc}$ evaluated for the companion star of
$\eta\;$Car. Comparing the acceleration due to orbital motion against
the radiative line force calculated for the companion star one sees
that the latter has a higher magnitude close to the star where the
mass-loss rate is set (right panel of
Fig.~\ref{fig:gorbs_peri}).

\begin{figure}
  \begin{center}
    \begin{tabular}{c}
\resizebox{80mm}{!}{\includegraphics{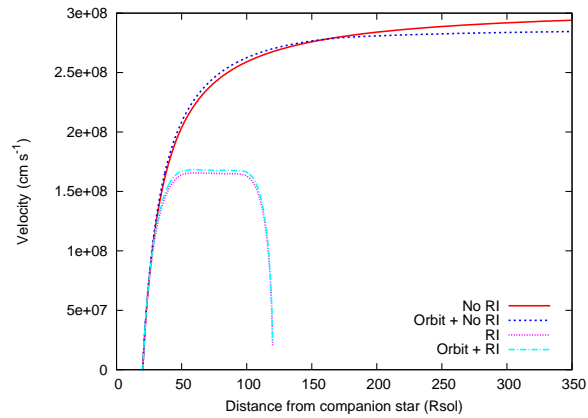}} \\
    \end{tabular}
    \caption{Companion star wind velocity along the line of centres
      between the stars. The primary star is situated at a distance of
      359\Rsol (corresponding to periastron separation when
      $e=0.9$). The following solutions are shown: without radiative
      inhibition and orbital motion (No RI)(i.e. single star), without
      RI and with orbital motion (Orbit + No RI), with RI and without
      orbital motion (RI), and with RI and orbital motion (Orbit +
      RI). Dynamically consistent FDCFs have been used in these
      calculations (c.f. Eqs.~\ref{eqn:FDCF_RI1} and
      \ref{eqn:FDCF_RI2}). The orbital parameters are noted in
      Table~\ref{tab:system_parameters}, and stellar and line driving
      parameters are noted in Table~\ref{tab:stellar_parameters}.}
    \label{fig:ri_orbit}
  \end{center}
\end{figure}

The influence of the orbital acceleration on the mass-loss rate and
velocity profile of the wind can be estimated by incorporating $g_{\rm
  cen}$ and $g_{\rm loc}$ into the equation of motion
\begin{equation}
  F(r, v, dv/dz) = \left(v - \frac{s^2}{v}  \right)\frac{dv}{dr} + \frac{d \psi}{dr} - \frac{2 s^2}{r} - g_{\rm rad} - g_{\rm orb},
\end{equation}
\noindent where $r$ is the distance from the centre of the star along
the line-of-centres, $s$ is the speed of sound, $\psi$ is the
gravitational potential, and $g_{\rm orb}= g_{\rm cen} + g_{\rm loc}$
is the orbital acceleration. The radiative line force, $g_{\rm rad}$
takes the form of Eq.~\ref{eqn:RI}. The value of $v_{\rm orb}$ at a
given radius will depend on the time at which that parcel of gas left
the star. However, as we are mainly interested in the affect of
orbital motion on the base of the wind (i.e. where we initiate the
winds in the hydrodynamic simulations), we neglect the radial
dependence of $v_{\rm orb}$, and take $v_{\rm orb}(r) \approx v_{\rm
  orb}$. As the orbital acceleration is applied to the base of the
wind we replace the $R$ in the denominator of Eqs.~\ref{eqn:gcen} and
\ref{eqn:gloc} with $R - R_{\ast}$. Note that we do not use $R - r$ in
the denominator as this would introduce an unrealistic singularity at
the centre of mass of the binary system. To proceed, we make the
coordinate transform using the substitution of variables
\citep{Abbott:1980},
\begin{equation}
  u = \frac{-2 GM_1(1 - \Gamma_1)}{r s^2};      w = \frac{v^2}{s^2};      w' = r^2 v \frac{dv}{dr} [G M_1 (1 -\Gamma_1)]^{-1},
\end{equation}
\noindent leading to
\begin{equation}
  F(u, w, w') = \left(1 - \frac{1}{w} \right)w' + h(u) - C B(u,w,w')w'^{\alpha},
\end{equation}
\noindent where
\begin{equation}
  B(u,w,w') = K_1(u,w,w') - \left(\frac{M_2 \Gamma_2}{M_1 \Gamma_1} \right)K_{2}(u,w,w')A(u),
\end{equation}
\noindent and
\begin{equation}
 A(u) = \left(\frac{u_{\rm d}}{u - u_{\rm d}} \right)^2
\end{equation}
\noindent with $C$ given by equation (13) of \cite{Stevens:1994}. The
FDCFs
\begin{equation}
  K_{1}(u,w,w') = \frac{1 - \left[1 - \left(\frac{u}{u_{\ast}}\right)^2 - \frac{2 w}{w' u}\left(\frac{u}{u_{\ast}}\right)^2\right]^{1 + \alpha}}{(1 + \alpha)\left[1 + \frac{2 w}{w' u}\right]\left(\frac{u}{u_{\ast}}\right)^{2}}, \label{eqn:FDCF_RI1}
\end{equation}
\noindent and 
\begin{equation} 
K_{2}(u,w,w') = \frac{1 - \left[1 - \left(\frac{u}{u_{\ast}}\right)^2A(u) + \frac{2 w}{w' u} \left(\frac{u}{u_{\ast}}\right)^2 A(u)^{3/2}\right]^{1 + \alpha}}{(1 + \alpha)\left[1 - \frac{2 w}{w' u}A(u)^{1/2}\right]\left(\frac{u}{u_{\ast}}\right)^2 A(u)} \label{eqn:FDCF_RI2}
\end{equation}
\noindent where $u_{\rm d} = u(d_{\rm sep})$, $u_{\rm \ast} =
u(R_{\ast})$. The orbital motion terms, which only depend on $r$, are
contained in the potential function, 
\begin{equation}
  h(u) = 1 + \frac{4}{u} - \frac{M_2(1 - \Gamma_2)}{M_1(1 - \Gamma_1)}A(u) + \frac{1}{u^2}\frac{u_{\rm orb}u_{\ast}}{u_{\ast} - u_{\rm orb}} \zeta,
\end{equation}
\noindent with 
\begin{equation}
  \zeta = \frac{2}{s^2} R g_{\rm orb} = \frac{2 }{s^2}\left(\frac{2 \pi a e v_{\rm orb} \cos \omega}{P \sqrt{1 - e^2}} + v_{\rm orb}^2   \right),
\end{equation}
\noindent where $u_{\rm orb} = u(R)$. The critical point of the wind
profile is solved for using: i) the equation of motion $F(u,w,w')=0$,
ii) the singularity condition $\partial F/ \partial w' = 0$, and iii)
the regularity condition $\partial F/ \partial u + w'\partial F/
\partial w = 0$. The velocity gradient terms present in the FDCFs
(Eqs.~\ref{eqn:FDCF_RI1} and \ref{eqn:FDCF_RI2}) require the solution
for the wind profile to be attained using an iterative process,
whereby monotonic FDCFs are used to attain an initial solution, then
velocity gradient terms are included in subsequent iterations
\citep{Pauldrach:1986}.

Fig.~\ref{fig:ri_orbit} shows wind solutions for $\eta~$Car
with/without radiative inhibition and orbital motion, and calculated
with a stellar separation corresponding to periastron. In these
calculations $g_{\rm rad}$ has the form of Eq.~\ref{eqn:RI}, whereas
in the hydrodynamic simulations $g_{\rm rad}$ has the form of
Eq.~\ref{eqn:RI2}. Hence, these calculations are intended to
illustrate the qualitative effect of orbital acceleration on the
wind. The inclusion of orbital motion slightly steepens the velocity
profile close to the star. This is the opposite effect to stellar
rotation, which causes a shallower velocity profile
\citep{Abbott:1980}. Interestingly, neglecting radiative inhibition,
the terminal wind velocity is lower when orbital motion is
included. However, when radiative inhibition is included, and
recalling that the WCR occurs at $\simeq 100\Rsol$ from the companion
star (at periastron and along the line of centres) in model Orbit-RD,
there is a minor increase in the preshock velocity. Examining the
critical point radius and mass-loss rate we find reduction factors of
0.12 \% and 2.3 \%, respectively. Therefore, orbital acceleration
contributes a negligible difference in comparison to the interplay
between the radiation fields and the gravity of the stars.

\label{lastpage}


\end{document}